\documentclass[%
 reprint,
nofootinbib,
 amsmath,amssymb,
 aps,
]{revtex4-2}

\usepackage[caption=false]{subfig}
\usepackage{graphicx}
\usepackage{dcolumn}
\usepackage{bm}
\usepackage[normalem]{ulem}
\usepackage{physics}
\usepackage{verbatim}
\usepackage{amsxtra}
\usepackage{stmaryrd}
\usepackage{mathrsfs}
\usepackage{psfrag}
\usepackage{rotating}
\usepackage{setspace}
\usepackage{soul}
\usepackage{url}
\usepackage{amsfonts}
\usepackage{amsbsy}
\usepackage{amscd}
\usepackage{amsthm}
\usepackage{color}
\usepackage{braket}
\usepackage{esint}
\definecolor{hellgruen}{rgb}{0.2,0.7,0.2}
\newcommand{\cb}{\color{black}}
\newcommand{\cred}{\color{black}}
\newcommand{\cn}{\color{black}}

\newcommand{\fintd}{\fint \displaylimits}

\newcommand{\bx}{\boldsymbol{\textbf{x}}}
\newcommand{\bxprime}{\boldsymbol{\textbf{x}^{\prime}}}

\newcommand{\bI}{\boldsymbol{I}}

\newcommand{\bX}{\boldsymbol{X}}

\newcommand{\bR}{\boldsymbol{R}}

\newcommand{\bhatH}{\boldsymbol{\hat{\textbf{H}}}}
\newcommand{\bbreveH}{\boldsymbol{\breve{\textbf{H}}}}

\newcommand{\bGam}{\boldsymbol{\Gamma}}
\newcommand{\bGamtilde}{\boldsymbol{\Gamma}_{\textrm{proj}}}
\newcommand{\bGamtildekappa}{\boldsymbol{\Gamma}_{\textrm{proj},{\textbf{k}}}}
\newcommand{\bGamtildesigma}{\boldsymbol{\Gamma}_{\textrm{proj},\sigma}}
\newcommand{\bGambreve}{\boldsymbol{\breve{\Gamma}}}

\newcommand{\Psitilderi}{{\widetilde{\Psi}}_{{\bf r},\,i}}

\newcommand{\bPsitildef}{\boldsymbol{\widetilde{\Psi}_{\bf f}}}

\newcommand{\bPsitildeR}{{\boldsymbol{\widetilde{\Psi}}}_{\bf r}}

\newcommand{\bPsitildeo}{{\boldsymbol{\widetilde{\Psi}}}_{\bf o}}

\newcommand{\bPsitildeoDagger}{{{\boldsymbol{\widetilde{\Psi}}}_{\bf o}}^{\dagger}}

\newcommand{\bPsitildeRsigma}{{\boldsymbol{\widetilde{\Psi}}}_{\bf r,\,\sigma}}
\newcommand{\bPsitildeRDagger}{{{\boldsymbol{\widetilde{\Psi}}}_{\bf r}}^{\dagger}}
\newcommand{\bPsitildeRsigmaDagger}{{{\boldsymbol{\widetilde{\Psi}}}_{\bf r,\,\sigma}}^{\dagger}}

\newcommand{\btH}{\boldsymbol{\widetilde{\textbf{H}}}}

\newcommand{\innerproductcomma}[2]{\langle #1, #2 \rangle}

\newcommand{\QE}{\texttt{QE}~}
\newcommand{\DFTFE}{\texttt{DFT-FE}~}

\begin{document}

\preprint{APS/123-QED}

\title{Accelerating self-consistent field iterations in Kohn-Sham density functional theory using a low rank approximation of the dielectric matrix}

\author{Sambit Das}
\affiliation{%
Department of Mechanical Engineering,\\ University of Michigan, Ann Arbor, USA
}%


\author{Vikram Gavini}
\email{vikramg@umich.edu}
\affiliation{%
Department of Mechanical Engineering, and \\
Department of Materials Science and Engineering,\\ University of Michigan, Ann Arbor, USA
}



\begin{abstract}

We present an efficient preconditioning technique for accelerating the fixed point iteration in real-space Kohn-Sham density functional theory (DFT) calculations. The preconditioner uses a low rank approximation of the dielectric matrix (LRDM) based on  G\^ateaux derivatives of the residual of fixed point iteration along appropriately chosen direction functions. We develop a computationally efficient method to evaluate these G\^ateaux derivatives in conjunction with the Chebyshev filtered subspace iteration procedure, an approach widely used in large-scale Kohn-Sham DFT calculations. Further, we propose a variant of LRDM preconditioner based on adaptive accumulation of low-rank approximations from previous SCF iterations, and also extend the LRDM preconditioner to spin-polarized Kohn-Sham DFT calculations. We demonstrate the robustness and efficiency of the LRDM preconditioner against other widely used preconditioners on a range of  benchmark systems with sizes ranging from $\sim$ 100-1100 atoms ($\sim$ 500--20,000 electrons). The benchmark systems include various combinations of metal-insulating-semiconducting heterogeneous material systems, nanoparticles with localized $d$ orbitals near the Fermi energy, nanofilm with metal dopants, and magnetic systems. In all benchmark systems, the LRDM preconditioner converges robustly within 20--30 iterations. In contrast, other widely used preconditioners show slow convergence in many cases, as well as divergence of the fixed point iteration in some cases. Finally, we demonstrate the computational efficiency afforded by the LRDM method, with up to 3.4$\times$ reduction in computational cost for the total ground-state calculation compared to other preconditioners.

\end{abstract}

\maketitle


\section{Introduction}
\label{sec:intro}
Electronic-structure calculations based on Kohn-Sham density functional theory (KS-DFT)~\cite{kohn65,kohn96} provide an excellent balance between accuracy and computational efficiency by reducing the many-body Schr\"odinger problem of interacting electrons into an equivalent problem of non-interacting electrons in an effective mean field that is governed by the electron-density.  This has led to KS-DFT being one of the most widely used electronic-structure method for the predictive modelling of materials and for gaining qualitative and quantitative insights into various materials properties. The significant increase in the computational resources over the last decade, including the advent of hybrid CPU-GPU architectures, has also played an important role in the wide adoption of KS-DFT. Furthermore, the simultaneous development of efficient and scalable numerical schemes in conjunction with systematically convergent real-space discretizations~\cite{zhou2006,RS-DFTcode,genovese2011daubechies,motamarri2020,SPARC,das2022dft}, including reduced-order scaling approaches (cf.~e.g.~\cite{Bowler, PEXSI, motamarri2014, lin2021a}), have advanced the ability to conduct fast and accurate DFT calculations using large-scale computing platforms. As a result, applications using KS-DFT are increasingly targeting larger as well as more complex heterogeneous material systems~\cite{norskov2011density,fernando2015quantum,cole2016applications,pham2017modelling,mis2018,das2019,motamarridna2020,ghosh2021,yao2022modulating}. However, existing numerical methods to solve the Kohn-Sham equations suffer from instabilities for heterogeneous systems, with the convergence worsening for larger system sizes. To elaborate, the ground-state solution in KS-DFT is often computed via the solution of the non-linear Kohn-Sham eigenvalue problem, that is  posed as a fixed point iteration---commonly referred to as the self-consistent field (SCF) iteration---in the electron-density, written as $\rho=F[V_{\textrm{eff}}[\rho]]$, where $\rho$ denotes the electron-density and $V_{\textrm{eff}}[\rho]$ is the Kohn-Sham effective mean field potential. As each step in the fixed point iteration  involves the computation of Kohn-Sham eigenstates that scales cubically with the number of electrons ($N_e$), the slow convergence is a serious computational bottleneck for large-scale DFT calculations. As will be discussed below, existing state-of-the-art methods for accelerating the Kohn-Sham fixed point iteration are either not suitable to generic heterogeneous material systems, or incur significant computational overheads.

The origin of the instabilities in the Kohn-Sham SCF iteration is due to the large condition number of the Jacobian operator corresponding to the residual of the fixed point iteration. The Jacobian operator, denoted by $J=\frac{\delta}{\delta \rho}\left(F[V_{\textrm{eff}}[\rho]]-\rho\right)$, is related to the physical dielectric operator of the material system, $J=-\epsilon^{\dagger}$.  Methods such as the HIJ method~\cite{Ho1982} and the extrapolar preconditioner~\cite{Anglade2008} have been proposed to directly compute $\epsilon$ that allow efficient convergence of the SCF iteration. However, the associated quartic-scaling computational cost of computing $\epsilon$  limit their application to small system sizes of $\sim $ 100 atoms~\cite{Anglade2008}. The quartic-scaling cost arises from the computation of the static susceptibility matrix ($\chi_0$), a portion of $\epsilon$, using the Alder-Wiser expression~\cite{Alder1962,Wiser1963} that involves a double summation over all occupied and many unoccupied eigenstates for each matrix element of $\chi_0$. Thus, DFT codes have primarily relied on cheaper quasi-Newton techniques with an approximation of $J$ as a preconditioner to accelerate the SCF iteration. Broadly, two groups of such methods have found wide usage in the DFT community, one using direct numerical approximation of $J$ from the history of previous SCF iterations and the other based on physically motivated approximations of $\epsilon$.  Concerning the former group of methods, the most widely used techniques are Anderson~\cite{anderson1965}, Pulay~\cite{PULAY1980393}, Broyden~\cite{Broyden1988} and DIIS~\cite{Kudin2002}. These techniques can be generically considered as multisecant approximations of $J$ or $J^{-1}$~\cite{fang2009two}, where the preconditioner is constructed to optimally satisfy the secant approximation of $J$ at multiple steps, simultaneously, based on a history of $\rho$ and $F[V_{\textrm{eff}}[\rho]]$ from previous SCF iterations. Further, such schemes have been shown to behave like Krylov subspace methods with Q-superlinear convergence near the ground-state solution~\cite{Dederichs1983,saad2003iterative}. However, as shown in previous studies~\cite{LinLin2013,Herbst_2020}, for large metallic systems and  heterogeneous systems with large condition numbers of $J$, multi-secant approaches demonstrate slow and system-size dependent convergence. These issues can be further compounded by potential strong nonlinearites in $F[V_{\textrm{eff}}[\rho]]$ in heterogeneous systems, resulting in divergence of the SCF iteration as will be demonstrated in this work. Given the limitations of the multi-secant methods, several physically motivated approximations of $\epsilon$ have been proposed, and they are typically combined with the aforementioned multi-secant methods such as Anderson, Pulay or Broyden. The Kerker preconditioner~\cite{Kerker1981} is one such widely used approximation, that is based on the Thomas-Fermi screening theory of homogeneous electron gas. Although Kerker preconditioner captures the long wavelength divergent eigenvalues of $\epsilon$ in bulk metallic systems, it is not suitable for semiconducting and insulating systems as it does not model the screening behavior in these systems. In order to better capture the screening effects in semiconducting and insulating systems, the Resta~\cite{Resta1977} and truncated-Kerker~\cite{Kresse1996} preconditioners have been proposed based on material specific parameterizations related to the static dielectric constant. However, these preconditioners are still not suitable for heterogeneous materials systems with spatially varying screening behavior. \cb We refer to ~\cite{woods2019computing} for a more in-depth review of the numerical convergence aspects of the above preconditioners, and their systematic comparison on robustness and efficiency measures assesed on a test suite of benchmarks that encapsulates various sources of ill-conditioning of the Kohn-Sham SCF iteration.\cn 

In order to address the aforementioned challenges posed by heterogeneous systems, preconditioners such as the Thomas-Fermi-von Weizsacker preconditioner (TFW)~\cite{Raczkowski2011}, elliptic preconditioner~\cite{LinLin2013} and local density of states (LDOS) based preconditioner~\cite{Herbst_2020} have recently been developed. The TFW preconditioner approximates the $\chi_0$ portion of $\epsilon$, relying on the equivalence between $\chi_0$ and the inverse of the double functional derivative of the non-interacting kinetic energy functional ($T_s[\rho]$) and using the TFW approximation for $T_s$. However, TFW functional approximation cannot accurately capture the complex dielectric response in general heterogeneous systems due to the semi-local nature of the functional, and thus limits its suitability to simpler metal-vacuum systems.  In the case of elliptic and LDOS preconditioners, a key limitation is that they consider only long-range eigenmodes of $\chi_0$ in the construction of the preconditioner. This prevents the elliptic and LDOS preconditioners from appropriately accounting for the eigenmodes of $\chi_0$ with large eigenvalues related to localized states near the Fermi energy~\cite{Herbst_2020} that would require resolving eigenmodes with atomic-scale wavelengths. Such localized states near the Fermi energy can occur for metallic elements with d and f valence orbitals. Further, strong nonlinearties in $F[V_{\textrm{eff}}[\rho]]$ are not accounted for in these preconditioners. \cred  In addition to the above discussed preconditioners for the Kohn-Sham fixed point iteration map, methods to directly minimize the Kohn-Sham finite-temperature free energy functional over the Kohn-Sham orbitals and fractional occupancies have also been developed, for example, the ensemble DFT method~\cite{marzari1997ensemble}. Due to its improved global convergence, in general, over fixed point iteration based methods, ensemble DFT has been demonstrated to converge robustly for challenging SCF problems. However, the computational cost associated with direct minimization can be substantially larger than fixed point iteration methods~\cite{woods2019computing}.\cn

In this work, we present a robust and computationally efficient preconditioning approach for the Kohn-Sham SCF iteration based on a low-rank approximation of $J$, or equivalently $\epsilon$. This approach, which we refer to as the low-rank dielectric matrix (LRDM) preconditioning approach, constructs an approximation of $J$ based on sum of rank-1 tensor products between direction functions corresponding to an approximate Krylov subspace of $J$ and the G\^ateaux derivative of the residual of the fixed point iteration, $F[V_{\textrm{eff}}[\rho]]-\rho$, along the direction functions. \cb We note that this preconditioning strategy was proposed in a recent work by Niklasson~\cite{niklasson2020krylov}, in the context of self-consistent charge density functional tight-binding theory employing a reduced order atom centered basis. In this work, we build upon this idea to~\cn apply the LRDM  method to Kohn-Sham DFT calculations employing systematically convergent  higher-order finite-element basis sets~\cite{motamarri2013}.  Further, we demonstrate the robustness and efficiency of LRDM using extensive large-scale heterogeneous benchmark systems, and comparing against widely used preconditioners. The primary challenge here is the efficient computation of the the G\^ateaux derivatives of $F[V_{\textrm{eff}}[\rho]]$ along the direction functions---referred to as the first-order density response functions---that constitutes the most computationally expensive portion of the LRDM method. Further, developing strategies to enable modest values of rank for a wide range of heterogeneous systems is critical to minimize the overheads associated with the method and boost the practical computational efficiency. Regarding the first aspect, one of the key contributions of the present work is to develop a computationally efficient method to approximate the first-order density response functions in real-space, and within the context of Chebyshev filtered subspace iteration (ChFSI) procedure~\cite{zhou2006}. We note that ChFSI is a computationally efficient and scalabale eigensolver that progressively approximates the eigensubspace corresponding to the occupied Kohn-Sham eigenstates instead of performing an exact diagonalization in each SCF iteration. ChFSI provides significant computational gains over other iterative eigensolvers for solution of the Kohn-Sham non-linear eigenvalue problem~\cite{zhou2006,motamarri2013}. As a result, many real-space DFT codes, based on finite-element basis~\cite{motamarri2020,das2022dft} or finite-difference discretization~\cite{SPARC,octopus2015,michaud2016rescu}, currently employ the ChFSI procedure. In the proposed work, the evaluation of the first-order density response functions in conjunction with the ChFSI procedure involves two steps. First, we compute  the first-order density-matrix response in the approximate eigensubspace of dimension $\sim N_e$, obtained in each iteration of the ChFSI procedure. Subsequently, we transform the density-matrix response to the space spanned by the finite-element basis, and obtain the density response functions from the diagonal of the density-matrix response. This results in a $\mathcal{O}(MN_e^2)$ scaling method ($M$ denoting the size of the finite-element basis) with a small computational prefactor, as will be demonstrated in this work. The second important aspect of the present work targets reduction of the average rank by developing an accumulated variant of LRDM, which adaptively accumulates the Jacobian approximation from previous SCF iterations. Our proposed numerical approach for accumulation entails taking new direction functions that are orthogonal to the ones from the previous SCF iterations in conjunction with an adaptive strategy that either continues or clears the accumulation based on numerical metrics informing the linearity of the  residual function with respect to the electron-density, and the low rank approximation error. Furthermore, in the present work, we extend the formulation of the LRDM preconditioner to collinear spin-polarized KS-DFT calculations.

We demonstrate the robustness and efficiency of the LRDM preconditioner, and compare it against Anderson, Kerker and TFW preconditioners on a comprehensive set of benchmark heterogeneous material systems ranging up to $\sim$ 1,100 atoms ($\sim$ 20,000 electrons), including spin-polarized magnetic systems. In all the benchmark systems LRDM demonstrates robust and system-size independent convergence. In contrast, other preconditioners studied in this work do not converge, in some cases, for larger-scale heterogeneous benchmark systems. We also demonstrate the advantages of the proposed accumulated LRDM variant in reducing the average rank. Finally, we demonstrate the computational efficiency of the LRDM preconditioner against Anderson/Kerker preconditioners, where we observe up to $3.4\times$ reduction in the computational times for full ground-state calculations on the benchmark problems.

The remainder of the paper is organized as follows. Section~\ref{sec:method} starts with a brief theoretical background on the convergence  aspects of the Kohn-Sham SCF iteration. Subsequently, in Section~\ref{sec:lrdm} we present the formulation of the LRDM preconditioner and its proposed accumulated variant. In Section~\ref{sec:densityResponse}, we develop the computational method for evaluation of the density response functions in real-space Kohn-Sham DFT calculations within the ChFSI procedure. Finally, in Section~\ref{sec:lrdmSpin} we extend the formulation to spin-polarized calculations. Section~\ref{sec:results} presents computational results on heterogeneous benchmark systems comparing the SCF convergence of the LRDM preconditioner against other widely used preconditioners. Section~\ref{sec:results} also demonstrates the computational efficiency of LRDM preconditioner on hybrid CPU-GPU architectures. We finally conclude with an outlook in Section~\ref{sec:conclusions}.

\section{Formulation}
\label{sec:method}
We first introduce the mathematical notation that will be used in the subsequent sections. We assume the electron-density, Kohn-Sham wavefunctions, and other electronic-fields appearing in the formulation to belong to an appropriate function space, $\Upsilon(\Omega)$. In the case of non-periodic calculations, $\Omega$ corresponds to a large enough domain containing the compact support of the electronic fields, and, in periodic calculations, it corresponds to a periodic domain. We denote the inner product between two functions as $\innerproductcomma{g_1}{g_2}=\int_{\Omega} g_1^{\ast}(\bx)g_2(\bx) d \bx$, and $\norm{.}$ to be the norm induced from this inner product.  In what follows, we denote the action of an infinite dimensional bounded linear operator $A:\, \Upsilon\rightarrow \Upsilon$ on $g$ as $A\, g:=\int_{\Omega} A(\bx,\bx^{\prime})\, g(\bxprime) d \bxprime$, where $A(\bx,\bx^{\prime})$ and $g(\bx)$ denote the real-space representations of $A$ and $g$, respectively. Finally, we denote the action of a non-linear operator $F$ on $g$ as $F[g]: \Upsilon\rightarrow \Upsilon$.
\subsection{Background on Kohn-Sham SCF convergence}
\label{sec:background}
We begin by considering the Kohn-Sham  fixed point iteration. For a materials system with $N_{at}$ nuclei and $N_e$ electrons, the spin restricted ground-state properties in Kohn-Sham density functional theory are given by solving the $N/2$ lowest eigenstates of the following non-linear eigenvalue problem~\cite{kohn65}:
\begin{equation}\label{eq:kseqcont}
\begin{gathered}
\left(-\frac{1}{2} \laplacian +  V_{\textrm{eff}}[\rho,\bR]\right){\psi}_{k} = \varepsilon_{k} \,{\psi}_{k},\notag\\
2\sum_{k}f(\epsilon_k,\mu) = N_e\,,\;\;\;\;
f(\epsilon,\mu) = \frac{1}{1 + \exp\left(\frac{\epsilon - \mu}{k_B T} \right)}\,,\notag\\
\rho(\bx) = 2\sum_{k}f(\varepsilon_k,\mu)|\psi_{k}(\bx)|^2\,,\notag\\
V_{\textrm{eff}}[\rho,\bR]=V_{\textrm{xc}}[\rho] + V_{\textrm{H}}[\rho] +V_{\textrm{ext}}(\bR)
\end{gathered}
\end{equation}
where $\bR = \{\bR_1,\,\bR_2,\,\cdots \bR_{N_{at}}\}$ denotes the positions of the $N_{at}$ nuclei, $\rho$ denotes the electron density, $V_{\text{eff}}[\rho,\bR]$ denotes the effective single electron Kohn-Sham potential, $\varepsilon_{k}$ and $\psi_{k}$ denote the eigenstates of the Kohn-Sham Hamiltonian (henceforth denoted by $\mathcal{H}[\rho]$), $f(\epsilon,\mu)$ denotes the Fermi-Dirac distribution with $\mu$ being the Fermi energy or the chemical potential. $V_{\text{eff}}[\rho,\bR]$ is composed of the exchange-correlation potential ($V_{\textrm{xc}}[\rho]$) accounting for the many-body quantum mechanical interactions, the Hartree electrostatic potential corresponding to the electron density ($V_{\textrm{H}}[\rho]$), and the external electrostatic potential from the nuclei ($V_{ext}(\bR)$). The above non-linear eigenvalue problem can be viewed as the fixed point iteration $\rho=F[V_{\textrm{eff}}[\rho]]$, which is commonly referred to as the self-consistent field (SCF) iteration. Within this SCF procedure, the evaluation of $F[V_{\textrm{eff}}[\rho]]$ entails solving a linear eigenvalue problem. Solution of this fixed point problem is equivalent to finding the root of the residual in the electron density, $R[\rho]=F[V_{\textrm{eff}}[\rho]]-\rho=0$. Further, in the neighbourhood of the ground-state, fast quadratic convergence in the residual can be achieved by using the Newton method:
\begin{equation}
\rho^{(n+1)}=\rho^{(n)}-J^{-1} (F[V_{\textrm{eff}}[\rho^{(n)}]]-\rho^{(n)})\,,
\end{equation}
where $n$ denotes the $n^{\textrm{th}}$ SCF iteration, and $J$ denotes the  Jacobian  corresponding to $R[\rho]$ given by
\begin{align}\label{eq:jacobian}
J=&\left.\frac{\delta F[V_{\textrm{eff}}]}{\delta V_{\textrm{eff}}}\frac{\delta V_{\textrm{eff}}[\rho]}{\delta \rho}\right|_{\rho=\rho^{(n)}} -I := \chi_0 K-I=-\epsilon^{\dagger}\,,\notag\\
K=& \left.\frac{\delta V_{\textrm{eff}}[\rho]}{\delta \rho}\right|_{\rho=\rho^{(n)}}=\left.\frac{\delta V_{\textrm{xc}}[\rho]}{\delta \rho}\right|_{\rho=\rho^{(n)}}+\left.\frac{\delta V_{\textrm{H}}[\rho]}{\delta\rho}\right|_{\rho=\rho^{(n)}}\notag\\
:= & K_{\textrm{xc}} + K_{\textrm{c}}\,.
\end{align}
In the above, $\chi_0$ is commonly referred to as the independent particle susceptibility operator, $\epsilon$ is the dielectric operator, $K_{\textrm{c}}$ is the Coulomb kernel, and $K_{\textrm{xc}}$ is the exchange-correlation kernel, which is usually a small contribution to $\epsilon$ compared to $K_{\textrm{c}}$ (random-phase approximation). However, as the exact evaluation of $\epsilon$, in particular the evaluation of $\chi_0$ is computationally expensive, most widely implemented SCF acceleration strategies rely on a quasi-Newton iteration
\begin{equation}\label{eq:quasinewton}
\rho^{(n+1)}=\rho^{(n)}-\alpha P (F[V_{\textrm{eff}}[\rho^{(n)}]]-\rho^{(n)})\,,
\end{equation}
where $\alpha \in (0,1]$ is the damping parameter, and $P$ is a linear operator that approximates the inverse Jacobian of the residual, thereby acting as a preconditioner for the quasi-Newton step. In the neighbourhood of the ground-state solution $\rho^{(\ast)}$, it can easily be shown that the necessary condition for convergence is
\begin{equation}\label{eq:convergence}
    s\left(I-\alpha P_{\ast} J_{\ast}\right) < 1\,,
\end{equation}
where $s(A)$ denotes the spectral radius of a diagonalizable operator $A$, $J_{\ast}$ is the Jacobian operator evaluated at $\rho^{(\ast)}$ and $P_{\ast}$ is the approximation to the inverse of $J_{\ast}$. The role of $\alpha$ in satisfying the above convergence condition can be understood as follows. Considering an appropriately constructed $P \approx J^{-1}$, such that $P_{\ast} J_{\ast}$ has a positive real eigenspectrum\footnote{For the simplest case of $P=-I$, it can be shown that $-J_{\ast}$ has a positive real eigenspectrum using the generally assumed random phase approximation~\cite{LinLin2013,Herbst_2020}.}, we can choose $0<\alpha \le 1 $ to satisfy the convergence condition in Eq.~\eqref{eq:convergence}. Further, an optimal value of $\alpha$ can be chosen to minimize $s$ or equivalently maximize the linear convergence rate. The corresponding minimal value of $s$ is given by 
\begin{equation}\label{eq:optimalrate} 
s_{opt}=\frac{\bar{\kappa}(P_{\ast}J_{\ast})-1}{\bar{\kappa}(P_{\ast}J_{\ast})+1}\,,
\end{equation}
where $\bar{\kappa}(A)$ denotes the condition number of  a diagonalizable operator $A$, given by the ratio of the largest to the smallest eigenvalue magnitude  of $A$.  Thus, the slow convergence in the Kohn-Sham SCF iteration arises from the large condition number of $J$, or, equivalently, the dielectric operator $\epsilon$.  Practically, an optimal value of $\alpha$ is difficult to estimate due to the high computational cost of obtaining  $\bar{\kappa}(J)$. Further, the lack of a suitable preconditioner $P$ will necessitate the use of small values of $\alpha$ resulting in slow convergence, or for larger values of $\alpha$ may result in divergence of the fixed point iteration. 

Large condition numbers of $\epsilon$ arises primarily from the following scenarios encountered in large heterogeneous materials systems. The first and the most common one is the divergence of the Coulomb kernel $K_c(\abs{{\bf q}}) \propto \frac{1}{\abs{{\bf q}}^2}$ as $\abs{{\bf q}} \rightarrow 0$ in the Fourier space. In metals, it can be  shown from the Alder-Wiser expression of $\chi_0$ in Fourier space, that $\chi_0(\abs{{\bf q}})$  converges to a finite value as $\abs{{\bf q}} \rightarrow 0$, overall resulting in divergence of $\chi_0 K_c(\abs{{\bf q}})$ as $\abs{{\bf q}} \rightarrow 0$. The divergence manifests as the well-known charge sloshing behavior observed in large metallic materials systems, with the  long-wavelength  eigenvalue of $\chi_0$ scaling as $L^2$, $L$ being the extent of the metallic region. However, in the case of insulators and semiconductors, the long-wavelength divergence of $\chi_0 K_c$ is suppressed as $\chi_0(\abs{{\bf q}})\propto \abs{{\bf q}}^2$ as $\abs{{\bf q}} \rightarrow 0$. We refer to~\cite{Herbst_2020} for a detailed discussion on the derivation of the long-wavelength limit screening behaviour. The second important source of instabilities are from the localized $d$ and $f$ orbitals with large density of states near the Fermi energy, which lead to large eigenvalues of $\chi_0$. We refer to \cite{Dederichs1983} for an insightful discussion on such instabilities using a model of a 3$d$ impurity in jellium. Importantly, both long-ranged and short-ranged wavelengths can be potentially associated with eigenmodes corresponding to the large eigenvalues of $\chi_0$~\cite{Herbst_2020}. However, current preconditioners primarily approximate the long-ranged modes. Overall, given that a material system can involve any combination of the above scenarios in different spatial regions, a robust and generic preconditioning strategy that appropriately and efficiently accounts for above sources of large eigenvalues of $\epsilon$ has remained a challenge.  In this work, we use a low rank approximation of $\epsilon$ to efficiently precondition the SCF iteration that provides robust convergence for  complex and large-scale heterogeneous materials systems.

\subsection{Approximate Krylov subspace based low rank approximation of $J$}\label{sec:lrdm}
 
We now present the low rank approximation of $J$, or equivalently $-\epsilon^{\dagger}$, in a continuous real-space setting, and with an adaptive determination of the rank based on an error indicator. The continuous real-space setting is useful to employ the proposed method in conjunction with the Chebyshev filtered subspace iteration (ChFSI) procedure, as will be discussed in Section~\ref{sec:densityResponse}. Furthermore, we propose an adaptively accumulated variant that reuses the low rank approximation from previous SCF iterations. As will be discussed below, our adaptive accumulation strategy relies on an additional numerical indicator of the linearity of $R[\rho]$ with respect to the electron-density. Additionally, we extend the formulation to the spin-polarized case, as will be presented in Section~\ref{sec:lrdmSpin}. We remark that the idea of using a low rank approximation of the dielectric matrix or related quantities has been explored in electronic structure calculations, with a recent work employing this idea~\cite{niklasson2020krylov} in the context of extended Lagrangian Born–Oppenheimer molecular dynamics. In~\cite{niklasson2020krylov}, which employed self-consistent charge density functional tight-binding theory using a reduced order basis, the use of a low-rank approximation of $J$ to accelerate the SCF iteration was also suggested. \cb However, it was only demonstrated on a single insulating system of a reactive nitromethane mixture system containing $\sim$ 50 atoms. The present work applies the low rank approximation approach to Kohn-Sham DFT using a systematically convergent basis set and demonstrates the robustness of the approach on a wide range of medium to large-scale benchmark systems that include combinations of metal-insulating-semiconducting heterogeneous systems and magnetic systems. Testing on large-scale systems with metallic regions is critical as long-range charge sloshing effects during SCF convergence primarily manifest in such systems. \cn

In the LRDM approach, we construct a rank-$r$ approximation $J^{\textrm{lr}}_r$ of $J$ in each 
SCF iteration (indexed by $n$), based on generalized directional (G\^ateaux) derivatives, $t_{i}$, of the residual $R[\rho^{(n)}]$ along orthonormal direction functions $u_i$: 
\begin{align}\label{eq:jlowrank}
       & J(\bx,\bxprime)  \approx J^{\textrm{lr}}_r(\bx,\bxprime) = \sum_{i=1}^{r} t_{i} (\bx)   u_i (\bxprime) \,,\notag\\ 
       & {\rm where}\quad t_{i}=\left.\frac{\partial R [\rho^{(n)}+\lambda u_i]}{\partial \lambda }\right|_{\lambda=0}\,\, {\rm with} \,\, 
       \innerproductcomma{u_i}{u_j}=\delta_{ij}\,.
\end{align}
 In the above, the direction functions $u_i$ are related to the functions in the Krylov subspace of $J$ as will be discussed subsequently. We also note that $t_i$s are related to the density response functions corresponding to the direction functions $u_i$s. We remark that the above expansion converges to $J$ as $r\rightarrow \infty$, based on analogy to canonical decomposition of tensors of arbitrary order as sum of tensor products of rank-1 components~\cite{Hitchcock1927}. We seek an approximate solution, $\Delta\bar{\rho}$, of the equation $J \Delta \rho = R[\rho^{(n)}]$ that provides the update in the quasi-Newton step $\rho^{(n+1)}=\rho^{(n)} - \alpha \Delta\bar{\rho}$. Since $J$ is not explicitly known, we construct the Krylov subspace progressively as follows as we build the low rank approximation. We start by choosing the first normalized direction function as $u_1=R[\rho^{(n)}]/\norm{R[\rho^{(n)}]}$,  and compute its corresponding G\^ateaux derivative: $t_1=\left.\frac{\partial R [\rho^{(n)}+\lambda u_1]}{\partial \lambda }\right|_{\lambda=0}$. Subsequently, we choose the remaining orthonormal direction functions $u_i\,(1<i \le r)$ using the following iterative procedure based on Gram-Schmidt orthonormalization:
\begin{align}\label{eq:krylovOrtho}
        \,\,u_i =\, & t_{i-1} \notag\\
        \,\,u_i =\, & u_i- \sum_{k=1}^{i-1} \innerproductcomma{u_i}{u_k}  u_k\,\quad \notag\\
        \,\,u_i =\, & u_i/\norm{u_i}\,.
\end{align}
 Next, we  solve $J^{\textrm{lr}}_r \Delta \bar{\rho}=R[\rho^{(n)}]$, whose solution is given by
\begin{align}\label{eq:plowrank}
    \Delta\bar{\rho}=P^{\textrm{lr}}_r R[\rho^{(n)}] = & \sum_{i,j=1}^{r}  u_i   S^{-1}_{ij}  \innerproductcomma{t_{j}}{R[\rho^{(n)}]} \,,\notag\\ &{\rm where}\quad S_{ij}=\innerproductcomma{t_i}{t_j}\,,
\end{align}
and $P^{\textrm{lr}}_r$ denotes the pseudo-inverse of $J^{\textrm{lr}}_r$.

Overall, in the above low rank formulation, we have two important parameters: $r$ and $\alpha$, that significantly control the convergence behaviour of the fixed point iteration. We now discuss strategies for choosing these parameters to achieve robust convergence without case by case manual tuning. First, considering $r$, we note that $J^{\textrm{lr}}_r P^{\textrm{lr}}_r = \sum_{i,j=1}^{r}  t_i S^{-1}_{ij} t_{j} = I_r^t$ represents the rank-$r$ resolution of identity operator in the non-orthogonal $t_i$ basis. $I_r^t$ tends to $I$ as $r \rightarrow \infty$. This aspect can be used to design an adaptive metric for deciding the rank $r$.  Thereby, we use following error metric based on the relative error of 
($I_r^t-I$)  applied to the current residual:
\begin{equation}\label{eq:relativeError}
s_{\textrm{rel}}= \norm{\sum_{i,j=1}^{r}  t_i S^{-1}_{ij} \innerproductcomma{t_{j}}{R[\rho^{(n)}]} - R[\rho^{(n)}]}/\norm{R[\rho^{(n)}]}\,,
\end{equation}
where $s_{\textrm{rel}} \rightarrow 0$ as $r\rightarrow \infty$. In each SCF iteration step, we increase $r$ until $s_{\textrm{rel}}$ goes below a set tolerance, $s_{\textrm{tol}}$. As will be demonstrated in Section~\ref{sec:results}, we find that a value of $s_{\textrm{tol}} \sim 0.3$ is  sufficient to achieve accelerated and system-size independent convergence for all the heterogeneous benchmark systems considered in this work. \cn

Next, turning our attention to the damping parameter $\alpha$, if $J^{\textrm{lr}}_r$ closely approximates $J$, one could use $\alpha=1$ to obtain a quadratically convergent Newton step ($s_{opt}=0$ from Eq.~\eqref{eq:optimalrate}). However, there are a few practical issues with such a strategy when applied to the SCF iterations in KS-DFT. First, the rank $r$ required to obtain a very close approximation to $J^{-1}$ could be very high in a complex heterogeneous systems, thus significantly increasing the computational overhead of the preconditioner.  Second, Newton step assumes the starting point $\rho^{(1)}$ to lie within the linear approximation zone of $R[\rho]$ about the solution $\rho^{(\ast)}$. However, this may not be true for typical starting initial guess, $\rho^{(1)}$, obtained as a superposition of single atomic electron-densities. Particularly, in condensed matter systems, $\rho^{(1)}$ can be quite far  from $\rho^{(\ast)}$.  Hence, in this work, we set a small value of $\alpha = 0.1$ for the initial steps until $\norm{R[\rho^{(n)}]}$ goes below a threshold of $\mathcal{O}(1)$, below which we increase $\alpha$ to $0.5$ to achieve accelerated convergence. This choice is determined based on our numerical experiments, and as will be demonstrated in Section~\ref{sec:results}, we obtain robust convergence for all the benchmark systems using this choice of $\alpha$. 

\subsubsection{Accumulated low rank approximation}\label{sec:lrdm-accumulated-method}
We now propose an accumulated variant of LRDM, referred to as LRDMA, that accumulates the low rank approximation of $J$ from previous SCF iteration steps. This could potentially reduce the average rank $r$ of LRDM during the SCF convergence, thereby reducing the computational overhead. The primary consideration here is that the density response functions of $F[V_{\rm eff}[\rho]]$ ($\frac{\partial}{\partial\lambda}(F[V_{\rm eff}[\rho+\lambda u_i]])|_{\lambda=0}=t_i+u_i$) from the previous SCF iteration steps can provide a good approximation to the density response functions of the present iteration, especially if the electron densities from the previous iterations are close to the current iteration. Such a condition is expected to exist close to the solution of the fixed point iteration $\rho^{(\ast)}$. However, even when $\rho$ is not necessarily close to $\rho^{(\ast)}$, such an accumulation can benefit when the densities in the preceding iterations are close to the density in the current iteration and within the linear approximation zone, i.e., the region of the function space where the linear term in the Taylor series is dominant. To this end, we develop a numerical indicator that can guide the adaptive accumulation procedure.

We consider the Taylor series expansion of $R[\rho]$ about $\rho^{(n-1)}$ (density in the $(n-1)^{\rm th}$ SCF iteration) to the linear order. The linear approximation of $R[\rho^{(n)}]$ is given by 
\begin{align}
     R^{\textrm{lin}}[\rho^{(n)}] = &  R[\rho^{(n-1)}] + J (\rho^{(n)}-\rho^{(n-1)}) \notag \\ 
     = & R[\rho^{(n-1)}] - \alpha J\Delta\bar{\rho} \notag \\ 
     = & R[\rho^{(n-1)}] - \alpha J P^{\textrm{lr}}_r R[\rho^{(n-1)}] \notag \, 
\end{align}
where $J$ is the Jacobian at $\rho=\rho^{(n-1)}$. Although $J$ is unknown, using Eq.~\eqref{eq:jlowrank} and~\eqref{eq:plowrank} we can express $R^{\textrm{lin}}[\rho^{(n)}]$ as 
\begin{align}\label{eq:linearapprox}
    R^{\textrm{lin}}[\rho^{(n)}]= & R[\rho^{(n-1)}]- \alpha   \sum_{i,j=1}^{r}  t_i S^{-1}_{ij} t_{j}  R[\rho^{(n-1)}]\notag\\
    = & (I-\alpha I_r^t) R[\rho^{(n-1)}]\,.
\end{align}
Using the above, we design the following linearity indicator based on the ratio of norms of the predicted and the actual residual at the $n^{\textrm{th}}$ step:
\begin{equation}\label{eq:linearityIndicator}
    \beta=\abs{\norm{R^{\textrm{lin}}[\rho^{(n)}]}/\norm{R[\rho^{(n)}]}-1}\,.
\end{equation}
We note that $\beta$ will be close to zero in the linear approximation zone, and $\beta$ correlates with the strength of the non-linear terms in the Taylor series expansion of $R[\rho]$. 

The proposed LRDMA approach uses $\beta$, $s_{\textrm{rel}}$ and $\norm{R[\rho^{(n)}]}$ to determine whether to use and further accumulate on the low rank approximation from the previous SCF iteration, or to clear any accumulation and construct the low rank approximation solely from the current SCF iteration. In order to determine if the density response functions from previous SCF iterations are useful in the current iteration, we first rely on the linearity indicator $\beta$. If $\beta>\beta_{\textrm{tol}}$ then there is sufficient non-linearity which suggests that using density response functions from previous iterations can have substantial errors and may not provide a good approximation to $J$ for the current iteration. Thus, we clear any rank accumulation from previous iteration (labelled as CL event in Figure~\ref{fig:rankplot} to denote clearing based on linearity indicator), and construct the low-rank approximation of $J$ afresh using the direction functions and density response functions from the current SCF iteration. Based on numerical experiments, we find $\beta_{\textrm{tol}} \sim 0.1$ is a good choice. If $\beta<\beta_{\textrm{tol}}$, then we check $s_{\textrm{rel}}$. If $s_{\textrm{rel}}<s_{\textrm{tol}}$, then the available low rank approximation of the Jacobian from the previous SCF iteration is also a good approximation for the current iteration and no further update to the Jacobian is needed. On the other hand, if $s_{\textrm{rel}}>s_{\textrm{tol}}$ we will need to decide whether it is beneficial to continue the accumulation of the rank-1 updates using the density response functions from the current iteration, or to purge/clear and construct the low-rank approximation afresh in the current iteration. We note that if the current iterate is far from the solution $\rho^{(\ast)}$, then the direction response functions computed from previous iterations may not be useful in the current iteration and may result in a very large rank if the accumulation continues. Thus, it is beneficial to clear the accumulation and construct the approximation to the Jacobian in the current iteration. To this end, we use the $\norm{R[\rho^{(n)}]}$ as a proxy for how far/close the iterate is to $\rho^{(\ast)}$. If $\norm{R[\rho^{(n)}]}>1.0$, then we clear the accumulation (labelled as CR event in Figure~\ref{fig:rankplot} to denote clearing based on residual indicator). If a clearing of the accumulation based on the linearity indicator and the residual indicator are not triggered, we use the low-rank approximation from the previous SCF iteration ($J^{\textrm{lr}}_{r,\,(n-1)}$) and continue the accumulation using rank-1 updates from the current iteration to improve the Jacobian approximation as      
\begin{align}\label{eq:accum}
     J^{\textrm{lr}}_{(r+r_{n}),\,(n)} (\bx,\bxprime)= & \sum_{i=1}^{r_{n}} t_{i}^{(n)} (\bx)  u_i^{(n)} (\bxprime) + J^{\textrm{lr}}_{r,\,(n-1)}(\bx,\bxprime)\,.
\end{align}
In the above, $r_{n}$ is the additional rank added in the $n^{\textrm{th}}$ SCF iteration and the direction functions $u_i^{(n)}$ are chosen from the Krylov subspace of $J$ at the current $n^{\textrm{th}}$ SCF iteration, but orthogonalized with respect to all the other direction functions. We note that $u_1^{(n)}$ is chosen to be the normalized orthogonal complement of $R[\rho^{(n)}]$ to the previous history of direction functions. We continue the accumulation until either $s_{\textrm{rel}}<s_{\textrm{tol}}$ or $r_{n}=r^{\textrm{max}}_{\textrm{iter}}$, with $r^{\textrm{max}}_{\textrm{iter}}=5$. If the tolerance condition on $s_{\textrm{rel}}$ is not reached, we also trigger an accumulation clearing event for the next SCF step (labelled as CT event in Figure~\ref{fig:rankplot}). We investigate the efficiency and robustness of the proposed LRDMA method in comparison to the LRDM method in Section~\ref{sec:rankreduction}, where the numerical results demonstrate a reduction in the average rank across most benchmark systems resulting in improved computational efficiency.

\subsection{First-order density response computation in Chebyshev filtered subspace}\label{sec:densityResponse}
We now develop an efficient numerical methodology for the computation of the first-order density response functions $\left(\left.\frac{\partial }{\partial \lambda }(F[[V_{\rm eff}[\rho^{(n)}+\lambda u_i]])\right|_{\lambda=0}\right)$, i.e. the G\^ateaux derivatives along the direction functions, in Kohn-Sham DFT. We focus on Kohn-Sham DFT calculations using systematically convergent complete basis sets, such as the finite-element basis, where the nonlinear Kohn-Sham eigenvalue problem (cf. Eq.~\eqref{eq:kseqcont}) is discretized with $M$ non-orthogonal real-space basis functions, with $M\gg N$. The main aspect of our numerical method is to approximately compute the first-order density response in a finite-dimensional subspace with a dimension that is much smaller compared to $M$, and further take advantage of the efficient canonical density-matrix perturbation approach~\cite{niklasson2015canonical} that avoids explicit computation of first-order perturbations in the wavefunctions. 

To begin, we first briefly discuss the ChFSI procedure in real-space Kohn-Sham DFT~\cite{zhou2006} to compute the electron-density, and subsequently detail our approach to compute the first-order density response. We choose our real-space basis functions to be higher-order spectral finite-elements (FE) $\{l_a\}_{(1\leq a \leq M)}$ that are strictly local piece-wise polynomial basis functions. Since the spectral FE basis is non-orthogonal, resulting in a generalized Hermitian eigenvalue problem (GHEP), the L\"owdin orthogonalized FE basis functions (denoted as $\{p_a\}_{(1\leq a \leq M)}$) are used to obtain a standard Hermitian eigenvalue problem (SHEP)~\cite{motamarri2013}:
\begin{equation}\label{eq:shep}
\btH[\rho] \boldsymbol{\tilde{\psi}}_k = \varepsilon^h_k \boldsymbol{\tilde{\psi}}_k\,,\;\;\; k = 1,2,\cdots N\;\;\;\text{with}\;\;\;N > \frac{N_e}{2}\,,
\end{equation}
where $\btH[\rho]$ is the discrete Kohn-Sham Hamiltonian in the $\{p_a\}$ basis, and $\boldsymbol{\tilde{\psi}}_k$ are the expansion coefficients of the single-electron Kohn-Sham eigenfunctions in the $\{p_a\}$ basis. In order to solve the nonlinear SHEP, we use the ChFSI procedure, which exploits the fact that we are only interested in the occupied eigensubspace that is a very small portion at the lower end of the spectrum of $\btH$.  Instead of solving Eq.~\eqref{eq:shep} exactly in every SCF step, the ChFSI procedure progressively approximates the occupied eigensubspace. In particular, this involves applying a Chebyshev polynomial filter of degree $c$, $T_c(\bar{\bf H})$, to a trial subspace ${\bf X}$ with dimension $N$. Here, $\bar{\bf H}$ denotes a scaled and shifted Hamiltonian constructed from $\btH$ such that the wanted (occupied) spectrum is mapped to $(-\infty,-1)$ and the unwanted spectrum is mapped to $[-1,1]$. As $T_c(y)$ monotonically and rapidly increases for decreasing values of $y<-1$, the action of $T_c(\bar{\bf H})$ on ${\bf X}$ results in a filtered subspace $\bPsitildef$ that is a close approximation to the eigensubspace corresponding to the occupied spectrum of $\btH[\rho]$. Subsequently, $\bPsitildef$ is orthonormalized to obtain $\bPsitildeo$, and the discrete SHEP is projected onto the subspace spanned by $\bPsitildeo$ to solve the eigendecomposition of the projected Hamiltonian:
\begin{equation}
    \bhatH {\bf Q}= {\bf Q} {\bf D}\,,\quad \textrm{where}\quad \bhatH= \bPsitildeoDagger \btH \bPsitildeo\,,
\end{equation} 
where ${\bf Q}$ is the matrix comprising of the eigenvectors and ${\bf D}$ is a diagonal matrix with the corresponding eigenvalues, represented in the Chebyshev filtered subspace. The corresponding eigenvectors in the L\"owdin orthogonalized FE basis are obtained from the transformation $\bPsitildeR=\bPsitildeo {\bf Q} $. Finally, the output electron-density ($F[V_{\textrm{eff}}[\rho]](\bx)=\rho_{\textrm{out}}(\bx)$) in each SCF iteration is computed as 
\begin{align}\label{eq:densityChFSI}
&\rho_{\textrm{out}}(\bx)=\notag\\&2 \,(\bm{n^{{e}}}({\bf x}))^{T}\left[{\bf M}^{-1/2} \,\bPsitildeR \,f({\bf D},\,\mu)\,\bPsitildeRDagger\, {\bf M}^{{-1/2}^{\dagger}}\right]\bm{n^e}({\bf x}),
\end{align}
where $\bm{n^e}({\bf x})={\left[l^e_1({\bf x}) \,l^e_2({\bf x}) \, \cdots \, l^e_{M_{cell}}({\bf x})\right]}^T$ denotes the FE basis functions associated with the given finite-element cell ($M_{cell}$ denotes the number of nodes in the cell), and ${\bf M}$ is the positive-definite and symmetric finite-element overlap matrix (${\bf M}_{ab}=\innerproductcomma{l_a}{l_b}$) in the original non-orthogonal FE basis. We refer to \cite{motamarri2013,motamarri2020,das2022dft} for more details on the algorithmic aspects of employing ChFSI in Kohn-Sham DFT calculations using the FE basis.

The electron-density expression in Eq.~\eqref{eq:densityChFSI} can also equivalently be obtained from $\bGamtilde=f(\bPsitildeR \bPsitildeRDagger \btH \bPsitildeR \bPsitildeRDagger,\,\mu)$, which denotes the density-matrix corresponding to the projection of $\btH$ onto $\bPsitildeR$. The resulting electron-density expression is given by
\begin{equation}\label{eq:densityChFSIFromDensityMat}
\rho_{\textrm{out}}(\bx)=2 \,(\bm{n^{{e}}}({\bf x}))^{T}\left[{\bf M}^{-1/2} \,\bGamtilde\, {\bf M}^{{-1/2}^{\dagger}}\right]\bm{n^e}({\bf x})\,.
\end{equation}
The equivalence between the two expressions for the electron-density can be seen from
\begin{align}\label{eq:equivalence}
    \bGamtilde=& f(\bPsitildeR \bPsitildeRDagger \btH \bPsitildeR \bPsitildeRDagger,\,\mu) \notag\\&=\bPsitildeR f(\bPsitildeRDagger \btH \bPsitildeR,\,\mu) \bPsitildeRDagger=\bPsitildeR f({\bf D},\,\mu)\bPsitildeRDagger \,, 
\end{align}
where the second equality follows from the power series  representation of the analytic function $f(\varepsilon,\,\mu)$ combined with the spectral decomposition of the Hermitian matrix $\bPsitildeR \bPsitildeRDagger \btH \bPsitildeR \bPsitildeRDagger$~\cite{motamarri2014}. 
In the above, we note that $\bGamtilde$ systematically approaches the density matrix corresponding to the discrete FE Hamiltonian, $\bGam=f(\btH[\rho],\mu)$, as the SCF approaches convergence. This follows from the ChFSI procedure, where $\bPsitildeR$ progressively approaches the occupied eigensubspace of $\btH[\rho]$ as the SCF approaches convergence, i.e. $\btH[\rho]\rightarrow\btH[\rho^{(\ast)}]$.

We now consider a perturbation to the input electron-density $\rho$, in a given SCF iteration step, along the direction function $u_i$ expressed as $\rho(\lambda)=\rho+\lambda u_i$. We seek to compute the response in $\rho_{\textrm{out}}$ to first order, corresponding to the perturbations. The first-order response of $\rho_{\textrm{out}}$ can be obtained from the explicit first-order response of all the occupied eigenfunctions and eigenvalues. While the first-order response in the eigenvalues can be obtained without significant computational overheads by using the first-order perturbation theory involving only the first-order response of the Hamiltonian~\cite{rmartin}, the first-order response in eigenfunctions requires a solution of the Sternheimer equations that are linear systems of equations of dimension $M$, one for each eigenfunction perturbation. The solution to Sternheimer equations involve computation of the perturbations in a large subspace of the full finite dimensional space that is orthogonal to the occupied eigensubspace~\cite{dfpt2001}. This results in a computational cost that is ${\mathcal{O}(MN_e^2)}$, and typically with high prefactors. Thus, in systematically convergent real space basis sets, where $M \gg N$, solution of Sternheimer equations can result in substantial computational overheads. In this work, we adopt a different approach. We consider the first-order perturbation response of the FE discretized Hamiltonian:
\begin{equation}
      \btH[\rho(\lambda)]=\btH[\rho]+\lambda \btH^{\prime}, 
      \quad \btH^{\prime}=\left.\frac{\partial \btH[\rho+\lambda u_i]}{\partial \lambda}\right|_{\lambda=0} \,, 
\end{equation}
and approximately compute the first-order density response corresponding to the orthogonal projection of $\btH[\rho(\lambda)]$ onto an unperturbed subspace of small dimension ($\ll M$), which we choose to be $\bPsitildeR$. Utilizing the density-matrix based representation of $\rho_{\textrm{out}}$ in Eq.~\eqref{eq:densityChFSIFromDensityMat}, we obtain the approximate density response as
\begin{align}\label{eq:densityResponse}
&\left. \frac{\partial}{\partial \lambda} \Big(  F\left[V_{\textrm{eff}}[\rho + \lambda u_i]\right]  \Big)\right|_{\lambda=0} \notag \\ 
& \approx 2 \,(\boldsymbol{n^{{e}}}({\bf x}))^{T}\left[{\bf M}^{-1/2} \,\bGamtilde^{(1)} \, {\bf M}^{{-1/2}^{\dagger}}\right]\boldsymbol{n^e}({\bf x})\,.
\end{align}
In the above, $\bGamtilde^{(1)}$ is the approximate first-order density-matrix response computed in the unperturbed subspace $\bPsitildeR$. Using Eq.~\eqref{eq:equivalence}, we arrive at
\begin{align}\label{eq:densityMatResponseApprox}
    \bGamtilde^{(1)} = &  \left. \frac{\partial}{\partial \lambda} \Big(  f\left(\bPsitildeR \bPsitildeRDagger (\btH + \lambda \btH^{\prime}) \bPsitildeR \bPsitildeRDagger ,\mu+\lambda \mu^{\prime}\right) \Big) \right|_{\lambda=0} \notag\\
    = &\, \bPsitildeR   \left. \frac{\partial}{\partial \lambda} 
    \Big( f\left(\bbreveH + \lambda \bbreveH^{\prime},\mu+\lambda \mu^{\prime}\right)  \Big) \right|_{\lambda=0}   \bPsitildeRDagger \,,
\end{align}
where $\bbreveH= \bPsitildeRDagger \btH \bPsitildeR={\bf D}$, $\bbreveH^{\prime}= \bPsitildeRDagger \btH^{\prime} \bPsitildeR$.
We remark that there are two controllable approximations in the computation of $\bGamtilde^{(1)}$. The first is from the consideration of an unperturbed subspace $\bPsitildeR$ of dimension $N\ll M$. Since $\bPsitildeR$ is an orthonormal basis, the associated approximation error can be systematically decreased by increasing the dimension $N$ of the subspace $\bPsitildeR$, albeit with an increase in the computational cost. Practically, based on our numerical experiments, we find that the additional buffer states already used in the ChFSI procedure~\cite{zhou2006,motamarri2013} provide a large enough subspace size to obtain a sufficiently accurate $\bGamtilde^{(1)}$ for the purpose of SCF preconditioning. The second approximation arises from the nature of the ChFSI procedure itself---we compute the density-matrix response $\bGamtilde^{(1)}$ about the Chebyshev-filtered density-matrix $\bGamtilde$ that closely approximates $\bGam$ in each SCF iteration step, with the approximation error systematically decreasing with SCF convergence. Thus, $\bGamtilde^{(1)}$ can be considered to progressively approximate the true density-matrix response $\bGam^{(1)}=\left.\frac{\partial }{\partial \lambda}\left( f(\btH[\rho(\lambda)],\mu(\lambda))\right)\right|_{\lambda=0}$, within the constraint of the small unperturbed subspace $\bPsitildeR$. Again, practically, we find that the the typical choices for the Chebyshev polynomial degree $c$ and the eigenpair residual norm tolerance for the occupied states, $\norm{\btH \Psitilderi- \varepsilon_i^h\Psitilderi}$, are sufficient for the purpose of preconditioning.

We now briefly discuss the approach for an efficient computation of the G\^ateaux derivative of the subspace projected density matrix appearing in Eq.~\eqref{eq:densityMatResponseApprox},
\begin{align}\label{eq:subspaceDensityMatResponse}
   \bGambreve^{(1)}= & \left. \frac{\partial}{\partial \lambda} \Big(  f\big(\bbreveH + \lambda \bbreveH^{\prime},\mu+\lambda \mu^{\prime}\big)  \Big)\right|_{\lambda=0}\,,
\end{align}
without performing additional eigendecomposition calculations that will result in significant overheads for the proposed low-rank preconditioning approach. Since $\bGambreve^{(1)}$ can be interpreted as the first-order perturbation response of  $f\left(\bbreveH[\rho(\lambda)],\,\mu(\lambda)\right)$, we employ canonical density-matrix perturbation theory~\cite{niklasson2015canonical}, which is based on perturbation of the recursive Fermi operator expansion of $f\left(\bbreveH[\rho(\lambda)],\,\mu(\lambda)\right)$:
\begin{align}\label{eq:recursiveFermiOperator}
&\left. \frac{\partial}{\partial \lambda} \Big(f \big(\bbreveH[\rho(\lambda)],\mu(\lambda)\big)\Big)\right|_{\lambda=0}\approx \notag\\ &\left. \frac{\partial}{\partial \lambda} \Big(\mathcal{P}_T(\mathcal{P}_{T-1}(\cdots\mathcal{P}_{0}(\bbreveH[\rho(\lambda)],\mu(\lambda))\ldots))\Big)\right|_{\lambda=0}\,,
\end{align}
where $T$ denotes the degree of the recursive expansion and
\begin{align}
\bX_{0}=\mathcal{P}_{0}(\bbreveH[\rho(\lambda)])=&0.5\bI-2^{-(T+2)}\frac{\bbreveH[\rho(\lambda)]-\mu(\lambda)\bI}{k_{B} T}\,,\notag\\\quad \bX_{n}=\mathcal{P}_n(\bX_{n-1})=&\frac{\bX_{n-1}^2}{\bX_{n-1}^2+(\bI-\bX_{n-1})^2}\,.
\end{align}
We note that $\mu^{\prime}$ in Eq.~\eqref{eq:subspaceDensityMatResponse} is obtained from the requirement of traceless density-matrix response, $\textrm{Tr}[\bGambreve^{(1)}]=0$,  for ensuring the conservation of number of electrons. The Pad\'e polynomial functions $\mathcal{P}_n(\bX_{n-1})$ used in the above recursive expansion enables rapid convergence in the density-matrix first-order response calculations~\cite{niklasson2015canonical,niklasson2020krylov}, with $T=8-10$ found to be sufficient in numerical experiments. This aspect is also supported by our numerical studies in this work\cb, where we find the SCF convergence for the various benchmark systems to be insensitive to values of $T$ beyond 10.~\cn Importantly, the above recursive expansion further exploits the fact that $\bbreveH[\rho(\lambda=0)]$ is a diagonal matrix, thereby rendering the algorithm to only consist of computationally cheap matrix-vector multiplications and diagonal matrix inversion operations~\cite{niklasson2020krylov}. Our numerical benchmarks indeed show that evaluation of Eq.~\eqref{eq:recursiveFermiOperator} is a negligible cost compared to the ${\mathcal{O}(MN_e^2)}$ scaling steps involving the computation of $\bbreveH^{\prime}= \bPsitildeRDagger \btH^{\prime} \bPsitildeR$ and $\bPsitildeR \bGambreve^{(1)}$, required for evaluating the density-response in Eq.~\eqref{eq:densityResponse}. However, since the above operations are performed only once for each direction response computation, they incur much lower computational prefactor compared to the iterative solution of Sternheimer equations requiring ${\mathcal{O}(MN_e^2)}$ computations in each iteration step. Further, due to the involvement of dense matrix-matrix multiplications, these operations can be performed very efficiently on hybrid CPU-GPU architectures as will be demonstrated in Section~\ref{sec:cost}. 

Finally, we mention that the following changes are required for extending the density response approach to periodic systems which entails integration over the first Brillouin zone (BZ). The first consideration is the choice of the unperturbed subspace for the response contribution corresponding to each k-point, $\textbf{k}$, in the BZ.  We choose this unperturbed subspace to be the approximate occupied eigensubspace of the $\textbf{k}$ dependent FE discretized Hamiltonian, $\btH_{\textbf{k}}[\rho(\lambda=0)]$, obtained from the ChFSI procedure. Second, the constraint on conservation of number of electrons is extended to BZ sampling:
\begin{align}\label{eq:tracelessCondition}
\fintd_{BZ} \, \int_{\Omega}\, 2 \,\boldsymbol{n^{{e}^{T}}}({\bf x})\left[{\bf M}^{-1/2} \,\bGamtildekappa^{(1)} \, {\bf M}^{{-1/2}^{\dagger}}\right] &\boldsymbol{n^e}({\bf x}) d\bx\,d\boldsymbol{\textbf{k}}\notag\\&=0 \,,
\end{align}
where the value $\mu^{\prime}$ is now determined using the above constraint. 

\cred We remark that this diagonalization free first-order perturbation response methodology can, in principle, be applied to other finite temperature smearing schemes,  such as Methfessel-Paxton smearing~\cite{MethfesselPaxton} and cold-smearing~\cite{marzari1999thermal}, in conjunction with appropriate expansions~\cite{liang2003improved}. Methfessel-Paxton and cold-smearing allow for usage of relatively higher smearing temperatures compared to Fermi-Dirac smearing due to their lower finite-temperature errors in ground-state internal energy and forces, and hence popularly used in various DFT codes to reduce charge-sloshing effects. We will demonstrate in Section~\ref{sec:results} the robust performance of the developed LRDM preconditioner for bulk and heterogeneous metallic systems with first BZ sampling, while employing low to modest Fermi-Dirac smearing temperatures of $T=10$ K and $T=500$ K. Thus, the requirement of higher smearing temperatures to accelerate SCF convergence can be avoided.~\cn

\subsection{Extension of low rank approximation of $J$ to spin-polarized Kohn-Sham DFT}\label{sec:lrdmSpin}
We develop an extension of the LRDM preconditioner discussed in Section~\ref{sec:lrdm} to collinear spin-polarized Kohn-Sham DFT calculations. To this end, we define the residual in the spin-polarized case to be given by
\begin{equation}
    {R[\left(\rho_{\uparrow},\rho_{\downarrow}\right)]}={F[V_{\textrm{eff}}[\left(\rho_{\uparrow},\rho_{\downarrow}\right)]]}- \left(\rho_{\uparrow},\rho_{\downarrow}\right)\,,
\end{equation}
where $\left(\rho_{\uparrow},\rho_{\downarrow}\right) \in \Upsilon \times \Upsilon$ denotes an ordered pair of spin-up and spin-down density functions belonging to the Cartesian product of the function space $\Upsilon$ with itself.  $F[V_{\textrm{eff}}[(\rho_{\uparrow},\rho_{\downarrow})]]:\Upsilon \times \Upsilon \rightarrow \Upsilon \times \Upsilon$ represents the Kohn-Sham input to output spin-polarized density map in each SCF iteration. The inner product between two ordered pair of functions $\left(f_1,f_2\right) \in \Upsilon \times \Upsilon$ and $\left(g_1,g_2\right) \in \Upsilon \times \Upsilon$ is defined as
\begin{equation}\label{eq:innerproductDirectsumHilbertspace}
  \innerproductcomma{\left(f_1,f_2\right)}{\left(g_1,g_2\right)}=  \innerproductcomma{f_1}{g_1} +  \innerproductcomma{f_2}{g_2}\,.
\end{equation}
The associated norm induced by the inner product for an ordered pair $\left(f_1,f_2\right)$ is given by $\norm{\left(f_1,f_2\right)}=\sqrt{\innerproductcomma{\left(f_1,f_2\right)}{\left(f_1,f_2\right)}}$.

We solve for $R[\left(\rho_{\uparrow},\rho_{\downarrow}\right)]=0_{\Upsilon \times \Upsilon }$ using a damped quasi-Newton iteration scheme:
\begin{align}
& \left(\rho_{\uparrow}^{(n+1)},\rho_{\downarrow}^{(n+1)}\right)=\left(\rho_{\uparrow}^{n},\rho_{\downarrow}^{n}\right) -  \alpha \, P \,{R[{\left(\rho_{\uparrow}^n,\rho_{\downarrow}^n\right)}]} \,,\\
         P &  \approx J^{-1} \quad {\rm with} \quad  
         J = \left.\frac{\partial R[\left(\rho_{\uparrow},\rho_{\downarrow}\right)]}{\partial \left(\rho_{\uparrow},\rho_{\downarrow}\right)}\right|_{\left(\rho_{\uparrow},\rho_{\downarrow}\right)={\left(\rho_{\uparrow}^n,\rho_{\downarrow}^n\right)}}\,,
\end{align}
where $\alpha \in (0,1]$, and $P: \Upsilon \times \Upsilon \rightarrow \Upsilon \times \Upsilon$ approximates the inverse Jacobian of the residual. Analogous to the the spin-restricted case, the Jacobian in the spin-polarized case also admits a canonical decomposition into sum of tensor products of rank-1 components. In particular, we consider the following rank-$r$ approximation form:
\begin{align}
     & J_r^{\textrm{lr}} (\bx,\bxprime,\sigma,\sigma^{\prime}) =\sum_{i=1}^{r} {\left(t_{\uparrow},t_{\downarrow}\right)}^i_{\sigma} (\bx) {\left(u_{\uparrow},u_{\downarrow}\right)}^i_{\sigma^{\prime}} (\bxprime) \,,\notag\\&\quad{\rm where}\quad {\left(t_{\uparrow},t_{\downarrow}\right)}^i=\left.\frac{\partial}{\partial \lambda } \left(R \left[\left(\rho_{\uparrow}^n,\rho_{\downarrow}^n\right)+\lambda {\left(u_{\uparrow},u_{\downarrow}\right)}^i\right]\right)\right|_{\lambda=0}\,.
\end{align}
In the above, ${\left(t_{\uparrow},t_{\downarrow}\right)}^i \in \Upsilon \times \Upsilon$ represents the generalized directional derivative of $R[\left(\rho_{\uparrow},\rho_{\downarrow}\right)]$ along ordered pair of orthonormal direction functions, ${\left(u_{\uparrow},u_{\downarrow}\right)}^i \in \Upsilon \times \Upsilon$. Further, $(,)_{\sigma}$ denotes the choice of one of the functions $\sigma=\uparrow/\downarrow$ from the ordered pair. The orthonormality condition is given by $\innerproductcomma{{\left(u_{\uparrow},u_{\downarrow}\right)}^i}{{\left(u_{\uparrow},u_{\downarrow}\right)}^j}=\delta_{ij}$. Following Section~\ref{sec:lrdm}, for efficient convergence of the preconditioner $P_r^{\textrm{lr}}$ with respect to $r$, the ordered pairs  ${\left(u_{\uparrow},u_{\downarrow}\right)}^i $ are obtained from an approximate $r$-rank Krylov subspace of $J$, where we first choose ${\left(u_{\uparrow},u_{\downarrow}\right)}^1= R\left[\left(\rho_{\uparrow}^n,\rho_{\downarrow}^n\right)\right]/\norm{R\left[\left(\rho_{\uparrow}^n,\rho_{\downarrow}^n\right)\right]}$, and subsequently use the following iterative procedure for $1<i\le r$ :
\begin{align}\label{eq:krylovOrthoSpin}
        \,\,{\left(u_{\uparrow},u_{\downarrow}\right)}^i=&{\left(t_{\uparrow},t_{\downarrow}\right)}^{i-1}\notag\\
        \,\,{\left(u_{\uparrow},u_{\downarrow}\right)}^i=&{\left(u_{\uparrow},u_{\downarrow}\right)}^i- \notag\\&\sum_{k=1}^{i-1} \innerproductcomma{{\left(u_{\uparrow},u_{\downarrow}\right)}^i}{{\left(u_{\uparrow},u_{\downarrow}\right)}^k}  {\left(u_{\uparrow},u_{\downarrow}\right)}^k\,\quad \notag\\
        \,\,{\left(u_{\uparrow},u_{\downarrow}\right)}^i=&{\left(u_{\uparrow},u_{\downarrow}\right)}^i/\norm{{\left(u_{\uparrow},u_{\downarrow}\right)}^i}\,.
\end{align}
The spin-polarized extension of the remaining aspects of the low rank formulation, including the accumulated variant, follow along similar lines as discussed in Section~\ref{sec:lrdm}.

Finally, we mention the additional considerations in extending the Chebyshev filtered subspace projected density response computation discussed in Section~\ref{sec:densityResponse} to the spin-polarized case. Adopting a similar approach as the spin-restricted case, we approximately evaluate the density-matrix response corresponding to the projection of  $\btH_{\sigma}\left[\left(\rho_{\uparrow},\rho_{\downarrow}\right)(\lambda)\right]$  onto  $\bPsitildeRsigma$ as
\begin{align}
     \bGamtildesigma^{(1)}= &\bPsitildeRsigma   \left. \frac{\partial }{\partial \lambda} \Big(  f\big(\bbreveH_{\sigma} + \lambda \bbreveH^{\prime}_{\sigma}\,,\,\mu+\lambda \mu^{\prime}\big)  \Big)\right|_{\lambda=0}   \bPsitildeRsigmaDagger \,,
\end{align}
 where   
\begin{equation}
 \bbreveH^{\prime}_{\sigma}=\bPsitildeRsigmaDagger\left.\frac{\partial}{\partial \lambda} \Big(\btH_{\sigma}[\left(\rho_{\uparrow},\rho_{\downarrow}\right)+\lambda {\left(u_{\uparrow},u_{\downarrow}\right)}^i]\Big)\right|_{\lambda=0} \bPsitildeRsigma\,
\end{equation}
is  the first-order response of $\btH_{\sigma}\left[\left(\rho_{\uparrow},\rho_{\downarrow}\right)(\lambda)\right]$ along ${\left(u_{\uparrow},u_{\downarrow}\right)}^i$ projected onto $\bPsitildeRsigma$. 
Subsequently, the desired directional derivative of $F\left[V_{\textrm{eff}}\left[\left(\rho_{\uparrow},\rho_{\downarrow}\right)\right]\right]$ along ${\left(u_{\uparrow},u_{\downarrow}\right)}^i$ is obtained as
\begin{align}
&\left.\frac{\partial }{\partial \lambda } \Big( F[V_{\textrm{eff}}[\left(\rho_{\uparrow},\rho_{\downarrow}\right)+\lambda {\left(u_{\uparrow},u_{\downarrow}\right)}^i]]_{\sigma}\Big)\right|_{\lambda=0} \approx  \notag\\ &\,(\boldsymbol{n^{{e}}}({\bf x}))^{T}\left[{\bf M}^{-1/2} \,\bGamtildesigma^{(1)} \, {\bf M}^{{-1/2}^{\dagger}}\right]\boldsymbol{n^e}({\bf x})
\end{align}

\cb Finally, we remark that above numerical method for the first-order density response calculations and the proposed spin-polarized LRDM formulation is adaptable to other discretizations (plane-waves, finite-difference, wavelets) and eigensolver strategies other than the ChFSI procedure (Davidson, RMM-DIIS) implemented in various Kohn-Sham DFT codes. For instance, using the Davidson iterative eigensolver implemented in the Quantum Espresso software (plane-wave basis), the smaller subspace required for the efficient first-order density response calculations can be constructed from the eigenpairs computed in the Davidson method. Other aspects of the LRDM method including the use of the canonical density-matrix perturbation approach is agnostic to the choice of the discretization and eigensolver strategy.
\cn

\section{Results and discussion}
\label{sec:results}
In this section, we demonstrate the robustness and computational efficiency of the LRDM preconditioner for Kohn-Sham DFT calculations. We compare the performance against three other widely used preconditioners, namely, Anderson mixing~\cite{anderson1965}, Anderson mixing with Kerker preconditioner~\cite{Kerker1981}, and Broyden mixing with Thomas-Fermi-von Weizsacker preconditioner (TFW)~\cite{Raczkowski2011}. We consider various heterogeneous benchmark systems with system sizes up to $\sim$1,100 atoms ($\sim$20,000 electrons) and study the convergence of LRDM preconditioner and the accumulated variant, LRDMA. Furthermore, we also demonstrate the robustness of the proposed extension of the LRDM preconditioner to spin-polarized DFT calculations. Finally,  we comment on the computational cost of the LRDM preconditioner, and compare against Anderson and Kerker preconditioners on hybrid CPU-GPU architectures.

\subsection{General calculation details}\label{sec:generalCalc}
In all the DFT calculations reported in this work, we use the PBE exchange correlation functional~\cite{pbe} and ONCV~\cite{ONCV2013} pseudopotentials from the  PseudoDojo~\cite{van2018pseudodojo} database. \cb Unless otherwise specified, \cn we use Fermi-Dirac smearing with $T = 500$ $K$ in all our simulations. Additionally for periodic benchmark systems, we use shifted Monkhorst-Pack k-point grids to sample the first Brillouin zone with the minimum k-point spacing chosen to be $\sim$0.3 $\textrm{\AA}^{-1}$.

We have implemented the LRDM preconditioner in the \DFTFE software~\cite{das2022dft,motamarri2020,motamarri2013}, a recently developed open-source code for massively parallel large-scale real-space Kohn-Sham DFT calculations based on a finite-element discretization. \DFTFE already has implementations of Anderson mixing and Anderson with Kerker preconditioning. Thus, in this work, simulations using Anderson and Kerker preconditioners are performed using \DFTFE, while simulations using TFW preconditioner are performed using the implementation available in Quantum Espresso software~\cite{qe2009,qe2017} (\,\QE) using the same ONCV pseudopotential input files\cn. All simulations using \DFTFE for benchmarking the LRDM, Anderson and Kerker preconditioners are performed till $10^{-5}$ stopping tolerance in $L_2$ norm of electron-density residual ($\norm{R[\rho]}$). The simulations using \QE for the TFW preconditioner employ $5 \times 10^{-8}$ Ha as stopping tolerance in the total energy between consecutive SCF iteration steps. We remark that these two stopping criteria are approximately equivalent for the range of system sizes considered in this work, as will be demonstrated in the convergence studies reported in Section~\ref{sec:convergenceStudies}. Further, we note that the finite-element and plane-wave discretization parameters for \DFTFE and \QE are chosen such that we obtain $\sim$10 meV/atom accuracy in the ground-state energies. \cb We additionally remark that the different eigensolver implementations---Chebyshev filtered subspace iteration (ChFSI) procedure in \DFTFE and the Davidson iteration diagonalization  in \QE---can lead to differences in the SCF convergence behaviour for Anderson multi-secant method and its coupling with Kerker. We refer to Table I in Supplemental Material~\cite{supplementary} for a numerical comparison of the multi-secant preconditioner performance, both with and without the Kerker preconditioner, between \DFTFE and \QE on a subset of the benchmark systems studied below. The benchmark results demonstrate similar trends between \DFTFE and \QE implementations.\cn

Next, we mention the choice of the damping parameter, $\alpha$, and other preconditioner specific choices used in the benchmark studies. First, in the case of LRDM preconditioner, we choose $s_{\textrm{tol}}=0.2$--$0.3$ for adaptively setting the rank $r$ in each SCF iteration step (cf. Eq.~\eqref{eq:relativeError}). Further, we choose the damping parameter $\alpha=0.1$ till $\norm{R[\rho]}$ goes below 2.0, and subsequently switch to $\alpha=0.5$.  Additionally, for the accumulated variant of LRDM preconditioner proposed in Section~\ref{sec:lrdm-accumulated-method} (LRDMA), we use the tolerance on the linearity indicator to be $\beta_{\textrm{tol}}=0.1$, maximum additional rank that can be accumulated in an SCF iteration step to be $r^{\textrm{max}}_{\textrm{iter}}=5$, and further allow accumulation only when $\norm{R[\rho]} < 1.0$. The above parameters are determined based on numerical experiments, and provide a robust convergence for a wide range of benchmark material systems with increasing complexity and system sizes considered in this work, without any system specific tuning in the parameter values. In the SCF convergence studies to be discussed below, we also use the LRDM preconditioner computational framework to approximately compute the condition number of $J$, $\kappa(J)$, at the converged ground-state solution $\rho^{(\ast)}$, to assess the difficulty of convergence of the SCF iteration for each materials system. Specifically, for estimating $\kappa(J)$, we use a stringent low-rank approximation tolerance of $s_{\textrm{tol}}=5 \times 10^{-4}$ along with employing a relatively larger Chebyshev filtered eigensubspace ($\sim$ 25\% buffer over $N_e/2$) to obtain a closer approximation $J^{\textrm{lr}}_r$ to  $J$, and finally perform a power iteration on $J^{\textrm{lr}}_r$ and ${P^{\textrm{lr}}_r}$ to obtain estimates of the highest and lowest eigenvalues. Second, in the case of Anderson mixing, we use values of $\alpha$ between 0.015--0.05 and a mixing history range between 20--50. The relatively small values of $\alpha$ for Anderson mixing are required due to the large condition numbers of $J$ for the heterogeneous material systems considered in this work. Third, in the case of Anderson mixing with Kerker preconditioner, we use $\alpha=0.5$, mixing history of 20, and Thomas Fermi screening wavevector value of 0.8 ${\textrm{Bohr}}^{-1}$. In the case of TFW preconditioner, the parameters are set to be $\alpha=0.5$--$0.7$ and a mixing history of 15. The above parameter choices for Kerker and TFW preconditioners are determined based on numerical experiments to be broadly optimal for all the benchmark systems considered in this work. Finally, we remark that for the LRDM, Anderson, and Kerker preconditioners implemented in \DFTFE, we perform multiple sweeps of the ChFSI procedure in each SCF iteration till the residual norm of the eigenpair closest to the Fermi energy is below a tolerance of $2 \times 10^{-3}$.


All the numerical simulations with computational times reported in this work were executed using the using the Phase 1 GPU-accelerated nodes of the NERSC Perlmutter supercomputer, with each node containing four NVIDIA A100 Tensor Core GPUs and a single AMD Milan CPU (64 physical cores). Some of the simulations reported in this work were also executed on the OLCF Summit and XSEDE Stampede2 supercomputers. Summit comprises of 4,608 IBM Power System AC922 nodes with two IBM POWER9 processors (42 physical cores) and six NVIDIA Volta V100 GPUs in each node.

\subsection{Comparison of SCF convergence}\label{sec:convergenceStudies}

\begin{table*}[htpb]
\centering
\small
\caption{\label{tab:comparisonNoSpin}\small{Comparison of the performance of LRDM against other preconditioners. All DFT calculations are spin restricted, and use k-point sampling of the Brillouin zone for semi-periodic systems. A stopping tolerance of $10^{-5}$ in $L_2$ norm of the electron-density residual ($\norm{R[\rho]}$) is used. Based on numerical experiments, this is approximately equivalent to a stopping tolerance of $5 \times 10^{-8}$ Ha in the total energy difference between consecutive SCF iteration steps. DNC denotes that calculation did not converge within 200--250 SCF iterations.}}
\begin{tabular}{c c c c c c c } 
 \hline
 \hline
 System & $N_{\textrm{at}}$ & $\kappa(J)$ & Anderson & Kerker & TFW & {LRDM} ($r_{\textrm{avg}}$) \\ 
 \hline
  Pt cubic-NP--$3\times3\times3$ & 172 & 170 & 53 & 182 & 78 & 24 (7.3)  \\
  \hline
   Pt cubic-NP--$5\times5\times5$ & 666 & 390 & 73 & 138 & 97 & 28 (10.0)  \\ 
  \hline
    Pt cubic-NP--$6\times6\times6$ & 1099 & 557 & 90 & DNC & - & 30  (11.0)\\ 
  \hline
    Si$\textrm{O}_\textrm{2}$H--20 layers & 98 & 1.7 & 16 & 58 & 37 & 22 (1)  \\ 
   \hline
    \cb Si$\textrm{O}_\textrm{2}$--20 layers \cn & \cb 90 \cn & \cb 382 \cn  & \cb 87 \cn & \cb 75 \cn & \cb 34 \cn & \cb 25 (6.1)\cn  \\    
  \hline
    $\textrm{Li}_{\textrm{10}}\textrm{Ge}\textrm{P}_{\textrm{2}}\textrm{S}_{\textrm{12}}$ layers & 400 & 133 & DNC & 69 & 85 & 23 (4.9)  \\     
 \hline
   Pt+Si$\textrm{O}_\textrm{2}$H--10 layers  & 93 & 552 & 112 & 68 & 35 & 23 (6.5) \\ 
  \hline
 Pt+GaAs+Si$\textrm{O}_\textrm{2}$H--10 layers & 133 & 454 & 92 & 61 & 39 & 23 (6.9) \\ 
 \hline
 Au+GaAs+Si$\textrm{O}_\textrm{2}$H--10 layers & 133 & 222 & 93 & 30 & 26 & 22 (4.9)  \\
 \hline
 Al+GaAs+Si$\textrm{O}_\textrm{2}$H--10 layers & 133 & 199 & 55 & 26 & 23 & 23 (4.8) \\
 \hline
 Pt+GaAs+Si$\textrm{O}_\textrm{2}$H--20 layers & 258 & 1662 & DNC & 126 & 78 & 25 (9.5)  \\ 
 \hline
 Au+GaAs+Si$\textrm{O}_\textrm{2}$H--20 layers & 258 & 888 & DNC & 46 & 39 & 24  (6.6) \\
 \hline
 Al+GaAs+Si$\textrm{O}_\textrm{2}$H--20 layers & 258 & 708 & DNC & 40 & 46 & 31 (7.8) \\
 \hline
 Pt+GaAs+Si$\textrm{O}_\textrm{2}$H--40 layers & 508 & 9841 & DNC & 248 & DNC & 28 (13.8)  \\
 \hline
 Au+GaAs+Si$\textrm{O}_\textrm{2}$H--40 layers & 508 & 1294 & DNC & 74 & DNC & 29 (11.0) \\
 \hline
 Al+GaAs+Si$\textrm{O}_\textrm{2}$H--40 layers & 508 & 2051 & DNC & 76 & 112 & 30 (9.7) \\
 \hline
 Hf$\textrm{O}_\textrm{2}$ nano-film w/Al stripes & 864 & 535 & 125 & 85 & 155 & 31 (6.2) \\
 \hline
 \hline
\end{tabular}
\end{table*}

\begin{figure*}[t!]
    \centering
    \subfloat[\label{subfig:a}]{\includegraphics[scale=0.3]{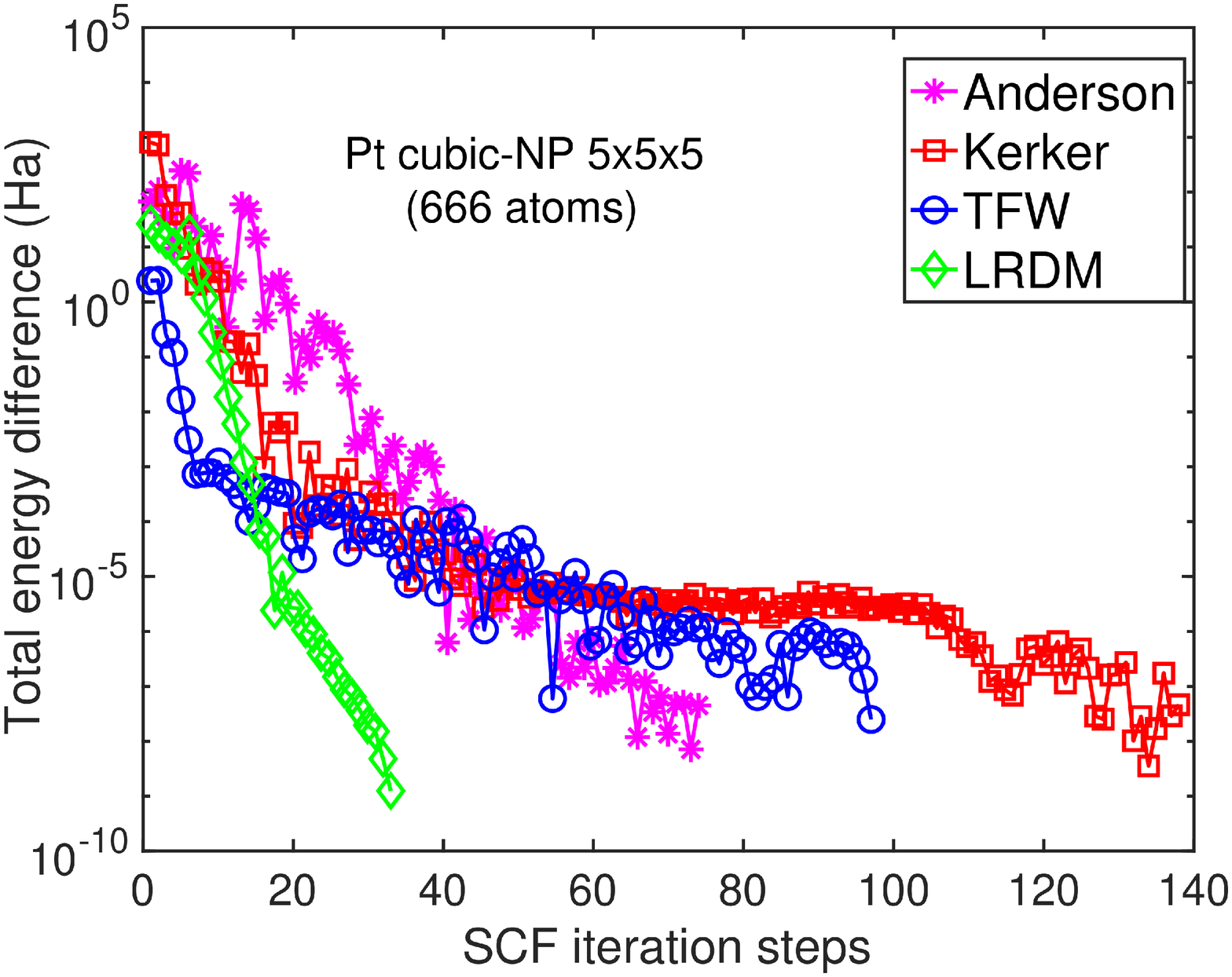}}    
    ~
       \subfloat[\label{subfig:b}]{\includegraphics[scale=0.3]{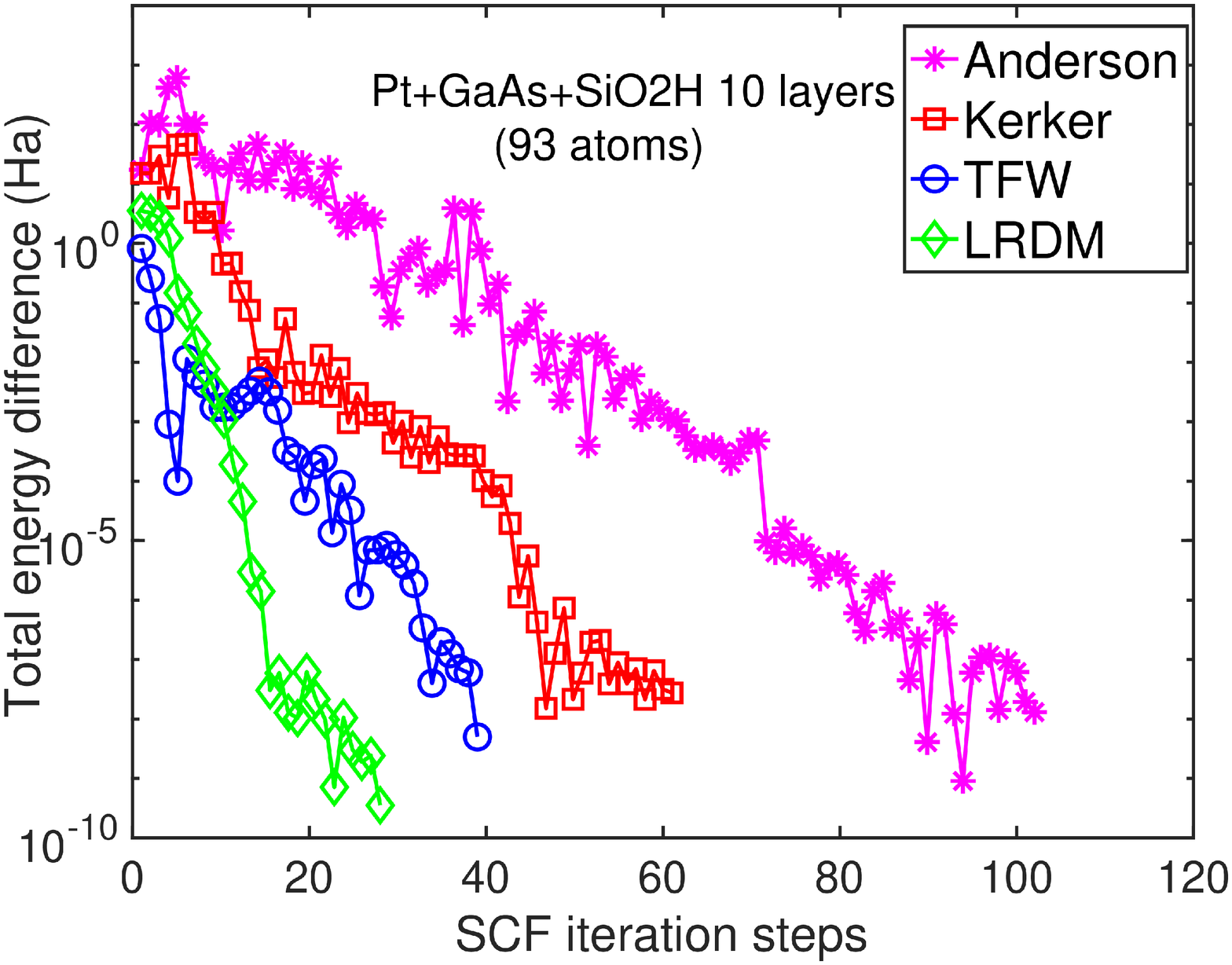}}\\
      \subfloat[\label{subfig:c}]{\includegraphics[scale=0.3]{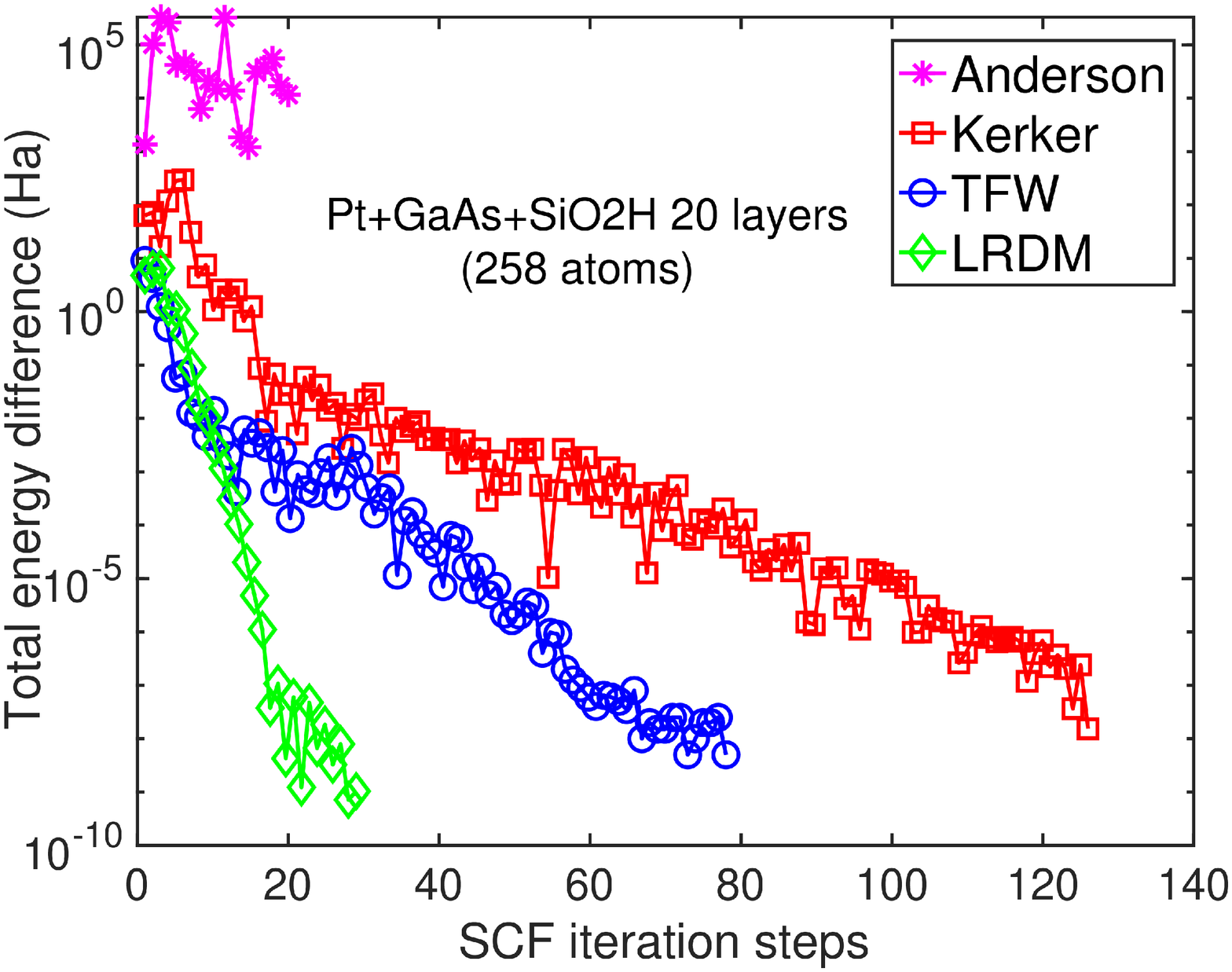}}
    ~
\subfloat[\label{subfig:d}]{\includegraphics[scale=0.3]{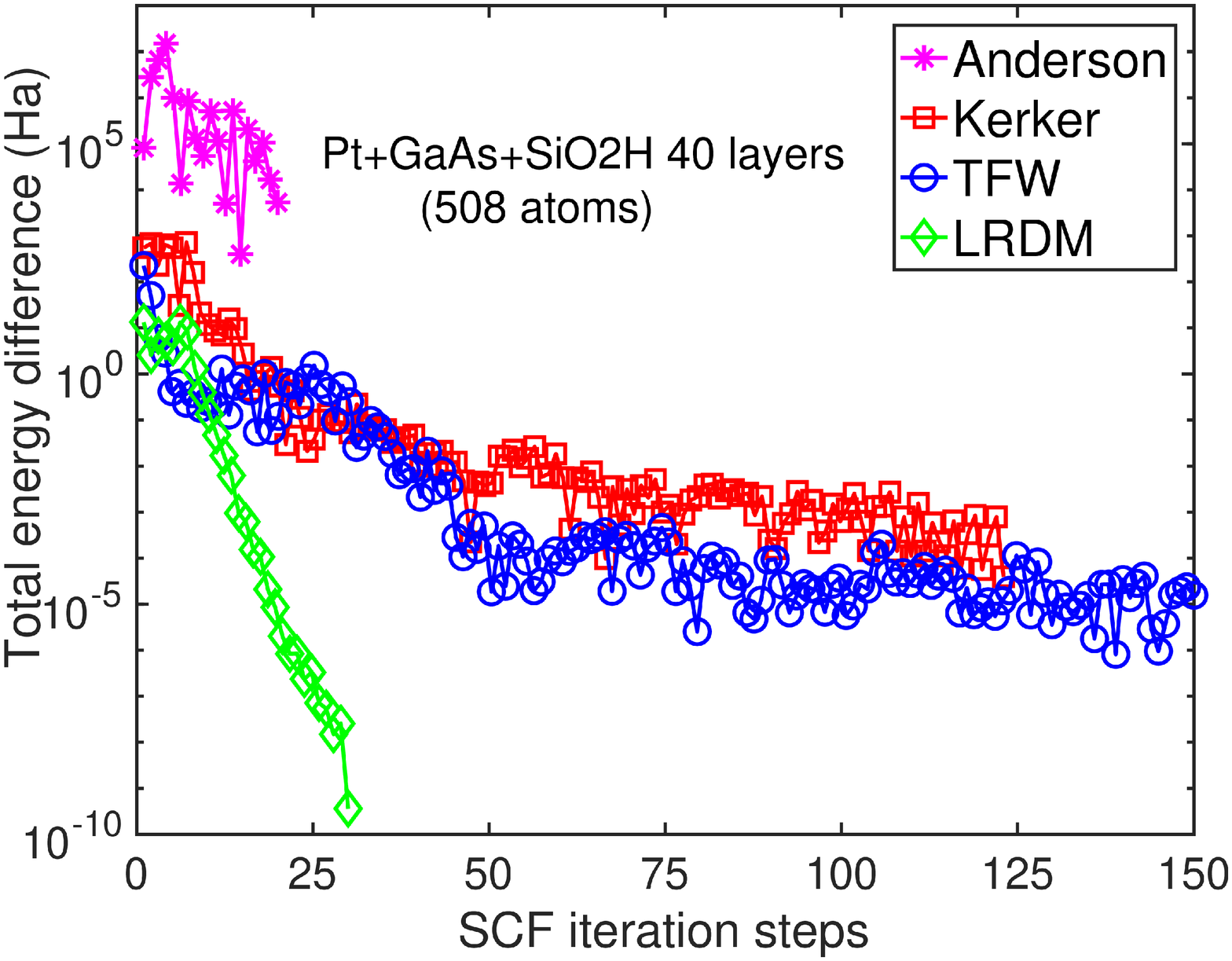}}
    \caption{\small{Convergence of the difference in total energy between successive Kohn-Sham SCF iterations for the various methods---Anderson mixing, Kerker preconditioner (Thomas-Fermi screening approximation of $\epsilon$), TFW preconditioner (Thomas-Fermi-von Weizsacker approximation of $\epsilon$), and the LRDM preconditioner. Absolute values of the total energy difference are plotted against the SCF iteration number (starting from 1).}}
    \label{fig:energyConvergence}
\end{figure*}

We first study the SCF convergence for the metal-vacuum type materials systems with increasing system sizes. As benchmark systems, we consider non-periodic cubic FCC Pt nanoparticles (NP) of increasing sizes, with the cube faces along the $\{100\}$ crystallographic planes and a vacuum layer of $\sim$15 \AA~around the nanoparticle. We note that the localized 5d orbitals of Pt near the Fermi energy can lead to large eigenvalues of $\chi_0$ that makes these calculations challenging. Table~\ref{tab:comparisonNoSpin} reports the approximate value of $\kappa(J)$ for three different sizes---$3\times3\times3$, $5\times5\times5$, and $6\times6\times6$ containing 172--1099 atoms (3,096--19,782 electrons)---and compares the SCF convergence of Anderson mixing, Kerker, TFW and LRDM preconditioners. We observe that the convergence of Anderson mixing deteriorates with system size requiring up to 90 SCF iteration steps for the Pt cubic-NP--$6\times6\times6$ system. This corresponds to the approximately 4$\times$ increase in $\kappa(J)$ as would be excepted from the $L^2$ scaling of the long wavelength eigenmode from the Coulomb kernel in the metallic region (cf. Section~\ref{sec:background}), where $L$ denotes the extent of the metallic region. We remark that the convergence of Anderson mixing did not improve upon increasing the mixing history from 20 to 50. The Kerker preconditioner performs worse than Anderson for this benchmark system, requiring around 150 SCF iterations or more. This can be attributed to the unsuitability of Thomas-Fermi screening for metal-vacuum systems. Likewise, the TFW preconditioner, which approximates a heterogeneous dielectric function, also requires a large number of SCF iterations of $\sim$100. We note that although the orbital-free Thomas-Fermi-von Weizsacker model can qualitatively capture the metal-vacuum transition region~\cite{Raczkowski2011}, it is not suited for systems with large eigenvalues of $\chi_0$ due to the localized states near the Fermi energy. Finally, considering the LRDM preconditioner, it achieves accelerated convergence within 24--30 SCF iteration steps for all the system sizes considered, up to 3$\times$ lesser iteration steps compared to Anderson, the next best performing method for these benchmark systems.  We note that even for the largest system size of 1099 atoms (19,782 electrons), $r_{\textrm{avg}}$ in LRDM is a modest value of  11, with further reduction in the rank achieved by using LRDMA as will be demonstrated in Section~\ref{sec:rankreduction}. We also refer to Figure~\ref{fig:energyConvergence}(a) where we observe a noticeable plateauing of the energy difference in successive SCF iterations beyond the initial 20--40 SCF iteration steps for Kerker and TFW preconditioners. In contrast, LRDM demonstrates close to an exponential convergence of energy differences.

Next, we consider semiconductor-vacuum benchmark systems. We consider two semi-periodic benchmark systems: (i) 20 $(110)$ plane mono-layers of  Si$\textrm{O}_\textrm{2}$ with the silica surfaces passivated by hydrogen, \cb (ii) the same silica layers system but without the surface passivation, and (iii) \cn a disordered $\textrm{Li}_{\textrm{10}}\textrm{Ge}\textrm{P}_{\textrm{2}}\textrm{S}_{\textrm{12}}$ (LGPS superionic conductor) slab. The benchmark system (i) is obtained from a previous work related to the development and testing of a local density of states based preconditioner (LDOS)~\cite{Herbst_2020}. The benchmark system (iii) is constructed based on the atomic coordinates and occupation factors obtained from X-ray structure analysis of LGPS crystal~\cite{kuhn2013single}, and further a vacuum layer is considered normal to the slab surfaces\footnote{This corresponds to a model system for investigations of surface effects in LGPS.}. First, analyzing the SCF convergence of silica layers \cb with surface passivation\cn, we observe that both the Anderson mixing and LRDM preconditioner converge efficiently taking $\sim$20 iterations, whereas Kerker and TFW preconditioners demonstrate a slow convergence despite a low $\kappa(J)\sim2$ for this system. The slower convergence is attributed to the well-known inability of Kerker and TFW preconditioners to qualitatively approximate the dielectric function of semiconductors, which can, in turn, even deteriorate the net conditioning of the preconditioned quasi-Newton problem (cf. Eq.~\eqref{eq:convergence}). \cb In case of the unpassivatied silica layers, the convergence of Anderson and Kerker deteriorates due to high $\kappa(J)\sim 400$ caused by the dangling bonds, whereas the LRDM preconditioner requires similar number of SCF iterations as the passivated case.~\cn Subsequently, analyzing the convergence for the LGPS system, which has a much larger value of $\kappa(J)$ compared to the silica layers, we observe that the Anderson mixing did not converge. Both the Kerker and TFW preconditioners also demonstrate slower convergence for LGPS compared to the silica system, taking 3--3.7$\times$ more SCF iteration steps compared to LRDM (23 iterations).

Next, we consider metal-semiconductor-insulator-vacuum heterogeneous benchmark systems. In particular, we consider semi-periodic layered systems of the type (FCC metal)+GaAs+${\textrm{SiO}}_{2}$H--$N_l$ layers, with $N_l$ denoting the number of mono-layers in each of the metal and semiconductor regions. The benchmark systems are adapted from previous work~\cite{Herbst_2020,ldosgithub}, wherein they are constructed by first arranging $N_l$ $(100)$ plane  mono-layers of the FCC metallic crystal, followed by $N_l$ $(110)$ plane mono-layers of GaAs and Si$\textrm{O}_\textrm{2}$. The metal and silica surfaces are exposed to the vacuum. Further, we note that the silica surfaces are passivated by hydrogen and all the interfaces are made to be coherent by applying appropriate affine deformations to the respective regions. We choose three sets of system sizes---$N_l=$ 10, 20, and 40 containing 133--508 atoms (up to $\sim$ 6,100 electrons) and three different metallic elements---Al, Au and Pt. We note that for $N_l=40$, we use 20 mono-layers for the Si$\textrm{O}_\textrm{2}$ region as the benchmark systems obtained from ~\cite{Herbst_2020,ldosgithub} are only available till $N_l=20$. Going to $N_l=40$ from $N_l=20$, we increased the layers of only the metallic and semiconducting layers to maintain the similar hydrogen passivated silica surfaces as $N_l=20$. Examining the SCF convergence of these benchmark systems from Table~\ref{tab:comparisonNoSpin} and Figures~\ref{fig:energyConvergence} (b)-(d) (for Pt), we observe that except in the case of Pt/Al/Au+GaAs+Si$\textrm{O}_\textrm{2}$H--10 layers, Anderson mixing did not converge despite using a small $\alpha=0.015$. We anticipate this to be a result of the large values of $\kappa(J)$ that increase with number of layers, reaching up to $\sim$ 10,000 for Pt+GaAs+Si$\textrm{O}_\textrm{2}$H--40 layers, in combination with the strong nonlinear effects in such highly heterogeneous systems, as will be numerically demonstrated in Section~\ref{sec:rankreduction}. We also remark that the values of $\kappa(J)$ are higher for Pt+GaAs+${\textrm{ SiO}}_{2}$H--$N_l$ compared to Au+GaAs+Si$\textrm{O}_\textrm{2}$H--$N_l$ due to the  partially filled 5d subshell of Pt in comparison to the fully filled 5d subshell in Au. Analyzing the SCF convergence of Kerker and TFW, we observe both approaches deteriorating with increasing $N_l$ as neither the Thomas-Fermi homogeneous screening or the Thomas-Fermi-von Weizsacker model can qualitatively capture the transition of the dielectric function between metal, semiconductor, insulator and vacuum regions. In contrast, the LRDM preconditioner performs remarkably well for these highly heterogeneous and large-scale benchmark systems, converging within $\sim$20--30 SCF iteration steps for all the 9 different benchmark systems considered here. Notably, for the Pt+GaAs+Si$\textrm{O}_\textrm{2}$H--40 layers system, excepting for LRDM preconditioner, all the other preconditioners exhibit significantly deteriorated convergence requiring $\sim$250 SCF iterations or more.  The relatively higher rank $r_{\textrm{avg}}=13.8$ required by LRDM preconditioner for the Pt+GaAs+Si$\textrm{O}_\textrm{2}$H--40 layers system indicates the complex polarization response of the dielectric matrix for such systems. 

Finally, we consider a large-scale semiconductor-metal-vacuum semi-periodic benchmark system, Hf$\textrm{O}_\textrm{2}$ nano-film with Al stripes, used to investigate the origins of ferro-electricity in Al-doped Hf$\textrm{O}_\textrm{2}$ nanofilms~\cite{yao2022modulating}. 
This benchmark system is constructed as a 3 nm thick tetragonal phase Hf$\textrm{O}_\textrm{2}$ (001) nanofilm with vacuum on both sides, and 18 Al dopant atoms distributed  along two separate (001) layers/stripes in the nanofilm. Overall, this system contains 864 atoms (6,750 electrons). The results from the SCF convergence study for this system in Table~\ref{tab:comparisonNoSpin} show slow convergence for Anderson, TFW and Kerker. In contrast, LRDM converges within 31 SCF iteration steps. The deteriorated performance for Anderson/TFW/Kerker, which use the previous SCF history, could be due to non-linear effects. This is corroborated by our findings in Section~\ref{sec:rankreduction} below, where we numerically demonstrate strong non-linear effects for the Hf$\textrm{O}_\textrm{2}$-Al benchmark system.

\subsubsection{SCF convergence comparison for magnetic systems}\label{sec:magneticsystemsConvergence}
We now test the performance of LRDM for magnetic benchmark systems using the spin-polarized extension of LRDM proposed in Section~\ref{sec:lrdmSpin}. We consider two benchmark systems: (i) ferromagnetic BCC Fe $3\times3\times3$ periodic supercell with a monovacancy, and (ii) ferromagnetic ${\textrm{Pt}}_3\textrm{Ni}$ cubic $2\times2\times2$ nanoparticle with a vacuum layer of $\sim$15 \AA~around the nanoparticle.  We remark that ${\textrm{Pt}}_3\textrm{Ni}$ nanoparticles have important applications as catalysts for oxygen reduction reaction in hydrogen fuel cells~\cite{shen2018deconvolution}.  Table~\ref{tab:comparisonSpin} reports the SCF convergence comparison of spin-polarized LRDM against other preconditioners for these benchmark systems. In the case of the close to homogeneous BCC $\textrm{Fe}_{\textrm{vac}}$--$3\times3\times3$ system, we observe  efficient convergence for Kerker, TFW and LRDM preconditioners requiring $\sim$20 SCF iteration steps, compared to Anderson which requires $\sim 2\times$ more iterations. Next, considering the ${\textrm{Pt}}_3\textrm{Ni}$ cubic-NP, we observe accelerated convergence only with the LRDM preconditioner requiring 25 SCF iteration steps, while Anderson, Kerker and TFW require more than 4$\times$ the iterations. The relatively higher $r_{\textrm{avg}}$ of around 15, compared to the spin-restricted  Pt metallic nanoparticle benchmarks discussed previously, indicates the complex nature of the dielectric matrix due to the additional spin-density response of $R[\left(\rho_{\uparrow},\rho_{\downarrow}\right)]$. Additionally, we note that the net magnetization at convergence obtained from LRDM is within 0.1 $\mu_B$ of the net magnetization obtained from other preconditioners.

\begin{table*}[htp]
\centering
\small
\caption{\label{tab:comparisonSpin}\small{Comparison of LRDM performance against other preconditioners for spin polarized DFT benchmark calculations with non-zero net magnetization. A stopping tolerance of $10^{-5}$ in $L_2$ norm of the electron-density residual ($\norm{R[\left(\rho_{\uparrow},\rho_{\downarrow}\right)]}$) is used. Based on numerical experiments, this is approximately equivalent to a stopping tolerance of $5 \times 10^{-8}$ Ha in the total energy difference between consecutive SCF iteration steps. DNC denotes the calculation did not converge within 200--250 SCF iteration steps. $M_{\textrm{tot}} $ denotes the net magnetization.}}
\begin{tabular}{c c c c c c c c} 
 \hline
 \hline
 System & $N_{\textrm{at}}$ & \cb $\kappa(J)$ \cn & Anderson & Kerker & TFW & {LRDM} ($r_{\textrm{avg}}$) & $M_{\textrm{tot}} (\mu_B)$\\ 
 \hline
    BCC $\textrm{Fe}_{\textrm{vac}}$--$3\times3\times3$ & 53 &\cb 64 \cn &  41 & 18 & 16 & 24 (5.0) & 125.8 \\ 
  \hline
   ${\textrm{Pt}}_3{\textrm{Ni}}$ cubic-NP--$2\times2\times2$ & 63 & \cb 128 \cn & 102 & DNC & 108 & 25 (14.8)  & 43.9  \\ 
  \hline
  \hline
\end{tabular}
\end{table*}

\cb
\subsubsection{SCF convergence comparison at lower smearing temperatures}\label{sec:lowerTempConvergence}
We now test the performance of LRDM for a lower Fermi-Dirac smearing temperature $T=10$ K compared to the 500 K used in the earlier benchmarks. Lower smearing temperature leads to larger total density of states at the Fermi energy for metallic systems, which can increase $\kappa(J)$ as $\chi_0(\abs{{\bf q}})$  is proportional to the total density of states at the Fermi energy as $\abs{{\bf q}} \rightarrow 0$. This can also be interpreted more physically as sharpening of the occupancies near the Fermi energy leading to pronounced occupancy sloshing across degenerate or close to degenerate states near the Fermi energy. We reconsider three benchmark systems with metallic character: (i) cubic FCC $3\times3\times3$ Pt NP, (ii) Pt+GaAs+Si$\textrm{O}_\textrm{2}$H layered metal-semiconductor-insulator-vacuum heterogeneous system, and (iii) ferromagnetic BCC Fe $3\times3\times3$ periodic supercell with a monovacancy. Table~\ref{tab:comparisonLowerTemp} reports the lower temperature SCF convergence comparison of LRDM against other preconditioners for these benchmark systems. Comparing the $T=10$~K  results to $T=500$~K results for benchmark systems (i) and (ii) (cf. Table~\ref{tab:comparisonNoSpin}), we first observe that $\kappa(J)$ has increased by 1.5--2$\times$. Overall this results in slower convergence for Anderson, Kerker and TFW preconditioners, whereas the LRDM preconditioner converges in same number of iterations as with  $T=500$ K, although with a slightly higher average rank due to increase in $\kappa(J)$. Subsequently, considering the ferromagnetic BCC Fe monovacancy system, we observe that both LRDM and Kerker preconditioner demonstrate robust convergence similar to the $T=500$~K benchmark results in Table~\ref{tab:comparisonSpin}, whereas the Anderson and TFW convergence is significantly deteriorated. We note that to obtain robust convergence for LRDM in the BCC Fe system, the damping parameter $\alpha$ was reduced from 0.5 to 0.4.

\begin{table*}[htpb]
\centering
\small
\caption{\label{tab:comparisonLowerTemp}\small{\cb Comparison of the performance of LRDM against other preconditioners at low Fermi-Dirac smearing temperature of $T=10$ K. DNC denotes the calculation did not converge within 200--250 SCF iteration steps.\cn}}
\begin{tabular}{c c c c c c c } 
 \hline
 \hline
 \cb System \cn & \cb $N_{\textrm{at}}$ \cn & \cb $\kappa(J)$ \cn & \cb Anderson \cn & \cb Kerker \cn & \cb TFW \cn & \cb {LRDM} ($r_{\textrm{avg}}$) \cn \\ 
 \hline
 \cb Pt cubic-NP--$3\times3\times3$ \cn & \cb 172 \cn & \cb 244 \cn & \cb 64 \cn  & \cb DNC \cn  & \cb 149 \cn  & \cb 24 (8.9) \cn \\
   \hline

 \cb Pt+GaAs+Si$\textrm{O}_\textrm{2}$H--10 layers \cn & \cb 133 \cn & \cb 875 \cn  & \cb 127 \cn & \cb 72 \cn & \cb 54 \cn & \cb 23 (6.9) \cn \\ 
 \hline
   \cb  BCC $\textrm{Fe}_{\textrm{vac}}$--$3\times3\times3$ \cn & \cb 53 \cn &  \cb 69 \cn & \cb 92 \cn & \cb 20 \cn & \cb DNC \cn  & \cb 33 (5.7) \cn \\ 
 \hline
  \hline
\end{tabular}
\end{table*}

\cn

\subsection{Results on $r_{\textrm{avg}}$ reduction using LRDMA}\label{sec:rankreduction}
\begin{figure*}[t!]
\subfloat[\label{subfig:energyconv1}]{
        \includegraphics[scale=0.27]{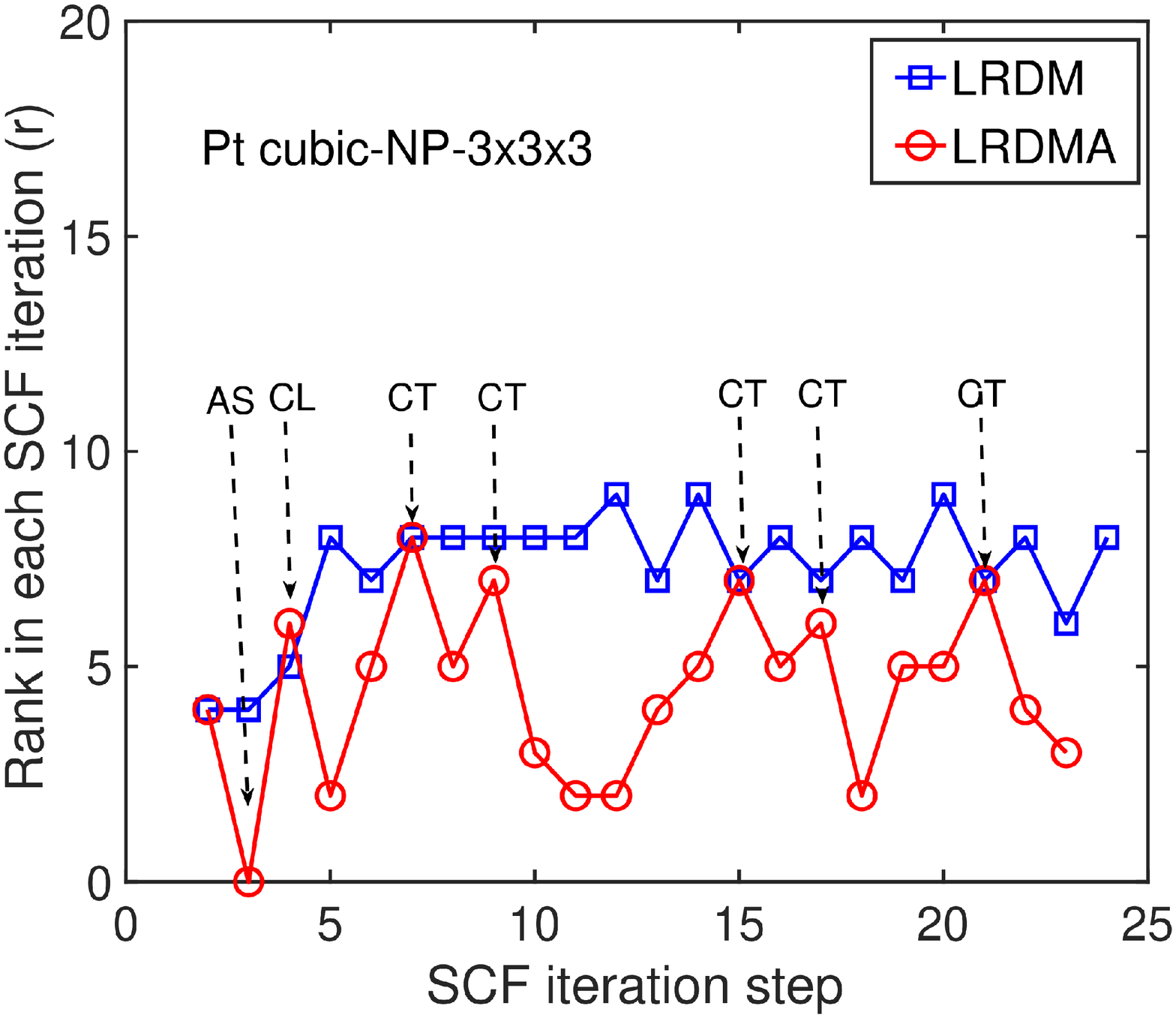}}
    ~
\subfloat[\label{subfig:energyconv2}]{
        \includegraphics[scale=0.27]{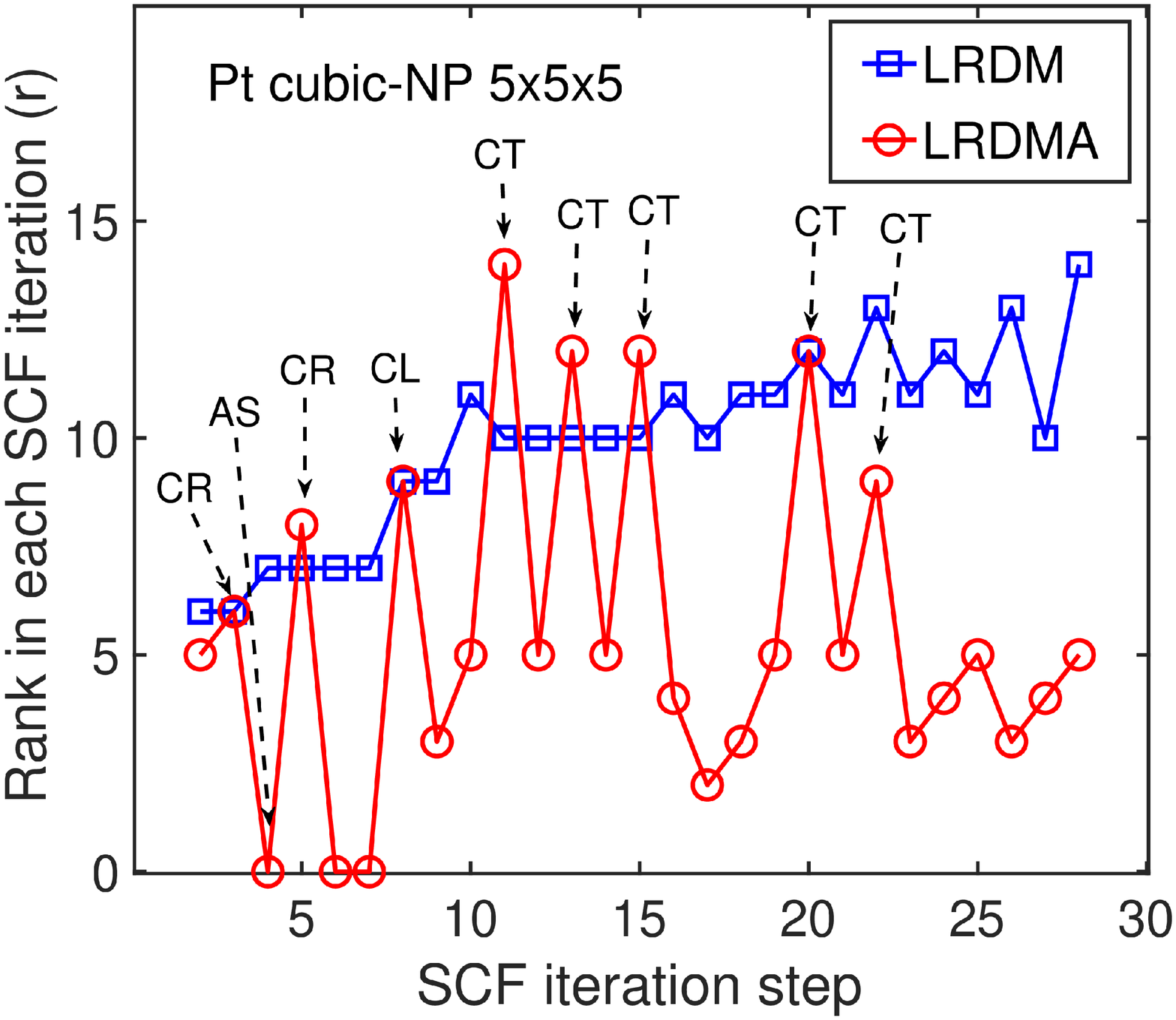}}
    \\
\subfloat[\label{subfig:energyconv3}]{
        \includegraphics[scale=0.27]{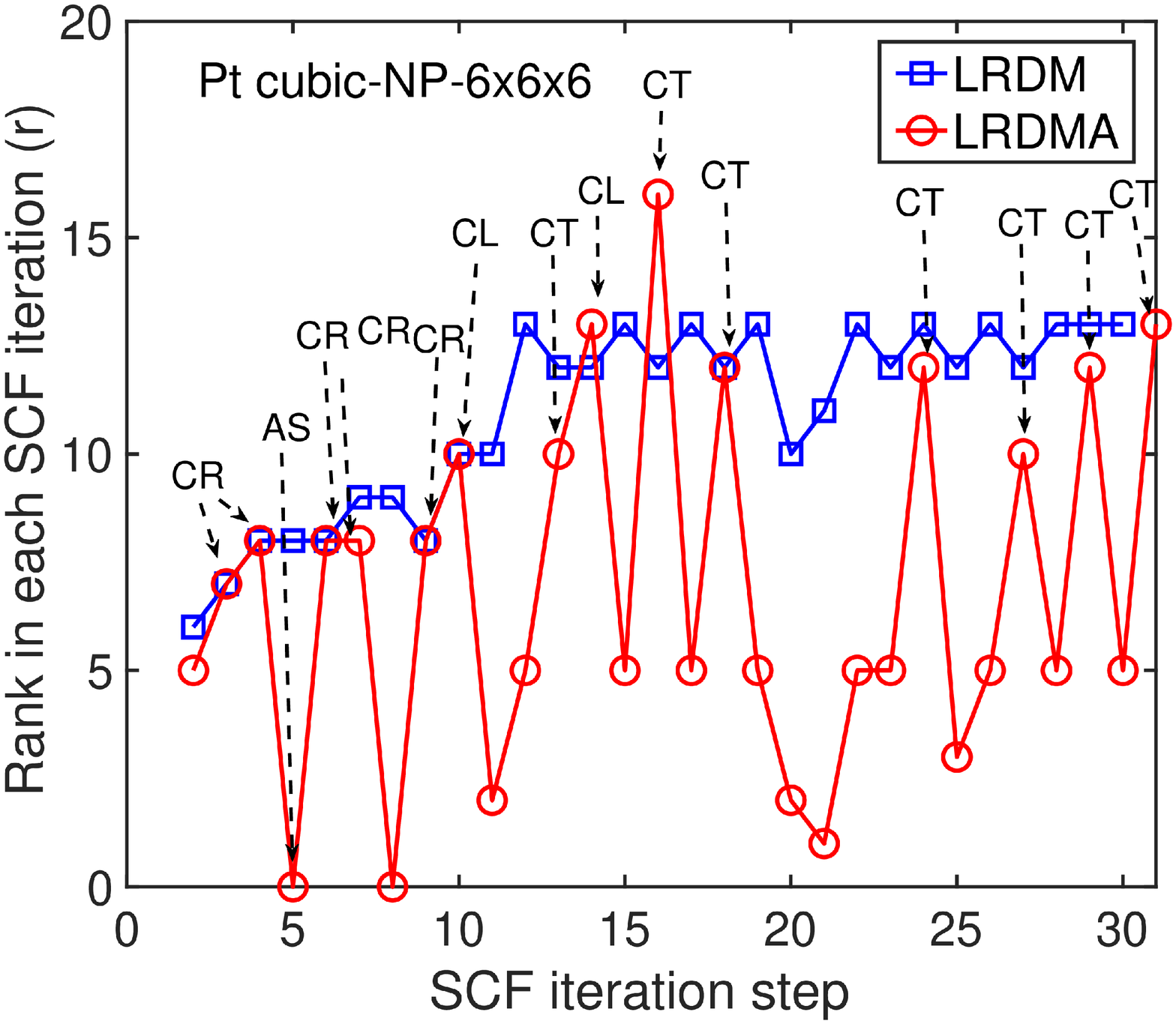}}
    ~
\subfloat[\label{subfig:energyconv4}]{
        \includegraphics[scale=0.27]{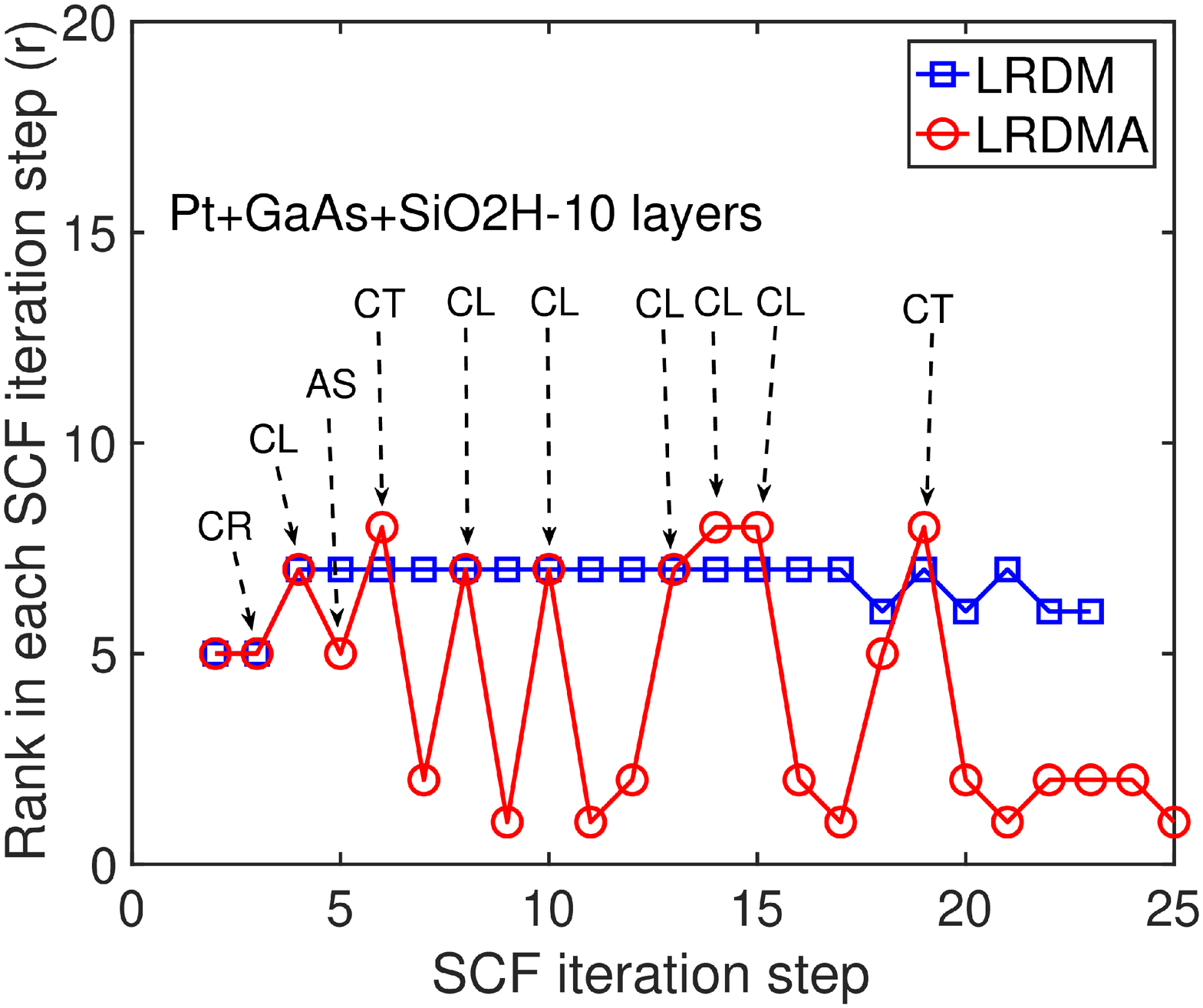}}\\
\subfloat[\label{subfig:energyconv5}]{
        \includegraphics[scale=0.27]{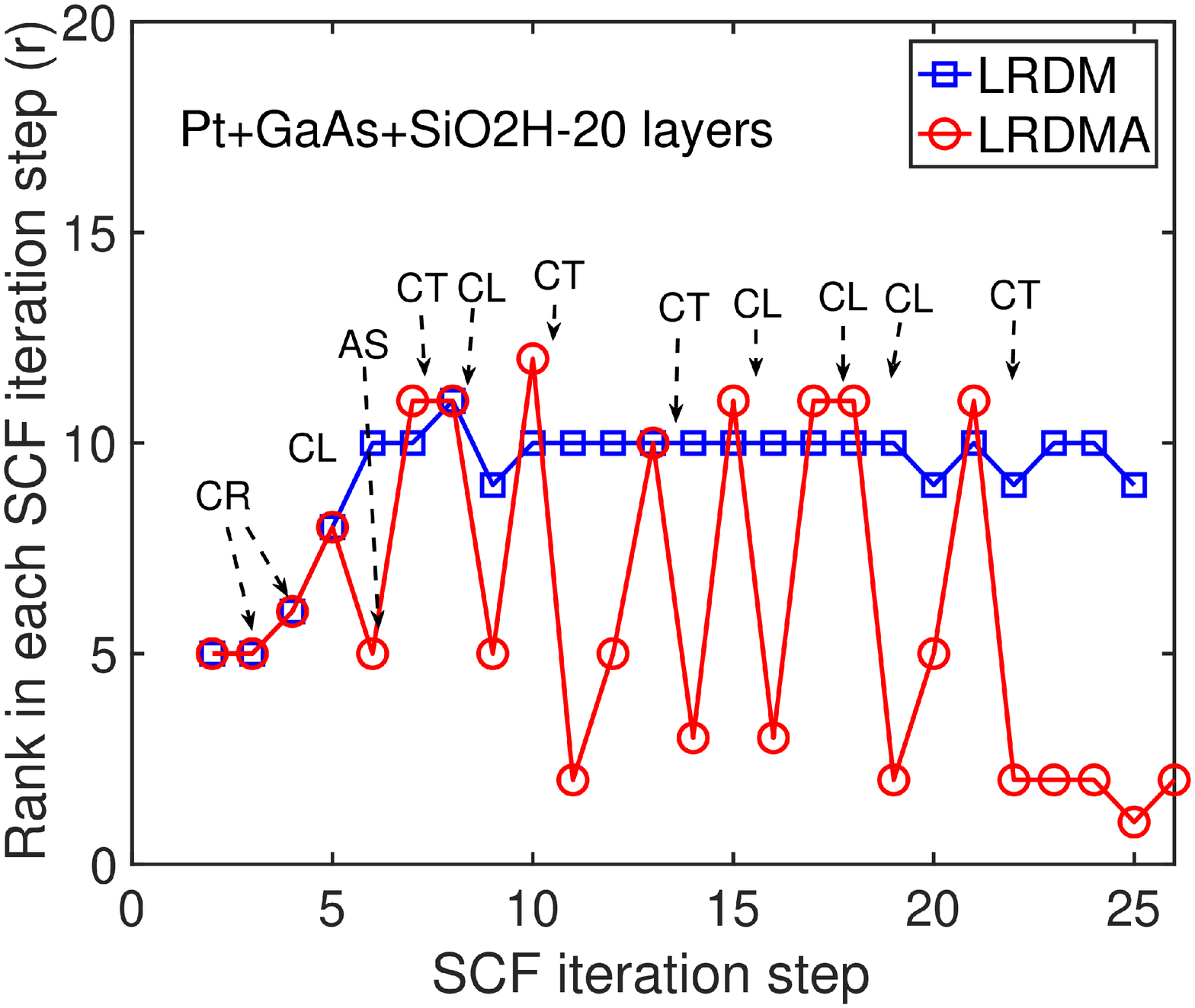}}        
   ~
\subfloat[\label{subfig:energyconv6}]{
        \includegraphics[scale=0.27]{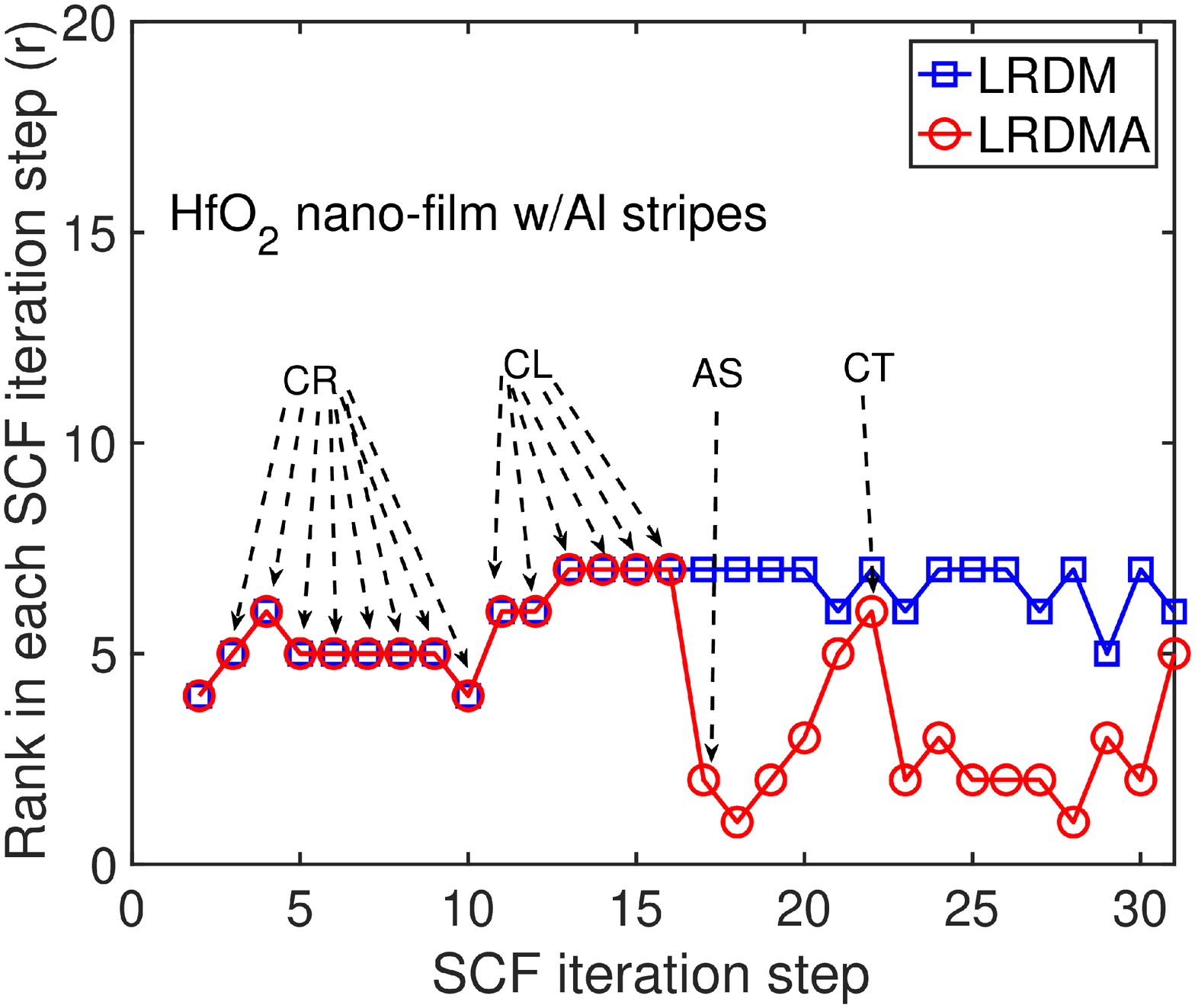}}   
    \caption{\small{LRDM and LRDMA ranks in each SCF iteration step for the various heterogeneous benchmark systems. The plot for LRDMA indicates the additional rank-1 updates in each SCF iteration. Further, in the case of LRDMA, the labels AS denotes the first accumulated SCF step, and labels CR, CL and CT denote clearing of accumulation from previous SCF iteration steps due to non-satisfaction of density residual norm criteria, linearity-indicator criteria and low-rank approximation error criteria, respectively.}}
    \label{fig:rankplot}
\end{figure*}

In the SCF convergence studies discussed in Section~\ref{sec:convergenceStudies} for the LRDM preconditioner, we find that $r_{\textrm{avg}}$ can be high for systems with large $\kappa(J)$. In Section~\ref{sec:lrdm-accumulated-method}, we proposed an improvement to LRDM that can potentially reduce $r_{\textrm{avg}}$ by accumulating the low rank approximation of $J^{\textrm{lr}}_r$ from previous SCF steps combined with an adaptive algorithm that controls the accumulation. We now demonstrate and analyze the $r_{\textrm{avg}}$ reduction achieved by the accumulated variant (LRDMA) over LRDM for a subset of the difficult heterogeneous benchmark systems discussed in Section~\ref{sec:convergenceStudies}, namely the Pt cubic nanoparticles, Pt+GaAs+Si$\textrm{O}_\textrm{2}$H--$N_l$ layers benchmark systems, and the Hf$\textrm{O}_\textrm{2}$ nano-film w/Al stripes benchmark system. Table~\ref{tab:comparisonAccumulated} reports the total SCF iterations and $r_{\textrm{avg}}$ required by LRDM and LRDMA approaches. We further analyze the rank accumulation in Figure~\ref{fig:rankplot}, which shows the adaptively determined rank in each SCF iteration step for the LRDM and LRDMA approaches, and also marks the accumulation clearing events with labels CL, CR and CT related to non-satisfaction of linearity-indicator criteria, residual norm criteria and low-rank approximation error criteria, respectively (cf. Section~\ref{sec:lrdm-accumulated-method}).

In the case of Pt cubic-NPs, we observe an appreciable $\sim$1.8$\times$ reduction in $r_{\textrm{avg}}$ for LRDMA compared to LRDM, while the number of SCF iterations are the same except for the largest system, Pt cubic-NP--$6\times6\times6$ taking an additional SCF step in LRDMA. Next, in the case of the more heterogeneous Pt+GaAs+Si$\textrm{O}_\textrm{2}$H--$N_l$ layers system and Hf$\textrm{O}_\textrm{2}$-Al system, we observe around 1.6$\times$ reduction in $r_{\textrm{avg}}$ with a slight increase in the number of SCF iterations by $\sim$10\% for the most difficult Pt+GaAs+Si$\textrm{O}_\textrm{2}$H--40 system. Although LRDMA performs robustly for the highly heterogeneous systems, the relatively lower advantage as compared to the Pt cubic-NPs is attributed to the stronger nonlinear effects in the more heterogeneous layered material systems when the SCF iteration is not close to convergence. The above reasoning is supported by Figure~\ref{fig:rankplot}, which shows that in the case of Pt cubic-NPs only a small number of CL labelled clearing events occur across all the LRDMA preconditioned SCF iterations. On the other hand, Pt+GaAs+Si$\textrm{O}_\textrm{2}$H--$N_l$ and Hf$\textrm{O}_\textrm{2}$-Al systems have a much larger number of CL clearing events when the SCF iteration is not close to convergence. Specifically, we observe successive CL labelled events in these systems, that correspond to non-linear regimes in $R[\rho]$ when considering the Taylor series expansion about $\rho$ at the current iterate. Relying on CT events only in such regimes leads to sub-optimal accumulation, which we have observed from our numerical experiments.   Overall, the above studies demonstrate that the LRDMA approach, in general, is an improvement over LRDM, with the $r_{\textrm{avg}}$ reduction dependent on the strength of the non-linearities of $F[V_{\textrm{eff}}[\rho]]$ in the materials system.

\begin{table}[htpb]
\centering
\small
\caption{\label{tab:comparisonAccumulated}\small{Comparison of performance of LRDM and LRDMA preconditioners.}}
\begin{tabular}{c  c c} 
 \hline
 \hline
 System &  LRDM ($r_{\textrm{avg}}$)  & LRDMA ($r_{\textrm{avg}}$) \\ 
 \hline
   Pt cubic-NP--$3\times3\times3$ &  24 (7.3) & 23 (3.7)  \\ 
   \hline
   Pt cubic-NP--$5\times5\times5$ &  28 (10.0) & 28 (5.5)    \\ 
  \hline
     Pt cubic-NP--$6\times6\times6$ &  30 (11.0) & 31 (6.8)  \\ 
  \hline
   Pt+GaAs+SiO2H--10 layers &  23 (6.9) & 25 (4.1)  \\     
  \hline
   Pt+GaAs+SiO2H--20 layers &  25 (9.5) & 26 (6.0)  \\    
  \hline
   Pt+GaAs+SiO2H--40 layers &  28 (13.8) & 32 (8.8)\\
 \hline
    Hf$\textrm{O}_\textrm{2}$ nano-film w/Al stripes &  31 (6.2) & 31 (4.2) \\
 \hline 
 \hline
\end{tabular}
\end{table}

\subsection{Computational cost of LRDM and its system size scaling}\label{sec:cost}
We now demonstrate the computational efficiency of the LRDM preconditioner and the accumulated variant (LRDMA) against Anderson and Kerker preconditioners on a subset of the benchmark systems considered in Section~\ref{sec:convergenceStudies}, with system sizes ranging from $\sim$ 3,000--20,000 electrons.
The benchmark calculations are performed using implementations of Anderson, Kerker, LRDM and LRDMA preconditioners in the \DFTFE code. All simulations are performed using multiple GPU nodes of NERSC Perlmutter supercomputer, with comparisons conducted between simulations using the same number of nodes. Further, we note that  the key computational steps in the LRDM preconditioner, in particular the first-order density response computation in the Chebyshev filtered subspace (cf. Section~\ref{sec:densityResponse}), have been efficiently ported to GPUs. On the other hand, the electron-density mixing operations required by Anderson and Anderson with Kerker preconditioners are performed on CPUs, as their overheads are less than 5\% compared to the ChFSI eigensolver cost in each SCF iteration in \DFTFE. We also remark that the ChFSI procedure in \DFTFE was recently GPU ported~\cite{das2022dft}, and achieves significant CPU-GPU speedups of  $\sim 15-20 \times$ on OLCF Summit.

\begin{table*}[t!]
\centering
\small
\caption{\label{tab:comparisonEfficiencyLRDMAnderson}\small{Comparison of computational efficiency between Anderson, LRDM and LRDMA approaches as implemented in the \DFTFE code for benchmark systems comprising of Pt cubic nanoparticle of various sizes.  Computational cost is reported in terms of NERSC Perlmutter GPU Node-hrs for the total SCF solve and the average cost per SCF iteration.  The first SCF iteration, which involves multiple passes of Chebyshev filtering, is excluded in measuring the average per SCF iteration cost. \cb The total SCF cost includes all SCF iteration steps.\cn}}
\begin{tabular}{c c c c c c c c c} 
 \hline
\hline
 System &   Anderson  &  LRDM  & LRDMA & Anderson &  LRDM   &   LRDMA   &LRDM-Anderson  &LRDMA-Anderson  \\ 
  &   total SCF & total SCF &  total SCF & per SCF &  per SCF & per SCF & ratio per SCF & ratio per SCF\\ 
 \hline
  Pt cubic-NP--$3\times3\times3$ & 0.78  & 0.71 & 0.56 & 0.013& 0.026 & 0.021  & 2.0  & 1.62\\
 \hline
  Pt cubic-NP--$5\times5\times5$ &  13.14 & 9.18 & 8.04 & 0.157 & 0.284 & 0.239 & 1.81 & 1.52 \\ 
 \hline
  Pt cubic-NP--$6\times6\times6$ &  56.94 & 40.85 & 36.0 & 0.564 & 1.174  & 1.0 & 2.08 & 1.77\\ 
  \hline
  \hline
\end{tabular}

\end{table*}

\begin{table*}[t!]
\centering
\small
\caption{\label{tab:comparisonEfficiencyLRDMKerker}\small{Comparison of computational efficiency between Kerker, LRDM and LRDMA approaches as implemented in the \DFTFE code for heterogeneous metal-insulator-semiconductor benchmark systems of various sizes.  Computational cost is reported in terms of NERSC Perlmutter GPU Node-hrs for the total SCF solve and the average cost per SCF iteration.  The first SCF iteration, which involves multiple passes of Chebyshev filtering, is excluded in measuring the average per SCF iteration cost. \cb The total SCF cost includes all SCF iteration steps.\cn}}
\begin{tabular}{c c c c c c c c c} 
 \hline
 \hline
 System &   Kerker  &  LRDM  & LRDMA & Kerker &  LRDM   &   LRDMA   &LRDM-Kerker  &LRDMA-Kerker  \\ 
  &   total SCF & total SCF &  total SCF & per SCF &  per SCF & per SCF & ratio per SCF & ratio per SCF\\ 
 \hline
  Pt+GaAs+SiO2H--10 layers & 3.06  & 1.86 & 1.74 & 0.047& 0.072 & 0.061 & 1.53  & 1.30\\
 \hline
  Pt+GaAs+SiO2H--20 layers &  22.31 & 8.66 & 7.43 &  0.166& 0.290 & 0.263 & 1.75 & 1.58 \\ 
 \hline
  Pt+GaAs+SiO2H--40 layers &  213.30 & 63.5 & 64.58 & 0.837 & 2.112 & 1.937 & 2.52 & 2.31\\ 
  \hline
HfO2 nano-film w/Al stripes &  6.30 & 4.04 & 3.50 & 0.069 & 0.122 & 0.104 & 1.77 & 1.51\\ 
  \hline  
  \hline
\end{tabular}
\end{table*}

We first consider the Pt cubic nanoparticle benchmark systems with system sizes ranging from $172-1099$ atoms ($3,096-19,782$ electrons). Table~\ref{tab:comparisonEfficiencyLRDMAnderson} reports the computational cost in GPU Node-hrs for Anderson, LRDM and LRDMA. We present both the computational cost of the full SCF convergence as well as the average per SCF iteration cost. We observe that the total SCF computational cost speedup of LRDM and LRDMA with respect to Anderson are $\sim1.4\times$ and $\sim1.6\times$, respectively, for the larger system sizes. The relatively higher speedup for LRDMA is due to the previously demonstrated lower $r_{\textrm{avg}}$ for LRDMA compared to LRDM for these benchmark systems. Furthermore, the per SCF computational cost ratio of LRDM to Anderson are quite modest, with an average value across the three system sizes of 2.0$\times$ and 1.6$\times$ for LRDM and LRDMA, respectively, where the computational benefit of LRDMA over LRDM is further evident. The above computational cost ratio also demonstrates a weak system size dependence, which is related to the  $\mathcal{O}(MN_e^2)$ scaling of LRDM. From Table~\ref{tab:comparisonEfficiencyLRDMAnderson}, it can be numerically verified that the additional per SCF computational cost for LRDM/LRDMA compared to Anderson scales close to cubically with respect to number of electrons when the system size is increased from Pt cubic-NP--$5\times5\times5$ (666 atoms) to Pt cubic-NP--$6\times6\times6$ (1099 atoms). Thus, for larger system sizes, given the asymptotic cubic scaling of KS-DFT and the weak dependence of $r_{\textrm{avg}}$ on the system size (cf. Table~\ref{tab:comparisonNoSpin}), the computational cost ratio can be expected to approach a constant factor for very large system sizes.

Next, we investigate the computational cost of LRDM/LRDMA for the more heterogeneous benchmark systems---Pt+GaAs+$\textrm{SiO}_2$H (metal-semiconductor-insulator-vacuum) layers with increasing sizes and a large-scale $\textrm{HfO}_2$ nano-film w/Al stripes (metal-semiconductor-vacuum) system. For these materials systems we compare against Kerker preconditioner, which is relatively more robust for such systems compared to Anderson.  Table~\ref{tab:comparisonEfficiencyLRDMKerker} reports the results for this study, where we observe LRDM to achieve 1.6--3.4$\times$ speedups in the total SCF cost compared to Kerker for the above benchmark systems. The per SCF computational cost ratio of LRDM to Kerker is 1.5--2.5$\times$, with up to a $\sim1.2\times$ reduction observed for the per SCF computational cost ratio while using LRDMA. We remark that LRDMA does not provide an improvement in the total SCF cost reduction relative to LRDM for the Pt+GaAs+$\textrm{SiO}_2$H-40 layers system due to the slight increase in the number of SCFs (cf. Section~\ref{sec:rankreduction}) that negates the reduction in the per SCF cost.

\section{Conclusions}
\label{sec:conclusions}

We present a robust and efficient approach to accelerate the self-consistent field iteration in Kohn-Sham DFT calculations. In particular, we use a low-rank approximation of the dielectric matrix (LRDM)---constructed as a sum of rank-1 tensor products---as the  preconditioner for solving the self-consistent fixed point iteration. We note that the low-rank approximation is adaptive and systematically convergent, which lends to the robustness of the approach for tackling generic heterogeneous materials systems, including systems with large condition number and strong non-linearities in the fixed point iteration. A key contribution of the present effort is the development of an efficient numerical method for computing the first-order density response in real-space that is critical to the efficiency of the LRDM procedure. To this end, we compute the density response in the Chebyshev filtered subspace iteration (ChFSI) procedure, which adaptively approximates the occupied eigensubspace in each SCF iteration step. The smaller dimension of the Chebyshev filtered subspace,  in conjunction with the use of canonical density-matrix perturbation theory, provides a computationally efficient approach for using the LRDM preconditioner. In order to further improve the computational efficiency of the LRDM method, we developed an accumulated variant of the formulation, which adaptively accumulates the low rank dielectric matrix approximation from the previous SCF iterations using indicators based on the extent of linearity of the fixed point iteration map and the residual norm. Additionally, we extended the LRDM preconditioner to collinear spin-polarized KS-DFT, using a generalization of the low rank formulation to spin densities and their corresponding response functions. We note that this generalization is also extensible to non-collinear spin-polarized KS-DFT, which is a topic for future studies. \cb We additionally remark that the proposed numerical method for the LRDM procedure, although developed in the context of finite-element basis and ChFSI eigensolver, is adaptable to other Kohn-Sham DFT implementations that use different  discretizations (e.g.~plane-waves, finite-difference, wavelets) and eigensolver strategies (e.g.~Davidson, RMM-DIIS).~\cn

We investigated the robustness and efficiency of the LRDM approach on a comprehensive set of heterogeneous non-periodic and periodic benchmark systems. To this end, we chose materials systems such as metallic and bi-metallic nanoparticles, layered materials with various combinations of metal-semiconductor-insulator systems, semiconducting nano-films with metal dopants, and magnetic systems with spin-polarization. The benchmark systems range from small to large system sizes comprising of $\sim$ 100--1100 atoms ($\sim$500--20,000 electrons). In all the benchmark systems we compared the SCF convergence of LRDM against three other widely used preconditioners---Anderson mixing~\cite{anderson1965}, Anderson mixing with Kerker preconditioner~\cite{Kerker1981}, and Broyden mixing with Thomas-Fermi-von Weizsacker preconditioner (TFW)~\cite{Raczkowski2011}. Our studies have demonstrated that LRDM achieves robust convergence within 20--30 SCF iterations, for all the benchmark systems considered here, with a weak dependence observed on the system size. In comparison, Anderson, Kerker and Broyden showed slow convergence for systems with high condition numbers, requiring $3-5\times$ more SCF iterations compared to LRDM. The accumulated variant (LRDMA) showed promise in reducing the average (over the SCF iterations) adaptive rank $r_{\textrm{avg}}$ by $1.5-1.8\times$ across various benchmark systems without affecting the robustness of the SCF iteration. We also found the LRDM/LRDMA approach to outperform Anderson and Kerker preconditioners in computational cost for full ground-state calculations by $\sim1.4-3.4\times$ for Pt cubic nanoparticles and layered heterogeneous benchmark systems.

\section*{Acknowledgements}
We gratefully acknowledge the support from the Department of Energy, Office of Basic Energy Sciences (Award number DE-SC0008637) and the Toyota Research Institute that funded this work. The GPU implementation was done under the auspices of Department of Energy, Office of Basic Energy Sciences Award number DE-SC0022241. This research used resources of the Oak Ridge Leadership Computing Facility, which is a DOE Office of Science User Facility supported under Contract DE-AC05-00OR22725. This research used resources of the National Energy Research Scientific Computing Center, a DOE Office of Science User Facility supported by the Office of Science of the U.S. Department of Energy under Contract No. DE-AC02-05CH11231. This work used the Extreme Science and Engineering Discovery Environment (XSEDE), which is supported by National Science Foundation Grant number ACI-1053575. 



\begin{thebibliography}{60}%
\makeatletter
\providecommand \@ifxundefined [1]{%
 \@ifx{#1\undefined}
}%
\providecommand \@ifnum [1]{%
 \ifnum #1\expandafter \@firstoftwo
 \else \expandafter \@secondoftwo
 \fi
}%
\providecommand \@ifx [1]{%
 \ifx #1\expandafter \@firstoftwo
 \else \expandafter \@secondoftwo
 \fi
}%
\providecommand \natexlab [1]{#1}%
\providecommand \enquote  [1]{``#1''}%
\providecommand \bibnamefont  [1]{#1}%
\providecommand \bibfnamefont [1]{#1}%
\providecommand \citenamefont [1]{#1}%
\providecommand \href@noop [0]{\@secondoftwo}%
\providecommand \href [0]{\begingroup \@sanitize@url \@href}%
\providecommand \@href[1]{\@@startlink{#1}\@@href}%
\providecommand \@@href[1]{\endgroup#1\@@endlink}%
\providecommand \@sanitize@url [0]{\catcode `\\12\catcode `\$12\catcode
  `\&12\catcode `\#12\catcode `\^12\catcode `\_12\catcode `\%12\relax}%
\providecommand \@@startlink[1]{}%
\providecommand \@@endlink[0]{}%
\providecommand \url  [0]{\begingroup\@sanitize@url \@url }%
\providecommand \@url [1]{\endgroup\@href {#1}{\urlprefix }}%
\providecommand \urlprefix  [0]{URL }%
\providecommand \Eprint [0]{\href }%
\providecommand \doibase [0]{https://doi.org/}%
\providecommand \selectlanguage [0]{\@gobble}%
\providecommand \bibinfo  [0]{\@secondoftwo}%
\providecommand \bibfield  [0]{\@secondoftwo}%
\providecommand \translation [1]{[#1]}%
\providecommand \BibitemOpen [0]{}%
\providecommand \bibitemStop [0]{}%
\providecommand \bibitemNoStop [0]{.\EOS\space}%
\providecommand \EOS [0]{\spacefactor3000\relax}%
\providecommand \BibitemShut  [1]{\csname bibitem#1\endcsname}%
\let\auto@bib@innerbib\@empty
\bibitem [{\citenamefont {Kohn}\ and\ \citenamefont {Sham}(1965)}]{kohn65}%
  \BibitemOpen
  \bibfield  {author} {\bibinfo {author} {\bibfnamefont {W.}~\bibnamefont
  {Kohn}}\ and\ \bibinfo {author} {\bibfnamefont {L.~J.}\ \bibnamefont
  {Sham}},\ }\bibfield  {title} {\bibinfo {title} {Self-consistent equations
  including exchange and correlation effects},\ }\href@noop {} {\bibfield
  {journal} {\bibinfo  {journal} {Physical Review}\ }\textbf {\bibinfo {volume}
  {140}},\ \bibinfo {pages} {A1133} (\bibinfo {year} {1965})}\BibitemShut
  {NoStop}%
\bibitem [{\citenamefont {Kohn}(1996)}]{kohn96}%
  \BibitemOpen
  \bibfield  {author} {\bibinfo {author} {\bibfnamefont {W.}~\bibnamefont
  {Kohn}},\ }\bibfield  {title} {\bibinfo {title} {Density functional and
  density matrix method scaling linearly with the number of atoms},\
  }\href@noop {} {\bibfield  {journal} {\bibinfo  {journal} {Physical Review
  Letters}\ }\textbf {\bibinfo {volume} {76}},\ \bibinfo {pages} {3168}
  (\bibinfo {year} {1996})}\BibitemShut {NoStop}%
\bibitem [{\citenamefont {Zhou}\ \emph {et~al.}(2006)\citenamefont {Zhou},
  \citenamefont {Saad}, \citenamefont {Tiago},\ and\ \citenamefont
  {Chelikowsky}}]{zhou2006}%
  \BibitemOpen
  \bibfield  {author} {\bibinfo {author} {\bibfnamefont {Y.}~\bibnamefont
  {Zhou}}, \bibinfo {author} {\bibfnamefont {Y.}~\bibnamefont {Saad}}, \bibinfo
  {author} {\bibfnamefont {M.~L.}\ \bibnamefont {Tiago}},\ and\ \bibinfo
  {author} {\bibfnamefont {J.~R.}\ \bibnamefont {Chelikowsky}},\ }\bibfield
  {title} {\bibinfo {title} {Self-consistent-field calculations using
  {C}hebyshev-filtered subspace iteration},\ }\href@noop {} {\bibfield
  {journal} {\bibinfo  {journal} {Journal of Computational Physics}\ }\textbf
  {\bibinfo {volume} {219}},\ \bibinfo {pages} {172} (\bibinfo {year}
  {2006})}\BibitemShut {NoStop}%
\bibitem [{\citenamefont {Iwata}\ \emph {et~al.}(2010)\citenamefont {Iwata},
  \citenamefont {Takahashi}, \citenamefont {Oshiyama}, \citenamefont {Boku},
  \citenamefont {Shiraishi}, \citenamefont {Okada},\ and\ \citenamefont
  {Yabana}}]{RS-DFTcode}%
  \BibitemOpen
  \bibfield  {author} {\bibinfo {author} {\bibfnamefont {J.-I.}\ \bibnamefont
  {Iwata}}, \bibinfo {author} {\bibfnamefont {D.}~\bibnamefont {Takahashi}},
  \bibinfo {author} {\bibfnamefont {A.}~\bibnamefont {Oshiyama}}, \bibinfo
  {author} {\bibfnamefont {T.}~\bibnamefont {Boku}}, \bibinfo {author}
  {\bibfnamefont {K.}~\bibnamefont {Shiraishi}}, \bibinfo {author}
  {\bibfnamefont {S.}~\bibnamefont {Okada}},\ and\ \bibinfo {author}
  {\bibfnamefont {K.}~\bibnamefont {Yabana}},\ }\bibfield  {title} {\bibinfo
  {title} {A massively-parallel electronic-structure calculations based on
  real-space density functional theory},\ }\href@noop {} {\bibfield  {journal}
  {\bibinfo  {journal} {Journal of Computational Physics}\ }\textbf {\bibinfo
  {volume} {229}},\ \bibinfo {pages} {2339} (\bibinfo {year}
  {2010})}\BibitemShut {NoStop}%
\bibitem [{\citenamefont {Genovese}\ \emph {et~al.}(2011)\citenamefont
  {Genovese}, \citenamefont {Videau}, \citenamefont {Ospici}, \citenamefont
  {Deutsch}, \citenamefont {Goedecker},\ and\ \citenamefont
  {M{\'e}haut}}]{genovese2011daubechies}%
  \BibitemOpen
  \bibfield  {author} {\bibinfo {author} {\bibfnamefont {L.}~\bibnamefont
  {Genovese}}, \bibinfo {author} {\bibfnamefont {B.}~\bibnamefont {Videau}},
  \bibinfo {author} {\bibfnamefont {M.}~\bibnamefont {Ospici}}, \bibinfo
  {author} {\bibfnamefont {T.}~\bibnamefont {Deutsch}}, \bibinfo {author}
  {\bibfnamefont {S.}~\bibnamefont {Goedecker}},\ and\ \bibinfo {author}
  {\bibfnamefont {J.-F.}\ \bibnamefont {M{\'e}haut}},\ }\bibfield  {title}
  {\bibinfo {title} {Daubechies wavelets for high performance electronic
  structure calculations: The {BigDFT} project},\ }\href@noop {} {\bibfield
  {journal} {\bibinfo  {journal} {Comptes Rendus M{\'e}canique}\ }\textbf
  {\bibinfo {volume} {339}},\ \bibinfo {pages} {149} (\bibinfo {year}
  {2011})}\BibitemShut {NoStop}%
\bibitem [{\citenamefont {Motamarri}\ \emph {et~al.}(2020)\citenamefont
  {Motamarri}, \citenamefont {Das}, \citenamefont {Rudraraju}, \citenamefont
  {Ghosh}, \citenamefont {Davydov},\ and\ \citenamefont
  {Gavini}}]{motamarri2020}%
  \BibitemOpen
  \bibfield  {author} {\bibinfo {author} {\bibfnamefont {P.}~\bibnamefont
  {Motamarri}}, \bibinfo {author} {\bibfnamefont {S.}~\bibnamefont {Das}},
  \bibinfo {author} {\bibfnamefont {S.}~\bibnamefont {Rudraraju}}, \bibinfo
  {author} {\bibfnamefont {K.}~\bibnamefont {Ghosh}}, \bibinfo {author}
  {\bibfnamefont {D.}~\bibnamefont {Davydov}},\ and\ \bibinfo {author}
  {\bibfnamefont {V.}~\bibnamefont {Gavini}},\ }\bibfield  {title} {\bibinfo
  {title} {{DFT-FE} -- {A} massively parallel adaptive finite-element code for
  large-scale density functional theory calculations},\ }\href@noop {}
  {\bibfield  {journal} {\bibinfo  {journal} {Computer Physics Communications}\
  }\textbf {\bibinfo {volume} {246}},\ \bibinfo {pages} {106853} (\bibinfo
  {year} {2020})}\BibitemShut {NoStop}%
\bibitem [{\citenamefont {Xu}\ \emph {et~al.}(2021)\citenamefont {Xu},
  \citenamefont {Sharma}, \citenamefont {Comer}, \citenamefont {Huang},
  \citenamefont {Chow}, \citenamefont {Medford}, \citenamefont {Pask},\ and\
  \citenamefont {Suryanarayana}}]{SPARC}%
  \BibitemOpen
  \bibfield  {author} {\bibinfo {author} {\bibfnamefont {Q.}~\bibnamefont
  {Xu}}, \bibinfo {author} {\bibfnamefont {A.}~\bibnamefont {Sharma}}, \bibinfo
  {author} {\bibfnamefont {B.}~\bibnamefont {Comer}}, \bibinfo {author}
  {\bibfnamefont {H.}~\bibnamefont {Huang}}, \bibinfo {author} {\bibfnamefont
  {E.}~\bibnamefont {Chow}}, \bibinfo {author} {\bibfnamefont {A.~J.}\
  \bibnamefont {Medford}}, \bibinfo {author} {\bibfnamefont {J.~E.}\
  \bibnamefont {Pask}},\ and\ \bibinfo {author} {\bibfnamefont
  {P.}~\bibnamefont {Suryanarayana}},\ }\bibfield  {title} {\bibinfo {title}
  {{SPARC}: Simulation package for ab-initio real-space calculations},\
  }\href@noop {} {\bibfield  {journal} {\bibinfo  {journal} {SoftwareX}\
  }\textbf {\bibinfo {volume} {15}},\ \bibinfo {pages} {100709} (\bibinfo
  {year} {2021})}\BibitemShut {NoStop}%
\bibitem [{\citenamefont {Das}\ \emph {et~al.}(2022)\citenamefont {Das},
  \citenamefont {Motamarri}, \citenamefont {Subramanian}, \citenamefont
  {Rogers},\ and\ \citenamefont {Gavini}}]{das2022dft}%
  \BibitemOpen
  \bibfield  {author} {\bibinfo {author} {\bibfnamefont {S.}~\bibnamefont
  {Das}}, \bibinfo {author} {\bibfnamefont {P.}~\bibnamefont {Motamarri}},
  \bibinfo {author} {\bibfnamefont {V.}~\bibnamefont {Subramanian}}, \bibinfo
  {author} {\bibfnamefont {D.~M.}\ \bibnamefont {Rogers}},\ and\ \bibinfo
  {author} {\bibfnamefont {V.}~\bibnamefont {Gavini}},\ }\bibfield  {title}
  {\bibinfo {title} {{DFT-FE} 1.0: A massively parallel hybrid {CPU-GPU}
  density functional theory code using finite-element discretization},\
  }\href@noop {} {\bibfield  {journal} {\bibinfo  {journal} {Computer Physics
  Communications}\ }\textbf {\bibinfo {volume} {280}},\ \bibinfo {pages}
  {108473} (\bibinfo {year} {2022})}\BibitemShut {NoStop}%
\bibitem [{\citenamefont {Bowler}\ and\ \citenamefont
  {Miyazaki}(2012)}]{Bowler}%
  \BibitemOpen
  \bibfield  {author} {\bibinfo {author} {\bibfnamefont {D.~R.}\ \bibnamefont
  {Bowler}}\ and\ \bibinfo {author} {\bibfnamefont {T.}~\bibnamefont
  {Miyazaki}},\ }\bibfield  {title} {\bibinfo {title} {$\mathcal{O}$({N})
  methods in electronic structure calculations},\ }\href@noop {} {\bibfield
  {journal} {\bibinfo  {journal} {Reports on Progress in Physics}\ }\textbf
  {\bibinfo {volume} {75}},\ \bibinfo {pages} {036503} (\bibinfo {year}
  {2012})}\BibitemShut {NoStop}%
\bibitem [{\citenamefont {Lin}\ \emph {et~al.}(2013)\citenamefont {Lin},
  \citenamefont {Chen}, \citenamefont {Yang},\ and\ \citenamefont
  {He}}]{PEXSI}%
  \BibitemOpen
  \bibfield  {author} {\bibinfo {author} {\bibfnamefont {L.}~\bibnamefont
  {Lin}}, \bibinfo {author} {\bibfnamefont {M.}~\bibnamefont {Chen}}, \bibinfo
  {author} {\bibfnamefont {C.}~\bibnamefont {Yang}},\ and\ \bibinfo {author}
  {\bibfnamefont {L.}~\bibnamefont {He}},\ }\bibfield  {title} {\bibinfo
  {title} {Accelerating atomic orbital-based electronic structure calculation
  via pole expansion and selected inversion},\ }\href@noop {} {\bibfield
  {journal} {\bibinfo  {journal} {Journal of Physics: Condensed Matter}\
  }\textbf {\bibinfo {volume} {25}},\ \bibinfo {pages} {295501} (\bibinfo
  {year} {2013})}\BibitemShut {NoStop}%
\bibitem [{\citenamefont {Motamarri}\ and\ \citenamefont
  {Gavini}(2014)}]{motamarri2014}%
  \BibitemOpen
  \bibfield  {author} {\bibinfo {author} {\bibfnamefont {P.}~\bibnamefont
  {Motamarri}}\ and\ \bibinfo {author} {\bibfnamefont {V.}~\bibnamefont
  {Gavini}},\ }\bibfield  {title} {\bibinfo {title} {Subquadratic-scaling
  subspace projection method for large-scale kohn-sham density functional
  theory calculations using spectral finite-element discretization},\
  }\href@noop {} {\bibfield  {journal} {\bibinfo  {journal} {Physical Review
  B}\ }\textbf {\bibinfo {volume} {90}},\ \bibinfo {pages} {115127} (\bibinfo
  {year} {2014})}\BibitemShut {NoStop}%
\bibitem [{\citenamefont {Lin}\ \emph {et~al.}(2021)\citenamefont {Lin},
  \citenamefont {Motamarri},\ and\ \citenamefont {Gavini}}]{lin2021a}%
  \BibitemOpen
  \bibfield  {author} {\bibinfo {author} {\bibfnamefont {C.-C.}\ \bibnamefont
  {Lin}}, \bibinfo {author} {\bibfnamefont {P.}~\bibnamefont {Motamarri}},\
  and\ \bibinfo {author} {\bibfnamefont {V.}~\bibnamefont {Gavini}},\
  }\bibfield  {title} {\bibinfo {title} {{Tensor-structured algorithm for
  reduced-order scaling large-scale Kohn–Sham density functional theory
  calculations}},\ }\href@noop {} {\bibfield  {journal} {\bibinfo  {journal}
  {npj Computational Materials}\ }\textbf {\bibinfo {volume} {7}},\ \bibinfo
  {pages} {50} (\bibinfo {year} {2021})}\BibitemShut {NoStop}%
\bibitem [{\citenamefont {N{\o}rskov}\ \emph {et~al.}(2011)\citenamefont
  {N{\o}rskov}, \citenamefont {Abild-Pedersen}, \citenamefont {Studt},\ and\
  \citenamefont {Bligaard}}]{norskov2011density}%
  \BibitemOpen
  \bibfield  {author} {\bibinfo {author} {\bibfnamefont {J.~K.}\ \bibnamefont
  {N{\o}rskov}}, \bibinfo {author} {\bibfnamefont {F.}~\bibnamefont
  {Abild-Pedersen}}, \bibinfo {author} {\bibfnamefont {F.}~\bibnamefont
  {Studt}},\ and\ \bibinfo {author} {\bibfnamefont {T.}~\bibnamefont
  {Bligaard}},\ }\bibfield  {title} {\bibinfo {title} {Density functional
  theory in surface chemistry and catalysis},\ }\href@noop {} {\bibfield
  {journal} {\bibinfo  {journal} {Proceedings of the National Academy of
  Sciences}\ }\textbf {\bibinfo {volume} {108}},\ \bibinfo {pages} {937}
  (\bibinfo {year} {2011})}\BibitemShut {NoStop}%
\bibitem [{\citenamefont {Fernando}\ \emph {et~al.}(2015)\citenamefont
  {Fernando}, \citenamefont {Weerawardene}, \citenamefont {Karimova},\ and\
  \citenamefont {Aikens}}]{fernando2015quantum}%
  \BibitemOpen
  \bibfield  {author} {\bibinfo {author} {\bibfnamefont {A.}~\bibnamefont
  {Fernando}}, \bibinfo {author} {\bibfnamefont {K.~D.~M.}\ \bibnamefont
  {Weerawardene}}, \bibinfo {author} {\bibfnamefont {N.~V.}\ \bibnamefont
  {Karimova}},\ and\ \bibinfo {author} {\bibfnamefont {C.~M.}\ \bibnamefont
  {Aikens}},\ }\bibfield  {title} {\bibinfo {title} {Quantum mechanical studies
  of large metal, metal oxide, and metal chalcogenide nanoparticles and
  clusters},\ }\href@noop {} {\bibfield  {journal} {\bibinfo  {journal}
  {Chemical reviews}\ }\textbf {\bibinfo {volume} {115}},\ \bibinfo {pages}
  {6112} (\bibinfo {year} {2015})}\BibitemShut {NoStop}%
\bibitem [{\citenamefont {Cole}\ and\ \citenamefont
  {Hine}(2016)}]{cole2016applications}%
  \BibitemOpen
  \bibfield  {author} {\bibinfo {author} {\bibfnamefont {D.~J.}\ \bibnamefont
  {Cole}}\ and\ \bibinfo {author} {\bibfnamefont {N.~D.}\ \bibnamefont
  {Hine}},\ }\bibfield  {title} {\bibinfo {title} {Applications of large-scale
  density functional theory in biology},\ }\href@noop {} {\bibfield  {journal}
  {\bibinfo  {journal} {Journal of Physics: Condensed Matter}\ }\textbf
  {\bibinfo {volume} {28}},\ \bibinfo {pages} {393001} (\bibinfo {year}
  {2016})}\BibitemShut {NoStop}%
\bibitem [{\citenamefont {Pham}\ \emph {et~al.}(2017)\citenamefont {Pham},
  \citenamefont {Ping},\ and\ \citenamefont {Galli}}]{pham2017modelling}%
  \BibitemOpen
  \bibfield  {author} {\bibinfo {author} {\bibfnamefont {T.~A.}\ \bibnamefont
  {Pham}}, \bibinfo {author} {\bibfnamefont {Y.}~\bibnamefont {Ping}},\ and\
  \bibinfo {author} {\bibfnamefont {G.}~\bibnamefont {Galli}},\ }\bibfield
  {title} {\bibinfo {title} {Modelling heterogeneous interfaces for solar water
  splitting},\ }\href@noop {} {\bibfield  {journal} {\bibinfo  {journal}
  {Nature materials}\ }\textbf {\bibinfo {volume} {16}},\ \bibinfo {pages}
  {401} (\bibinfo {year} {2017})}\BibitemShut {NoStop}%
\bibitem [{\citenamefont {Das}\ and\ \citenamefont
  {Mahapatra}(2018)}]{mis2018}%
  \BibitemOpen
  \bibfield  {author} {\bibinfo {author} {\bibfnamefont {B.}~\bibnamefont
  {Das}}\ and\ \bibinfo {author} {\bibfnamefont {S.}~\bibnamefont
  {Mahapatra}},\ }\bibfield  {title} {\bibinfo {title} {An atom-to-circuit
  modeling approach to {all-2D} metal--insulator--semiconductor field-effect
  transistors},\ }\href@noop {} {\bibfield  {journal} {\bibinfo  {journal} {npj
  2D Materials and Applications}\ }\textbf {\bibinfo {volume} {2}},\ \bibinfo
  {pages} {1} (\bibinfo {year} {2018})}\BibitemShut {NoStop}%
\bibitem [{\citenamefont {Das}\ \emph {et~al.}(2019)\citenamefont {Das},
  \citenamefont {Motamarri}, \citenamefont {Gavini}, \citenamefont {Turcksin},
  \citenamefont {Li},\ and\ \citenamefont {Leback}}]{das2019}%
  \BibitemOpen
  \bibfield  {author} {\bibinfo {author} {\bibfnamefont {S.}~\bibnamefont
  {Das}}, \bibinfo {author} {\bibfnamefont {P.}~\bibnamefont {Motamarri}},
  \bibinfo {author} {\bibfnamefont {V.}~\bibnamefont {Gavini}}, \bibinfo
  {author} {\bibfnamefont {B.}~\bibnamefont {Turcksin}}, \bibinfo {author}
  {\bibfnamefont {Y.~W.}\ \bibnamefont {Li}},\ and\ \bibinfo {author}
  {\bibfnamefont {B.}~\bibnamefont {Leback}},\ }\bibfield  {title} {\bibinfo
  {title} {Fast, scalable and accurate finite-element based ab initio
  calculations using mixed precision computing: {46 PFLOPS} simulation of a
  metallic dislocation system},\ }in\ \href@noop {} {\emph {\bibinfo
  {booktitle} {Proceedings of the International Conference for High Performance
  Computing, Networking, Storage and Analysis}}}\ (\bibinfo {year} {2019})\
  pp.\ \bibinfo {pages} {1--11}\BibitemShut {NoStop}%
\bibitem [{\citenamefont {Zhuravel}\ \emph {et~al.}(2020)\citenamefont
  {Zhuravel}, \citenamefont {Huang}, \citenamefont {Polycarpou}, \citenamefont
  {Polydorides}, \citenamefont {Motamarri}, \citenamefont {Katrivas},
  \citenamefont {Rotem}, \citenamefont {Sperling}, \citenamefont {Zotti},
  \citenamefont {Kotlyar}, \citenamefont {Cuevas}, \citenamefont {Gavini},
  \citenamefont {Skourtis},\ and\ \citenamefont {Porath}}]{motamarridna2020}%
  \BibitemOpen
  \bibfield  {author} {\bibinfo {author} {\bibfnamefont {R.}~\bibnamefont
  {Zhuravel}}, \bibinfo {author} {\bibfnamefont {H.}~\bibnamefont {Huang}},
  \bibinfo {author} {\bibfnamefont {G.}~\bibnamefont {Polycarpou}}, \bibinfo
  {author} {\bibfnamefont {S.}~\bibnamefont {Polydorides}}, \bibinfo {author}
  {\bibfnamefont {P.}~\bibnamefont {Motamarri}}, \bibinfo {author}
  {\bibfnamefont {L.}~\bibnamefont {Katrivas}}, \bibinfo {author}
  {\bibfnamefont {D.}~\bibnamefont {Rotem}}, \bibinfo {author} {\bibfnamefont
  {J.}~\bibnamefont {Sperling}}, \bibinfo {author} {\bibfnamefont {L.~A.}\
  \bibnamefont {Zotti}}, \bibinfo {author} {\bibfnamefont {A.~B.}\ \bibnamefont
  {Kotlyar}}, \bibinfo {author} {\bibfnamefont {J.~C.}\ \bibnamefont {Cuevas}},
  \bibinfo {author} {\bibfnamefont {V.}~\bibnamefont {Gavini}}, \bibinfo
  {author} {\bibfnamefont {S.~S.}\ \bibnamefont {Skourtis}},\ and\ \bibinfo
  {author} {\bibfnamefont {D.}~\bibnamefont {Porath}},\ }\bibfield  {title}
  {\bibinfo {title} {Backbone charge transport in double-stranded {DNA}},\
  }\href@noop {} {\bibfield  {journal} {\bibinfo  {journal} {Nature
  Nanotechnology}\ }\textbf {\bibinfo {volume} {15}},\ \bibinfo {pages} {836}
  (\bibinfo {year} {2020})}\BibitemShut {NoStop}%
\bibitem [{\citenamefont {Ghosh}\ \emph {et~al.}(2021)\citenamefont {Ghosh},
  \citenamefont {Ma}, \citenamefont {Onizhuk}, \citenamefont {Gavini},\ and\
  \citenamefont {Galli}}]{ghosh2021}%
  \BibitemOpen
  \bibfield  {author} {\bibinfo {author} {\bibfnamefont {K.}~\bibnamefont
  {Ghosh}}, \bibinfo {author} {\bibfnamefont {H.}~\bibnamefont {Ma}}, \bibinfo
  {author} {\bibfnamefont {M.}~\bibnamefont {Onizhuk}}, \bibinfo {author}
  {\bibfnamefont {V.}~\bibnamefont {Gavini}},\ and\ \bibinfo {author}
  {\bibfnamefont {G.}~\bibnamefont {Galli}},\ }\bibfield  {title} {\bibinfo
  {title} {Spin–spin interactions in defects in solids from mixed
  all-electron and pseudopotential first-principles calculations},\ }\href@noop
  {} {\bibfield  {journal} {\bibinfo  {journal} {npj Computational Materials}\
  }\textbf {\bibinfo {volume} {7}},\ \bibinfo {pages} {123} (\bibinfo {year}
  {2021})}\BibitemShut {NoStop}%
\bibitem [{\citenamefont {Yao}\ \emph {et~al.}(2022)\citenamefont {Yao},
  \citenamefont {Das}, \citenamefont {Liu}, \citenamefont {Cheng},
  \citenamefont {Gavini},\ and\ \citenamefont {Xiao}}]{yao2022modulating}%
  \BibitemOpen
  \bibfield  {author} {\bibinfo {author} {\bibfnamefont {L.}~\bibnamefont
  {Yao}}, \bibinfo {author} {\bibfnamefont {S.}~\bibnamefont {Das}}, \bibinfo
  {author} {\bibfnamefont {X.}~\bibnamefont {Liu}}, \bibinfo {author}
  {\bibfnamefont {Y.}~\bibnamefont {Cheng}}, \bibinfo {author} {\bibfnamefont
  {V.}~\bibnamefont {Gavini}},\ and\ \bibinfo {author} {\bibfnamefont
  {B.}~\bibnamefont {Xiao}},\ }\bibfield  {title} {\bibinfo {title} {Modulating
  the microscopic lattice distortions through the {Al}-rich layers for boosting
  the ferroelectricity in {Al}: {$\text{HfO}_2$} nanofilms},\ }\href@noop {}
  {\bibfield  {journal} {\bibinfo  {journal} {Journal of Physics D: Applied
  Physics}\ } (\bibinfo {year} {2022})}\BibitemShut {NoStop}%
\bibitem [{\citenamefont {Ho}\ \emph {et~al.}(1982)\citenamefont {Ho},
  \citenamefont {Ihm},\ and\ \citenamefont {Joannopoulos}}]{Ho1982}%
  \BibitemOpen
  \bibfield  {author} {\bibinfo {author} {\bibfnamefont {K.-M.}\ \bibnamefont
  {Ho}}, \bibinfo {author} {\bibfnamefont {J.}~\bibnamefont {Ihm}},\ and\
  \bibinfo {author} {\bibfnamefont {J.~D.}\ \bibnamefont {Joannopoulos}},\
  }\bibfield  {title} {\bibinfo {title} {Dielectric matrix scheme for fast
  convergence in self-consistent electronic-structure calculations},\
  }\href@noop {} {\bibfield  {journal} {\bibinfo  {journal} {Physical Review
  B}\ }\textbf {\bibinfo {volume} {25}},\ \bibinfo {pages} {4260} (\bibinfo
  {year} {1982})}\BibitemShut {NoStop}%
\bibitem [{\citenamefont {Anglade}\ and\ \citenamefont
  {Gonze}(2008)}]{Anglade2008}%
  \BibitemOpen
  \bibfield  {author} {\bibinfo {author} {\bibfnamefont {P.-M.}\ \bibnamefont
  {Anglade}}\ and\ \bibinfo {author} {\bibfnamefont {X.}~\bibnamefont
  {Gonze}},\ }\bibfield  {title} {\bibinfo {title} {Preconditioning of
  self-consistent-field cycles in density-functional theory: The extrapolar
  method},\ }\href@noop {} {\bibfield  {journal} {\bibinfo  {journal} {Physical
  Review B}\ }\textbf {\bibinfo {volume} {78}},\ \bibinfo {pages} {045126}
  (\bibinfo {year} {2008})}\BibitemShut {NoStop}%
\bibitem [{\citenamefont {Adler}(1962)}]{Alder1962}%
  \BibitemOpen
  \bibfield  {author} {\bibinfo {author} {\bibfnamefont {S.~L.}\ \bibnamefont
  {Adler}},\ }\bibfield  {title} {\bibinfo {title} {Quantum theory of the
  dielectric constant in real solids},\ }\href@noop {} {\bibfield  {journal}
  {\bibinfo  {journal} {Physical Review}\ }\textbf {\bibinfo {volume} {126}},\
  \bibinfo {pages} {413} (\bibinfo {year} {1962})}\BibitemShut {NoStop}%
\bibitem [{\citenamefont {Wiser}(1963)}]{Wiser1963}%
  \BibitemOpen
  \bibfield  {author} {\bibinfo {author} {\bibfnamefont {N.}~\bibnamefont
  {Wiser}},\ }\bibfield  {title} {\bibinfo {title} {Dielectric constant with
  local field effects included},\ }\href@noop {} {\bibfield  {journal}
  {\bibinfo  {journal} {Physical Review}\ }\textbf {\bibinfo {volume} {129}},\
  \bibinfo {pages} {62} (\bibinfo {year} {1963})}\BibitemShut {NoStop}%
\bibitem [{\citenamefont {Anderson}(1965)}]{anderson1965}%
  \BibitemOpen
  \bibfield  {author} {\bibinfo {author} {\bibfnamefont {D.~G.}\ \bibnamefont
  {Anderson}},\ }\bibfield  {title} {\bibinfo {title} {Iterative procedures for
  nonlinear integral equations},\ }\href@noop {} {\bibfield  {journal}
  {\bibinfo  {journal} {Journal of the Association for Computing Machinery}\
  }\textbf {\bibinfo {volume} {12}},\ \bibinfo {pages} {547} (\bibinfo {year}
  {1965})}\BibitemShut {NoStop}%
\bibitem [{\citenamefont {Pulay}(1980)}]{PULAY1980393}%
  \BibitemOpen
  \bibfield  {author} {\bibinfo {author} {\bibfnamefont {P.}~\bibnamefont
  {Pulay}},\ }\bibfield  {title} {\bibinfo {title} {Convergence acceleration of
  iterative sequences. the case of scf iteration},\ }\href@noop {} {\bibfield
  {journal} {\bibinfo  {journal} {Chemical Physics Letters}\ }\textbf {\bibinfo
  {volume} {73}},\ \bibinfo {pages} {393} (\bibinfo {year} {1980})}\BibitemShut
  {NoStop}%
\bibitem [{\citenamefont {Johnson}(1988)}]{Broyden1988}%
  \BibitemOpen
  \bibfield  {author} {\bibinfo {author} {\bibfnamefont {D.~D.}\ \bibnamefont
  {Johnson}},\ }\bibfield  {title} {\bibinfo {title} {Modified {Broyden's}
  method for accelerating convergence in self-consistent calculations},\
  }\href@noop {} {\bibfield  {journal} {\bibinfo  {journal} {Physical Review
  B}\ }\textbf {\bibinfo {volume} {38}},\ \bibinfo {pages} {12807} (\bibinfo
  {year} {1988})}\BibitemShut {NoStop}%
\bibitem [{\citenamefont {Kudin}\ \emph {et~al.}(2002)\citenamefont {Kudin},
  \citenamefont {Scuseria},\ and\ \citenamefont {Cancès}}]{Kudin2002}%
  \BibitemOpen
  \bibfield  {author} {\bibinfo {author} {\bibfnamefont {K.~N.}\ \bibnamefont
  {Kudin}}, \bibinfo {author} {\bibfnamefont {G.~E.}\ \bibnamefont
  {Scuseria}},\ and\ \bibinfo {author} {\bibfnamefont {E.}~\bibnamefont
  {Cancès}},\ }\bibfield  {title} {\bibinfo {title} {A black-box
  self-consistent field convergence algorithm: One step closer},\ }\href@noop
  {} {\bibfield  {journal} {\bibinfo  {journal} {The Journal of Chemical
  Physics}\ }\textbf {\bibinfo {volume} {116}},\ \bibinfo {pages} {8255}
  (\bibinfo {year} {2002})}\BibitemShut {NoStop}%
\bibitem [{\citenamefont {Fang}\ and\ \citenamefont
  {Saad}(2009)}]{fang2009two}%
  \BibitemOpen
  \bibfield  {author} {\bibinfo {author} {\bibfnamefont {H.-r.}\ \bibnamefont
  {Fang}}\ and\ \bibinfo {author} {\bibfnamefont {Y.}~\bibnamefont {Saad}},\
  }\bibfield  {title} {\bibinfo {title} {Two classes of multisecant methods for
  nonlinear acceleration},\ }\href@noop {} {\bibfield  {journal} {\bibinfo
  {journal} {Numerical linear algebra with applications}\ }\textbf {\bibinfo
  {volume} {16}},\ \bibinfo {pages} {197} (\bibinfo {year} {2009})}\BibitemShut
  {NoStop}%
\bibitem [{\citenamefont {Dederichs}\ and\ \citenamefont
  {Zeller}(1983)}]{Dederichs1983}%
  \BibitemOpen
  \bibfield  {author} {\bibinfo {author} {\bibfnamefont {P.~H.}\ \bibnamefont
  {Dederichs}}\ and\ \bibinfo {author} {\bibfnamefont {R.}~\bibnamefont
  {Zeller}},\ }\bibfield  {title} {\bibinfo {title} {Self-consistency
  iterations in electronic-structure calculations},\ }\href@noop {} {\bibfield
  {journal} {\bibinfo  {journal} {Physical Review B}\ }\textbf {\bibinfo
  {volume} {28}},\ \bibinfo {pages} {5462} (\bibinfo {year}
  {1983})}\BibitemShut {NoStop}%
\bibitem [{\citenamefont {Saad}(2003)}]{saad2003iterative}%
  \BibitemOpen
  \bibfield  {author} {\bibinfo {author} {\bibfnamefont {Y.}~\bibnamefont
  {Saad}},\ }\href@noop {} {\emph {\bibinfo {title} {Iterative methods for
  sparse linear systems}}}\ (\bibinfo  {publisher} {SIAM},\ \bibinfo {year}
  {2003})\BibitemShut {NoStop}%
\bibitem [{\citenamefont {Lin}\ and\ \citenamefont {Yang}(2013)}]{LinLin2013}%
  \BibitemOpen
  \bibfield  {author} {\bibinfo {author} {\bibfnamefont {L.}~\bibnamefont
  {Lin}}\ and\ \bibinfo {author} {\bibfnamefont {C.}~\bibnamefont {Yang}},\
  }\bibfield  {title} {\bibinfo {title} {Elliptic preconditioner for
  accelerating the self-consistent field iteration in kohn--sham density
  functional theory},\ }\href@noop {} {\bibfield  {journal} {\bibinfo
  {journal} {SIAM Journal on Scientific Computing}\ }\textbf {\bibinfo {volume}
  {35}},\ \bibinfo {pages} {S277} (\bibinfo {year} {2013})}\BibitemShut
  {NoStop}%
\bibitem [{\citenamefont {Herbst}\ and\ \citenamefont
  {Levitt}(2020)}]{Herbst_2020}%
  \BibitemOpen
  \bibfield  {author} {\bibinfo {author} {\bibfnamefont {M.~F.}\ \bibnamefont
  {Herbst}}\ and\ \bibinfo {author} {\bibfnamefont {A.}~\bibnamefont
  {Levitt}},\ }\bibfield  {title} {\bibinfo {title} {Black-box inhomogeneous
  preconditioning for self-consistent field iterations in density functional
  theory},\ }\href@noop {} {\bibfield  {journal} {\bibinfo  {journal} {Journal
  of Physics: Condensed Matter}\ }\textbf {\bibinfo {volume} {33}},\ \bibinfo
  {pages} {085503} (\bibinfo {year} {2020})}\BibitemShut {NoStop}%
\bibitem [{\citenamefont {Kerker}(1981)}]{Kerker1981}%
  \BibitemOpen
  \bibfield  {author} {\bibinfo {author} {\bibfnamefont {G.~P.}\ \bibnamefont
  {Kerker}},\ }\bibfield  {title} {\bibinfo {title} {Efficient iteration scheme
  for self-consistent pseudopotential calculations},\ }\href@noop {} {\bibfield
   {journal} {\bibinfo  {journal} {Physical Review B}\ }\textbf {\bibinfo
  {volume} {23}},\ \bibinfo {pages} {3082} (\bibinfo {year}
  {1981})}\BibitemShut {NoStop}%
\bibitem [{\citenamefont {Resta}(1977)}]{Resta1977}%
  \BibitemOpen
  \bibfield  {author} {\bibinfo {author} {\bibfnamefont {R.}~\bibnamefont
  {Resta}},\ }\bibfield  {title} {\bibinfo {title} {{Thomas-Fermi} dielectric
  screening in semiconductors},\ }\href@noop {} {\bibfield  {journal} {\bibinfo
   {journal} {Physical Review B}\ }\textbf {\bibinfo {volume} {16}},\ \bibinfo
  {pages} {2717} (\bibinfo {year} {1977})}\BibitemShut {NoStop}%
\bibitem [{\citenamefont {Kresse}\ and\ \citenamefont
  {Furthm\"uller}(1996)}]{Kresse1996}%
  \BibitemOpen
  \bibfield  {author} {\bibinfo {author} {\bibfnamefont {G.}~\bibnamefont
  {Kresse}}\ and\ \bibinfo {author} {\bibfnamefont {J.}~\bibnamefont
  {Furthm\"uller}},\ }\bibfield  {title} {\bibinfo {title} {Efficient iterative
  schemes for ab initio total-energy calculations using a plane-wave basis
  set},\ }\href@noop {} {\bibfield  {journal} {\bibinfo  {journal} {Physical
  Review B}\ }\textbf {\bibinfo {volume} {54}},\ \bibinfo {pages} {11169}
  (\bibinfo {year} {1996})}\BibitemShut {NoStop}%
\bibitem [{\citenamefont {Woods}\ \emph {et~al.}(2019)\citenamefont {Woods},
  \citenamefont {Payne},\ and\ \citenamefont {Hasnip}}]{woods2019computing}%
  \BibitemOpen
  \bibfield  {author} {\bibinfo {author} {\bibfnamefont {N.~D.}\ \bibnamefont
  {Woods}}, \bibinfo {author} {\bibfnamefont {M.}~\bibnamefont {Payne}},\ and\
  \bibinfo {author} {\bibfnamefont {P.}~\bibnamefont {Hasnip}},\ }\bibfield
  {title} {\bibinfo {title} {Computing the self-consistent field in kohn--sham
  density functional theory},\ }\href@noop {} {\bibfield  {journal} {\bibinfo
  {journal} {Journal of Physics: Condensed Matter}\ }\textbf {\bibinfo {volume}
  {31}},\ \bibinfo {pages} {453001} (\bibinfo {year} {2019})}\BibitemShut
  {NoStop}%
\bibitem [{\citenamefont {Raczkowski}\ \emph {et~al.}(2001)\citenamefont
  {Raczkowski}, \citenamefont {Canning},\ and\ \citenamefont
  {Wang}}]{Raczkowski2011}%
  \BibitemOpen
  \bibfield  {author} {\bibinfo {author} {\bibfnamefont {D.}~\bibnamefont
  {Raczkowski}}, \bibinfo {author} {\bibfnamefont {A.}~\bibnamefont
  {Canning}},\ and\ \bibinfo {author} {\bibfnamefont {L.~W.}\ \bibnamefont
  {Wang}},\ }\bibfield  {title} {\bibinfo {title} {{Thomas-Fermi} charge mixing
  for obtaining self-consistency in density functional calculations},\
  }\href@noop {} {\bibfield  {journal} {\bibinfo  {journal} {Physical Review
  B}\ }\textbf {\bibinfo {volume} {64}},\ \bibinfo {pages} {121101} (\bibinfo
  {year} {2001})}\BibitemShut {NoStop}%
\bibitem [{\citenamefont {Marzari}\ \emph {et~al.}(1997)\citenamefont
  {Marzari}, \citenamefont {Vanderbilt},\ and\ \citenamefont
  {Payne}}]{marzari1997ensemble}%
  \BibitemOpen
  \bibfield  {author} {\bibinfo {author} {\bibfnamefont {N.}~\bibnamefont
  {Marzari}}, \bibinfo {author} {\bibfnamefont {D.}~\bibnamefont
  {Vanderbilt}},\ and\ \bibinfo {author} {\bibfnamefont {M.~C.}\ \bibnamefont
  {Payne}},\ }\bibfield  {title} {\bibinfo {title} {Ensemble density-functional
  theory for ab initio molecular dynamics of metals and finite-temperature
  insulators},\ }\href@noop {} {\bibfield  {journal} {\bibinfo  {journal}
  {Physical review letters}\ }\textbf {\bibinfo {volume} {79}},\ \bibinfo
  {pages} {1337} (\bibinfo {year} {1997})}\BibitemShut {NoStop}%
\bibitem [{\citenamefont {Niklasson}(2020)}]{niklasson2020krylov}%
  \BibitemOpen
  \bibfield  {author} {\bibinfo {author} {\bibfnamefont {A.~M.~N.}\
  \bibnamefont {Niklasson}},\ }\bibfield  {title} {\bibinfo {title} {{Extended
  Lagrangian Born–Oppenheimer molecular dynamics using a Krylov subspace
  approximation}},\ }\href@noop {} {\bibfield  {journal} {\bibinfo  {journal}
  {The Journal of Chemical Physics}\ }\textbf {\bibinfo {volume} {152}},\
  \bibinfo {pages} {104103} (\bibinfo {year} {2020})}\BibitemShut {NoStop}%
\bibitem [{\citenamefont {Motamarri}\ \emph {et~al.}(2013)\citenamefont
  {Motamarri}, \citenamefont {Nowak}, \citenamefont {Leiter}, \citenamefont
  {Knap},\ and\ \citenamefont {Gavini}}]{motamarri2013}%
  \BibitemOpen
  \bibfield  {author} {\bibinfo {author} {\bibfnamefont {P.}~\bibnamefont
  {Motamarri}}, \bibinfo {author} {\bibfnamefont {M.}~\bibnamefont {Nowak}},
  \bibinfo {author} {\bibfnamefont {K.}~\bibnamefont {Leiter}}, \bibinfo
  {author} {\bibfnamefont {J.}~\bibnamefont {Knap}},\ and\ \bibinfo {author}
  {\bibfnamefont {V.}~\bibnamefont {Gavini}},\ }\bibfield  {title} {\bibinfo
  {title} {Higher-order adaptive finite-element methods for {K}ohn-{S}ham
  density functional theory},\ }\href@noop {} {\bibfield  {journal} {\bibinfo
  {journal} {Journal of Computational Physics}\ }\textbf {\bibinfo {volume}
  {253}},\ \bibinfo {pages} {308} (\bibinfo {year} {2013})}\BibitemShut
  {NoStop}%
\bibitem [{\citenamefont {Andrade}\ \emph {et~al.}(2015)\citenamefont
  {Andrade}, \citenamefont {Strubbe}, \citenamefont {De~Giovannini},
  \citenamefont {Larsen}, \citenamefont {Oliveira}, \citenamefont
  {Alberdi-Rodriguez}, \citenamefont {Varas}, \citenamefont {Theophilou},
  \citenamefont {Helbig}, \citenamefont {Verstraete}, \citenamefont {Stella},
  \citenamefont {Nogueira}, \citenamefont {Aspuru-Guzik}, \citenamefont
  {Castro}, \citenamefont {Marques},\ and\ \citenamefont
  {Rubio}}]{octopus2015}%
  \BibitemOpen
  \bibfield  {author} {\bibinfo {author} {\bibfnamefont {X.}~\bibnamefont
  {Andrade}}, \bibinfo {author} {\bibfnamefont {D.}~\bibnamefont {Strubbe}},
  \bibinfo {author} {\bibfnamefont {U.}~\bibnamefont {De~Giovannini}}, \bibinfo
  {author} {\bibfnamefont {A.~H.}\ \bibnamefont {Larsen}}, \bibinfo {author}
  {\bibfnamefont {M.~J.~T.}\ \bibnamefont {Oliveira}}, \bibinfo {author}
  {\bibfnamefont {J.}~\bibnamefont {Alberdi-Rodriguez}}, \bibinfo {author}
  {\bibfnamefont {A.}~\bibnamefont {Varas}}, \bibinfo {author} {\bibfnamefont
  {I.}~\bibnamefont {Theophilou}}, \bibinfo {author} {\bibfnamefont
  {N.}~\bibnamefont {Helbig}}, \bibinfo {author} {\bibfnamefont {M.~J.}\
  \bibnamefont {Verstraete}}, \bibinfo {author} {\bibfnamefont
  {L.}~\bibnamefont {Stella}}, \bibinfo {author} {\bibfnamefont
  {F.}~\bibnamefont {Nogueira}}, \bibinfo {author} {\bibfnamefont
  {A.}~\bibnamefont {Aspuru-Guzik}}, \bibinfo {author} {\bibfnamefont
  {A.}~\bibnamefont {Castro}}, \bibinfo {author} {\bibfnamefont {M.~A.~L.}\
  \bibnamefont {Marques}},\ and\ \bibinfo {author} {\bibfnamefont
  {A.}~\bibnamefont {Rubio}},\ }\bibfield  {title} {\bibinfo {title}
  {Real-space grids and the {O}ctopus code as tools for the development of new
  simulation approaches for electronic systems},\ }\href@noop {} {\bibfield
  {journal} {\bibinfo  {journal} {Physical Chemistry Chemical Physics}\
  }\textbf {\bibinfo {volume} {17}},\ \bibinfo {pages} {31371} (\bibinfo {year}
  {2015})}\BibitemShut {NoStop}%
\bibitem [{\citenamefont {Michaud-Rioux}\ \emph {et~al.}(2016)\citenamefont
  {Michaud-Rioux}, \citenamefont {Zhang},\ and\ \citenamefont
  {Guo}}]{michaud2016rescu}%
  \BibitemOpen
  \bibfield  {author} {\bibinfo {author} {\bibfnamefont {V.}~\bibnamefont
  {Michaud-Rioux}}, \bibinfo {author} {\bibfnamefont {L.}~\bibnamefont
  {Zhang}},\ and\ \bibinfo {author} {\bibfnamefont {H.}~\bibnamefont {Guo}},\
  }\bibfield  {title} {\bibinfo {title} {Rescu: A real space electronic
  structure method},\ }\href@noop {} {\bibfield  {journal} {\bibinfo  {journal}
  {Journal of Computational Physics}\ }\textbf {\bibinfo {volume} {307}},\
  \bibinfo {pages} {593} (\bibinfo {year} {2016})}\BibitemShut {NoStop}%
\bibitem [{\citenamefont {Hitchcock}(1927)}]{Hitchcock1927}%
  \BibitemOpen
  \bibfield  {author} {\bibinfo {author} {\bibfnamefont {F.~L.}\ \bibnamefont
  {Hitchcock}},\ }\bibfield  {title} {\bibinfo {title} {The expression of a
  tensor or a polyadic as a sum of products},\ }\href@noop {} {\bibfield
  {journal} {\bibinfo  {journal} {Journal of Mathematics and Physics}\ }\textbf
  {\bibinfo {volume} {6}},\ \bibinfo {pages} {164} (\bibinfo {year}
  {1927})}\BibitemShut {NoStop}%
\bibitem [{\citenamefont {Niklasson}\ \emph {et~al.}(2015)\citenamefont
  {Niklasson}, \citenamefont {Cawkwell}, \citenamefont {Rubensson},\ and\
  \citenamefont {Rudberg}}]{niklasson2015canonical}%
  \BibitemOpen
  \bibfield  {author} {\bibinfo {author} {\bibfnamefont {A.~M.}\ \bibnamefont
  {Niklasson}}, \bibinfo {author} {\bibfnamefont {M.~J.}\ \bibnamefont
  {Cawkwell}}, \bibinfo {author} {\bibfnamefont {E.~H.}\ \bibnamefont
  {Rubensson}},\ and\ \bibinfo {author} {\bibfnamefont {E.}~\bibnamefont
  {Rudberg}},\ }\bibfield  {title} {\bibinfo {title} {Canonical density matrix
  perturbation theory},\ }\href@noop {} {\bibfield  {journal} {\bibinfo
  {journal} {Physical Review E}\ }\textbf {\bibinfo {volume} {92}},\ \bibinfo
  {pages} {063301} (\bibinfo {year} {2015})}\BibitemShut {NoStop}%
\bibitem [{\citenamefont {Martin}(2004)}]{rmartin}%
  \BibitemOpen
  \bibfield  {author} {\bibinfo {author} {\bibfnamefont {R.~M.}\ \bibnamefont
  {Martin}},\ }\href@noop {} {\emph {\bibinfo {title} {Electronic structure:
  basic theory and practical methods}}}\ (\bibinfo  {publisher} {Cambridge
  university press},\ \bibinfo {address} {Cambridge, UK},\ \bibinfo {year}
  {2004})\BibitemShut {NoStop}%
\bibitem [{\citenamefont {Baroni}\ \emph {et~al.}(2001)\citenamefont {Baroni},
  \citenamefont {de~Gironcoli}, \citenamefont {Dal~Corso},\ and\ \citenamefont
  {Giannozzi}}]{dfpt2001}%
  \BibitemOpen
  \bibfield  {author} {\bibinfo {author} {\bibfnamefont {S.}~\bibnamefont
  {Baroni}}, \bibinfo {author} {\bibfnamefont {S.}~\bibnamefont
  {de~Gironcoli}}, \bibinfo {author} {\bibfnamefont {A.}~\bibnamefont
  {Dal~Corso}},\ and\ \bibinfo {author} {\bibfnamefont {P.}~\bibnamefont
  {Giannozzi}},\ }\bibfield  {title} {\bibinfo {title} {Phonons and related
  crystal properties from density-functional perturbation theory},\ }\href@noop
  {} {\bibfield  {journal} {\bibinfo  {journal} {Reviews of Modern Physics}\
  }\textbf {\bibinfo {volume} {73}},\ \bibinfo {pages} {515} (\bibinfo {year}
  {2001})}\BibitemShut {NoStop}%
\bibitem [{\citenamefont {Methfessel}\ and\ \citenamefont
  {Paxton}(1989)}]{MethfesselPaxton}%
  \BibitemOpen
  \bibfield  {author} {\bibinfo {author} {\bibfnamefont {M.}~\bibnamefont
  {Methfessel}}\ and\ \bibinfo {author} {\bibfnamefont {A.~T.}\ \bibnamefont
  {Paxton}},\ }\bibfield  {title} {\bibinfo {title} {High-precision sampling
  for brillouin-zone integration in metals},\ }\href@noop {} {\bibfield
  {journal} {\bibinfo  {journal} {Phys. Rev. B}\ }\textbf {\bibinfo {volume}
  {40}},\ \bibinfo {pages} {3616} (\bibinfo {year} {1989})}\BibitemShut
  {NoStop}%
\bibitem [{\citenamefont {Marzari}\ \emph {et~al.}(1999)\citenamefont
  {Marzari}, \citenamefont {Vanderbilt}, \citenamefont {De~Vita},\ and\
  \citenamefont {Payne}}]{marzari1999thermal}%
  \BibitemOpen
  \bibfield  {author} {\bibinfo {author} {\bibfnamefont {N.}~\bibnamefont
  {Marzari}}, \bibinfo {author} {\bibfnamefont {D.}~\bibnamefont {Vanderbilt}},
  \bibinfo {author} {\bibfnamefont {A.}~\bibnamefont {De~Vita}},\ and\ \bibinfo
  {author} {\bibfnamefont {M.}~\bibnamefont {Payne}},\ }\bibfield  {title}
  {\bibinfo {title} {Thermal contraction and disordering of the al (110)
  surface},\ }\href@noop {} {\bibfield  {journal} {\bibinfo  {journal}
  {Physical review letters}\ }\textbf {\bibinfo {volume} {82}},\ \bibinfo
  {pages} {3296} (\bibinfo {year} {1999})}\BibitemShut {NoStop}%
\bibitem [{\citenamefont {Liang}\ \emph {et~al.}(2003)\citenamefont {Liang},
  \citenamefont {Saravanan}, \citenamefont {Shao}, \citenamefont {Baer},
  \citenamefont {Bell},\ and\ \citenamefont {Head-Gordon}}]{liang2003improved}%
  \BibitemOpen
  \bibfield  {author} {\bibinfo {author} {\bibfnamefont {W.}~\bibnamefont
  {Liang}}, \bibinfo {author} {\bibfnamefont {C.}~\bibnamefont {Saravanan}},
  \bibinfo {author} {\bibfnamefont {Y.}~\bibnamefont {Shao}}, \bibinfo {author}
  {\bibfnamefont {R.}~\bibnamefont {Baer}}, \bibinfo {author} {\bibfnamefont
  {A.~T.}\ \bibnamefont {Bell}},\ and\ \bibinfo {author} {\bibfnamefont
  {M.}~\bibnamefont {Head-Gordon}},\ }\bibfield  {title} {\bibinfo {title}
  {Improved fermi operator expansion methods for fast electronic structure
  calculations},\ }\href@noop {} {\bibfield  {journal} {\bibinfo  {journal}
  {The Journal of chemical physics}\ }\textbf {\bibinfo {volume} {119}},\
  \bibinfo {pages} {4117} (\bibinfo {year} {2003})}\BibitemShut {NoStop}%
\bibitem [{\citenamefont {Perdew}\ \emph {et~al.}(1996)\citenamefont {Perdew},
  \citenamefont {Burke},\ and\ \citenamefont {Ernzerhof}}]{pbe}%
  \BibitemOpen
  \bibfield  {author} {\bibinfo {author} {\bibfnamefont {J.~P.}\ \bibnamefont
  {Perdew}}, \bibinfo {author} {\bibfnamefont {K.}~\bibnamefont {Burke}},\ and\
  \bibinfo {author} {\bibfnamefont {M.}~\bibnamefont {Ernzerhof}},\ }\bibfield
  {title} {\bibinfo {title} {{Generalized Gradient Approximation} made
  simple},\ }\href@noop {} {\bibfield  {journal} {\bibinfo  {journal} {Physical
  Review Letters}\ }\textbf {\bibinfo {volume} {77}},\ \bibinfo {pages} {3865}
  (\bibinfo {year} {1996})}\BibitemShut {NoStop}%
\bibitem [{\citenamefont {Hamann}(2013)}]{ONCV2013}%
  \BibitemOpen
  \bibfield  {author} {\bibinfo {author} {\bibfnamefont {D.~R.}\ \bibnamefont
  {Hamann}},\ }\bibfield  {title} {\bibinfo {title} {Optimized norm-conserving
  {V}anderbilt pseudopotentials},\ }\href@noop {} {\bibfield  {journal}
  {\bibinfo  {journal} {Physical Review B}\ }\textbf {\bibinfo {volume} {88}},\
  \bibinfo {pages} {085117} (\bibinfo {year} {2013})}\BibitemShut {NoStop}%
\bibitem [{\citenamefont {van Setten}\ \emph {et~al.}(2018)\citenamefont {van
  Setten}, \citenamefont {Giantomassi}, \citenamefont {Bousquet}, \citenamefont
  {Verstraete}, \citenamefont {Hamann}, \citenamefont {Gonze},\ and\
  \citenamefont {Rignanese}}]{van2018pseudodojo}%
  \BibitemOpen
  \bibfield  {author} {\bibinfo {author} {\bibfnamefont {M.~J.}\ \bibnamefont
  {van Setten}}, \bibinfo {author} {\bibfnamefont {M.}~\bibnamefont
  {Giantomassi}}, \bibinfo {author} {\bibfnamefont {E.}~\bibnamefont
  {Bousquet}}, \bibinfo {author} {\bibfnamefont {M.~J.}\ \bibnamefont
  {Verstraete}}, \bibinfo {author} {\bibfnamefont {D.~R.}\ \bibnamefont
  {Hamann}}, \bibinfo {author} {\bibfnamefont {X.}~\bibnamefont {Gonze}},\ and\
  \bibinfo {author} {\bibfnamefont {G.-M.}\ \bibnamefont {Rignanese}},\
  }\bibfield  {title} {\bibinfo {title} {The {PseudoDojo}: Training and grading
  a 85 element optimized norm-conserving pseudopotential table},\ }\href@noop
  {} {\bibfield  {journal} {\bibinfo  {journal} {Computer Physics
  Communications}\ }\textbf {\bibinfo {volume} {226}},\ \bibinfo {pages} {39}
  (\bibinfo {year} {2018})}\BibitemShut {NoStop}%
\bibitem [{\citenamefont {Giannozzi}\ \emph {et~al.}(2009)\citenamefont
  {Giannozzi}, \citenamefont {Baroni}, \citenamefont {Bonini}, \citenamefont
  {Calandra}, \citenamefont {Car}, \citenamefont {Cavazzoni}, \citenamefont
  {Ceresoli}, \citenamefont {Chiarotti}, \citenamefont {Cococcioni},
  \citenamefont {Dabo}, \citenamefont {{Dal Corso}}, \citenamefont
  {de~Gironcoli}, \citenamefont {Fabris}, \citenamefont {Fratesi},
  \citenamefont {Gebauer}, \citenamefont {Gerstmann}, \citenamefont
  {Gougoussis}, \citenamefont {Kokalj}, \citenamefont {Lazzeri}, \citenamefont
  {Martin-Samos}, \citenamefont {Marzari}, \citenamefont {Mauri}, \citenamefont
  {Mazzarello}, \citenamefont {Paolini}, \citenamefont {Pasquarello},
  \citenamefont {Paulatto}, \citenamefont {Sbraccia}, \citenamefont {Scandolo},
  \citenamefont {Sclauzero}, \citenamefont {Seitsonen}, \citenamefont
  {Smogunov}, \citenamefont {Umari},\ and\ \citenamefont
  {Wentzcovitch}}]{qe2009}%
  \BibitemOpen
  \bibfield  {author} {\bibinfo {author} {\bibfnamefont {P.}~\bibnamefont
  {Giannozzi}}, \bibinfo {author} {\bibfnamefont {S.}~\bibnamefont {Baroni}},
  \bibinfo {author} {\bibfnamefont {N.}~\bibnamefont {Bonini}}, \bibinfo
  {author} {\bibfnamefont {M.}~\bibnamefont {Calandra}}, \bibinfo {author}
  {\bibfnamefont {R.}~\bibnamefont {Car}}, \bibinfo {author} {\bibfnamefont
  {C.}~\bibnamefont {Cavazzoni}}, \bibinfo {author} {\bibfnamefont
  {D.}~\bibnamefont {Ceresoli}}, \bibinfo {author} {\bibfnamefont {G.~L.}\
  \bibnamefont {Chiarotti}}, \bibinfo {author} {\bibfnamefont {M.}~\bibnamefont
  {Cococcioni}}, \bibinfo {author} {\bibfnamefont {I.}~\bibnamefont {Dabo}},
  \bibinfo {author} {\bibfnamefont {A.}~\bibnamefont {{Dal Corso}}}, \bibinfo
  {author} {\bibfnamefont {S.}~\bibnamefont {de~Gironcoli}}, \bibinfo {author}
  {\bibfnamefont {S.}~\bibnamefont {Fabris}}, \bibinfo {author} {\bibfnamefont
  {G.}~\bibnamefont {Fratesi}}, \bibinfo {author} {\bibfnamefont
  {R.}~\bibnamefont {Gebauer}}, \bibinfo {author} {\bibfnamefont
  {U.}~\bibnamefont {Gerstmann}}, \bibinfo {author} {\bibfnamefont
  {C.}~\bibnamefont {Gougoussis}}, \bibinfo {author} {\bibfnamefont
  {A.}~\bibnamefont {Kokalj}}, \bibinfo {author} {\bibfnamefont
  {M.}~\bibnamefont {Lazzeri}}, \bibinfo {author} {\bibfnamefont
  {L.}~\bibnamefont {Martin-Samos}}, \bibinfo {author} {\bibfnamefont
  {N.}~\bibnamefont {Marzari}}, \bibinfo {author} {\bibfnamefont
  {F.}~\bibnamefont {Mauri}}, \bibinfo {author} {\bibfnamefont
  {R.}~\bibnamefont {Mazzarello}}, \bibinfo {author} {\bibfnamefont
  {S.}~\bibnamefont {Paolini}}, \bibinfo {author} {\bibfnamefont
  {A.}~\bibnamefont {Pasquarello}}, \bibinfo {author} {\bibfnamefont
  {L.}~\bibnamefont {Paulatto}}, \bibinfo {author} {\bibfnamefont
  {C.}~\bibnamefont {Sbraccia}}, \bibinfo {author} {\bibfnamefont
  {S.}~\bibnamefont {Scandolo}}, \bibinfo {author} {\bibfnamefont
  {G.}~\bibnamefont {Sclauzero}}, \bibinfo {author} {\bibfnamefont {A.~P.}\
  \bibnamefont {Seitsonen}}, \bibinfo {author} {\bibfnamefont {A.}~\bibnamefont
  {Smogunov}}, \bibinfo {author} {\bibfnamefont {P.}~\bibnamefont {Umari}},\
  and\ \bibinfo {author} {\bibfnamefont {R.~M.}\ \bibnamefont {Wentzcovitch}},\
  }\bibfield  {title} {\bibinfo {title} {{QUANTUM ESPRESSO}: a modular and
  open-source software project for quantum simulations of materials},\
  }\href@noop {} {\bibfield  {journal} {\bibinfo  {journal} {Journal of
  Physics: Condensed Matter}\ }\textbf {\bibinfo {volume} {21}},\ \bibinfo
  {pages} {395502} (\bibinfo {year} {2009})}\BibitemShut {NoStop}%
\bibitem [{\citenamefont {Giannozzi}\ \emph {et~al.}(2017)\citenamefont
  {Giannozzi}, \citenamefont {Andreussi}, \citenamefont {Brumme}, \citenamefont
  {Bunau}, \citenamefont {Nardelli}, \citenamefont {Calandra}, \citenamefont
  {Car}, \citenamefont {Cavazzoni}, \citenamefont {Ceresoli}, \citenamefont
  {Cococcioni}, \citenamefont {Colonna}, \citenamefont {Carnimeo},
  \citenamefont {Corso}, \citenamefont {de~Gironcoli}, \citenamefont {Delugas},
  \citenamefont {Jr}, \citenamefont {Ferretti}, \citenamefont {Floris},
  \citenamefont {Fratesi}, \citenamefont {Fugallo}, \citenamefont {Gebauer},
  \citenamefont {Gerstmann}, \citenamefont {Giustino}, \citenamefont {Gorni},
  \citenamefont {Jia}, \citenamefont {Kawamura}, \citenamefont {Ko},
  \citenamefont {Kokalj}, \citenamefont {Küçükbenli}, \citenamefont
  {Lazzeri}, \citenamefont {Marsili}, \citenamefont {Marzari}, \citenamefont
  {Mauri}, \citenamefont {Nguyen}, \citenamefont {Nguyen}, \citenamefont {de-la
  Roza}, \citenamefont {Paulatto}, \citenamefont {Poncé}, \citenamefont
  {Rocca}, \citenamefont {Sabatini}, \citenamefont {Santra}, \citenamefont
  {Schlipf}, \citenamefont {Seitsonen}, \citenamefont {Smogunov}, \citenamefont
  {Timrov}, \citenamefont {Thonhauser}, \citenamefont {Umari}, \citenamefont
  {Vast}, \citenamefont {Wu},\ and\ \citenamefont {Baroni}}]{qe2017}%
  \BibitemOpen
  \bibfield  {author} {\bibinfo {author} {\bibfnamefont {P.}~\bibnamefont
  {Giannozzi}}, \bibinfo {author} {\bibfnamefont {O.}~\bibnamefont
  {Andreussi}}, \bibinfo {author} {\bibfnamefont {T.}~\bibnamefont {Brumme}},
  \bibinfo {author} {\bibfnamefont {O.}~\bibnamefont {Bunau}}, \bibinfo
  {author} {\bibfnamefont {M.~B.}\ \bibnamefont {Nardelli}}, \bibinfo {author}
  {\bibfnamefont {M.}~\bibnamefont {Calandra}}, \bibinfo {author}
  {\bibfnamefont {R.}~\bibnamefont {Car}}, \bibinfo {author} {\bibfnamefont
  {C.}~\bibnamefont {Cavazzoni}}, \bibinfo {author} {\bibfnamefont
  {D.}~\bibnamefont {Ceresoli}}, \bibinfo {author} {\bibfnamefont
  {M.}~\bibnamefont {Cococcioni}}, \bibinfo {author} {\bibfnamefont
  {N.}~\bibnamefont {Colonna}}, \bibinfo {author} {\bibfnamefont
  {I.}~\bibnamefont {Carnimeo}}, \bibinfo {author} {\bibfnamefont {A.~D.}\
  \bibnamefont {Corso}}, \bibinfo {author} {\bibfnamefont {S.}~\bibnamefont
  {de~Gironcoli}}, \bibinfo {author} {\bibfnamefont {P.}~\bibnamefont
  {Delugas}}, \bibinfo {author} {\bibfnamefont {R.~A.~D.}\ \bibnamefont {Jr}},
  \bibinfo {author} {\bibfnamefont {A.}~\bibnamefont {Ferretti}}, \bibinfo
  {author} {\bibfnamefont {A.}~\bibnamefont {Floris}}, \bibinfo {author}
  {\bibfnamefont {G.}~\bibnamefont {Fratesi}}, \bibinfo {author} {\bibfnamefont
  {G.}~\bibnamefont {Fugallo}}, \bibinfo {author} {\bibfnamefont
  {R.}~\bibnamefont {Gebauer}}, \bibinfo {author} {\bibfnamefont
  {U.}~\bibnamefont {Gerstmann}}, \bibinfo {author} {\bibfnamefont
  {F.}~\bibnamefont {Giustino}}, \bibinfo {author} {\bibfnamefont
  {T.}~\bibnamefont {Gorni}}, \bibinfo {author} {\bibfnamefont
  {J.}~\bibnamefont {Jia}}, \bibinfo {author} {\bibfnamefont {M.}~\bibnamefont
  {Kawamura}}, \bibinfo {author} {\bibfnamefont {H.-Y.}\ \bibnamefont {Ko}},
  \bibinfo {author} {\bibfnamefont {A.}~\bibnamefont {Kokalj}}, \bibinfo
  {author} {\bibfnamefont {E.}~\bibnamefont {Küçükbenli}}, \bibinfo {author}
  {\bibfnamefont {M.}~\bibnamefont {Lazzeri}}, \bibinfo {author} {\bibfnamefont
  {M.}~\bibnamefont {Marsili}}, \bibinfo {author} {\bibfnamefont
  {N.}~\bibnamefont {Marzari}}, \bibinfo {author} {\bibfnamefont
  {F.}~\bibnamefont {Mauri}}, \bibinfo {author} {\bibfnamefont {N.~L.}\
  \bibnamefont {Nguyen}}, \bibinfo {author} {\bibfnamefont {H.-V.}\
  \bibnamefont {Nguyen}}, \bibinfo {author} {\bibfnamefont {A.~O.}\
  \bibnamefont {de-la Roza}}, \bibinfo {author} {\bibfnamefont
  {L.}~\bibnamefont {Paulatto}}, \bibinfo {author} {\bibfnamefont
  {S.}~\bibnamefont {Poncé}}, \bibinfo {author} {\bibfnamefont
  {D.}~\bibnamefont {Rocca}}, \bibinfo {author} {\bibfnamefont
  {R.}~\bibnamefont {Sabatini}}, \bibinfo {author} {\bibfnamefont
  {B.}~\bibnamefont {Santra}}, \bibinfo {author} {\bibfnamefont
  {M.}~\bibnamefont {Schlipf}}, \bibinfo {author} {\bibfnamefont {A.~P.}\
  \bibnamefont {Seitsonen}}, \bibinfo {author} {\bibfnamefont {A.}~\bibnamefont
  {Smogunov}}, \bibinfo {author} {\bibfnamefont {I.}~\bibnamefont {Timrov}},
  \bibinfo {author} {\bibfnamefont {T.}~\bibnamefont {Thonhauser}}, \bibinfo
  {author} {\bibfnamefont {P.}~\bibnamefont {Umari}}, \bibinfo {author}
  {\bibfnamefont {N.}~\bibnamefont {Vast}}, \bibinfo {author} {\bibfnamefont
  {X.}~\bibnamefont {Wu}},\ and\ \bibinfo {author} {\bibfnamefont
  {S.}~\bibnamefont {Baroni}},\ }\bibfield  {title} {\bibinfo {title} {Advanced
  capabilities for materials modelling with {QUANTUM ESPRESSO}},\ }\href@noop
  {} {\bibfield  {journal} {\bibinfo  {journal} {Journal of Physics: Condensed
  Matter}\ }\textbf {\bibinfo {volume} {29}},\ \bibinfo {pages} {465901}
  (\bibinfo {year} {2017})}\BibitemShut {NoStop}%
\bibitem [{\citenamefont {Das}\ and\ \citenamefont
  {Gavini}(2023)}]{supplementary}%
  \BibitemOpen
  \bibfield  {author} {\bibinfo {author} {\bibfnamefont {S.}~\bibnamefont
  {Das}}\ and\ \bibinfo {author} {\bibfnamefont {V.}~\bibnamefont {Gavini}},\
  }\href@noop {} {\bibinfo {title} {Supplemental material: Accelerating
  self-consistent field iterations in kohn-sham density functional theory using
  a low rank approximation of the dielectric matrix}} (\bibinfo {year}
  {2023})\BibitemShut {NoStop}%
\bibitem [{\citenamefont {Kuhn}\ \emph {et~al.}(2013)\citenamefont {Kuhn},
  \citenamefont {K{\"o}hler},\ and\ \citenamefont {Lotsch}}]{kuhn2013single}%
  \BibitemOpen
  \bibfield  {author} {\bibinfo {author} {\bibfnamefont {A.}~\bibnamefont
  {Kuhn}}, \bibinfo {author} {\bibfnamefont {J.}~\bibnamefont {K{\"o}hler}},\
  and\ \bibinfo {author} {\bibfnamefont {B.~V.}\ \bibnamefont {Lotsch}},\
  }\bibfield  {title} {\bibinfo {title} {Single-crystal {X-ray} structure
  analysis of the superionic conductor {Li10GeP2S12}},\ }\href@noop {}
  {\bibfield  {journal} {\bibinfo  {journal} {Physical Chemistry Chemical
  Physics}\ }\textbf {\bibinfo {volume} {15}},\ \bibinfo {pages} {11620}
  (\bibinfo {year} {2013})}\BibitemShut {NoStop}%
\bibitem [{\citenamefont {Herbst}(2020)}]{ldosgithub}%
  \BibitemOpen
  \bibfield  {author} {\bibinfo {author} {\bibfnamefont {M.~F.}\ \bibnamefont
  {Herbst}},\ }\href@noop {} {\bibinfo {title} {Black-box inhomogeneous
  preconditioning for self-consistent field iterations in density functional
  theory}},\ \bibinfo {howpublished}
  {\url{https://github.com/mfherbst/supporting-ldos-preconditioning}} (\bibinfo
  {year} {2020})\BibitemShut {NoStop}%
\bibitem [{\citenamefont {Shen}\ \emph {et~al.}(2018)\citenamefont {Shen},
  \citenamefont {Zhang}, \citenamefont {Zhang}, \citenamefont {Dai},
  \citenamefont {Zhang}, \citenamefont {Ge}, \citenamefont {Pan}, \citenamefont
  {Sharkey}, \citenamefont {Graham}, \citenamefont {Hunt} \emph
  {et~al.}}]{shen2018deconvolution}%
  \BibitemOpen
  \bibfield  {author} {\bibinfo {author} {\bibfnamefont {X.}~\bibnamefont
  {Shen}}, \bibinfo {author} {\bibfnamefont {C.}~\bibnamefont {Zhang}},
  \bibinfo {author} {\bibfnamefont {S.}~\bibnamefont {Zhang}}, \bibinfo
  {author} {\bibfnamefont {S.}~\bibnamefont {Dai}}, \bibinfo {author}
  {\bibfnamefont {G.}~\bibnamefont {Zhang}}, \bibinfo {author} {\bibfnamefont
  {M.}~\bibnamefont {Ge}}, \bibinfo {author} {\bibfnamefont {Y.}~\bibnamefont
  {Pan}}, \bibinfo {author} {\bibfnamefont {S.~M.}\ \bibnamefont {Sharkey}},
  \bibinfo {author} {\bibfnamefont {G.~W.}\ \bibnamefont {Graham}}, \bibinfo
  {author} {\bibfnamefont {A.}~\bibnamefont {Hunt}}, \emph {et~al.},\
  }\bibfield  {title} {\bibinfo {title} {Deconvolution of octahedral
  {${\textrm{Pt}}_3\textrm{Ni}$} nanoparticle growth pathway from in situ
  characterizations},\ }\href@noop {} {\bibfield  {journal} {\bibinfo
  {journal} {Nature Communications}\ }\textbf {\bibinfo {volume} {9}},\
  \bibinfo {pages} {1} (\bibinfo {year} {2018})}\BibitemShut {NoStop}%
\end{thebibliography}%
%

\end{document}